\documentclass[11pt,a4paper]{article}
\usepackage{jheppub}
\usepackage{appendix}
\usepackage{ifthen} 
\usepackage{overpic} %
\usepackage[utf8]{inputenc}

\newboolean{pdflatex}
\setboolean{pdflatex}{true} 

\newboolean{articletitles}
\setboolean{articletitles}{true} 

\newboolean{uprightparticles}
\setboolean{uprightparticles}{false} 

\title{A complete description of P- and S-wave contributions to the \texorpdfstring{\BdKpillbold}{B->Kpill} decay}

\author[1,2]{Marcel Alguer\'o,}
\author[3]{Paula Alvarez Cartelle,}
\author[4,5]{Alexander Mclean Marshall,}
\author[1,2]{Pere Masjuan,}
\author[1,2]{Joaquim Matias,}
\author[4]{Michael Andrew McCann,}
\author[4]{Mitesh Patel,}
\author[5]{Konstantinos A. Petridis,}
\author[4]{Mark Smith}

\affiliation[1]{Universitat Aut\`onoma de Barcelona, E-08193 Bellaterra (Barcelona), Spain}
\affiliation[2]{
Institut de Física d’Altes Energies and The Barcelona Institute of Science and Technology, Spain.
}
\affiliation[3]{Cavendish Laboratory, Cambridge, UK}
\affiliation[4]{Imperial College London, London, UK}
\affiliation[5]{University of Bristol, Bristol, UK}

\abstract{
In this paper we present a detailed study of the four-body decay \BdKpill, where tensions with the Standard Model predictions have been observed. Our analysis of the decay with P- and S-wave contributions to the $\Kp\pim$ system develops a complete understanding of the symmetries of the distribution, in the case of massless and massive leptons. In both cases, the symmetries determine relations between the observables in the \BdKpill decay distribution. This enables us to define the complete set of observables accessible to experiments, including several that have not previously been identified. The new observables arise when the decay rate is written differentially with respect to $m_{K\pi}$.
  We demonstrate that experiments will be able to fit this full decay distribution with currently available data sets and investigate the sensitivity to new physics scenarios given the experimental precision that is expected in the future. 

 The symmetry relations provide a unique handle to explore the behaviour of S-wave observables by expressing them in terms of P-wave observables, therefore minimising the dependence on poorly-known S-wave form factors. Using this approach, we construct two theoretically clean S-wave observables and explore their sensitivity to new physics. By further exploiting the symmetry relations, we obtain the first bounds on the S-wave observables using two different methods and highlight how these relations may be used as cross-checks of the experimental methodology. We identify a zero-crossing point that would be at a common dilepton invariant mass for a subset of P- and S-wave observables, and explore the information on new physics and hadronic effects 
 that this zero point can provide. 
}

\usepackage{microtype}
\usepackage{lineno}  
\usepackage{xspace} 

\usepackage{graphicx}  
\usepackage{color}
\usepackage{colortbl}
\graphicspath{{./figs/}} 
\DeclareGraphicsExtensions{.pdf,.PDF,png,.PNG}

\usepackage{amsmath} 
\usepackage{amssymb}
\usepackage{amsfonts}
\usepackage{upgreek} 

\newcommand*\patchAmsMathEnvironmentForLineno[1]{%
\expandafter\let\csname old#1\expandafter\endcsname\csname #1\endcsname
\expandafter\let\csname oldend#1\expandafter\endcsname\csname
end#1\endcsname
 \renewenvironment{#1}%
   {\linenomath\csname old#1\endcsname}%
   {\csname oldend#1\endcsname\endlinenomath}%
}
\newcommand*\patchBothAmsMathEnvironmentsForLineno[1]{%
  \patchAmsMathEnvironmentForLineno{#1}%
  \patchAmsMathEnvironmentForLineno{#1*}%
}
\AtBeginDocument{%
\patchBothAmsMathEnvironmentsForLineno{equation}%
\patchBothAmsMathEnvironmentsForLineno{align}%
\patchBothAmsMathEnvironmentsForLineno{flalign}%
\patchBothAmsMathEnvironmentsForLineno{alignat}%
\patchBothAmsMathEnvironmentsForLineno{gather}%
\patchBothAmsMathEnvironmentsForLineno{multline}%
\patchBothAmsMathEnvironmentsForLineno{eqnarray}%
}

\usepackage{hyperxmp}

\usepackage[colorinlistoftodos,textsize=scriptsize]{todonotes}

\usepackage[all]{hypcap} 

\usepackage{xspace} 
\usepackage{upgreek}
\newcommand{\offsetoverline}[2][0.1em]{\kern #1\overline{\kern -#1 #2}}%

\def\qsq {\ensuremath{q^{2}}\xspace}
\def\mB {\ensuremath{m_{\Bz}}\xspace}
\def\ctl {\ensuremath{\cos\theta_{\ell}}\xspace}
\def\ctk {\ensuremath{\cos\theta_{K}}\xspace}
\def\mkpi {\ensuremath{m_{K\pi}}\xspace}

\ifthenelse{\boolean{uprightparticles}}%
{

 \def\Pmu         {\ensuremath{\upmu}\xspace}

 \def\Ppi         {\ensuremath{\uppi}\xspace}

 \def\Ppsi        {\ensuremath{\uppsi}\xspace}

 \def\PDelta      {\ensuremath{\Delta}\xspace}                 
 \def\PXi         {\ensuremath{\Xi}\xspace}                 
 \def\PLambda     {\ensuremath{\Lambda}\xspace}                 
 \def\PSigma      {\ensuremath{\Sigma}\xspace}                 
 \def\POmega      {\ensuremath{\Omega}\xspace}                 
 \def\PUpsilon    {\ensuremath{\Upsilon}\xspace}

 \def\PB      {\ensuremath{\mathrm{B}}\xspace}

 \def\PJ      {\ensuremath{\mathrm{J}}\xspace}                 
 \def\PK      {\ensuremath{\mathrm{K}}\xspace}

 \def\Pi      {\ensuremath{\mathrm{i}}\xspace}

}
{

 \def\Pmu         {\ensuremath{\mu}\xspace}

 \def\Ppi         {\ensuremath{\pi}\xspace}

 \def\Ppsi        {\ensuremath{\psi}\xspace}                 
                  
 \mathchardef\PDelta="7101
 \mathchardef\PXi="7104
 \mathchardef\PLambda="7103
 \mathchardef\PSigma="7106
 \mathchardef\POmega="710A
 \mathchardef\PUpsilon="7107
                  
 \def\PB      {\ensuremath{B}\xspace}

 \def\PJ      {\ensuremath{J}\xspace}                 
 \def\PK      {\ensuremath{K}\xspace}

 \def\Pi      {\ensuremath{i}\xspace}

}

\makeatletter
\ifcase \@ptsize \relax
  \newcommand{\miniscule}{\@setfontsize\miniscule{4}{5}}
\or
  \newcommand{\miniscule}{\@setfontsize\miniscule{5}{6}}
\or
  \newcommand{\miniscule}{\@setfontsize\miniscule{5}{6}}
\fi
\makeatother

\DeclareRobustCommand{\optbar}[1]{\shortstack{{\miniscule (\rule[.5ex]{1.25em}{.18mm})}
  \\ [-.7ex] $#1$}}

\def\mumu       {{\ensuremath{\Pmu^+\Pmu^-}}\xspace}

\def\pion   {{\ensuremath{\Ppi}}\xspace}

\def\pim    {{\ensuremath{\pion^-}}\xspace}

\def\kaon    {{\ensuremath{\PK}}\xspace}
  \def\Kbar    {{\kern 0.2em\overline{\kern -0.2em \PK}{}}\xspace}

\def\KorKbar {\kern 0.18em\optbar{\kern -0.18em K}{}\xspace}

\def\Kp      {{\ensuremath{\kaon^+}}\xspace}

\def\Kstarz  {{\ensuremath{\kaon^{*0}}}\xspace}

\def\B       {{\ensuremath{\PB}}\xspace}
\def\Bbar    {{\ensuremath{\kern 0.18em\overline{\kern -0.18em \PB}{}}}\xspace}

\def\BorBbar    {\kern 0.18em\optbar{\kern -0.18em B}{}\xspace}
\def\Bz      {{\ensuremath{\B^0}}\xspace}
\def\Bzb     {{\ensuremath{\Bbar{}^0}}\xspace}

\def\Bd      {{\ensuremath{\B^0}}\xspace}

\def\jpsi     {{\ensuremath{{\PJ\mskip -3mu/\mskip -2mu\Ppsi\mskip 2mu}}}\xspace}

\newcommand{\decay}[2]{\mbox{\ensuremath{#1\!\to #2}}\xspace}         

\def\to                 {\ensuremath{\rightarrow}\xspace}

\def\qsq       {{\ensuremath{q^2}}\xspace}
\def\CP                {{\ensuremath{C\!P}}\xspace}

\def\BdKpill  {\decay{\Bd}{\Kp\pim\ell^+ \ell^-}}
\def\BdKpillbold  {\boldmath{\decay{\Bd}{\Kp\pim\ell^+ \ell^-}}}

\def\ctl       {\ensuremath{\cos{\theta_\ell}}\xspace}
\def\ctk       {\ensuremath{\cos{\theta_K}}\xspace}

\def\C#1      {\ensuremath{\mathcal{C}_{#1}}\xspace}                       
\def\Cp#1     {\ensuremath{\mathcal{C}_{#1}^{'}}\xspace}                    
\def\Ceff#1   {\ensuremath{\mathcal{C}_{#1}^{\mathrm{(eff)}}}\xspace}        
\def\Cpeff#1  {\ensuremath{\mathcal{C}_{#1}^{'\mathrm{(eff)}}}\xspace}       
\def\Ope#1    {\ensuremath{\mathcal{O}_{#1}}\xspace}                       
\def\Opep#1   {\ensuremath{\mathcal{O}_{#1}^{'}}\xspace}                    

\newcommand{\aunit}[1]{\ensuremath{\text{\,#1}}}       

\newcommand{\tev}{\aunit{Te\kern -0.1em V}\xspace}
\newcommand{\gev}{\aunit{Ge\kern -0.1em V}\xspace}
\newcommand{\gevgev}{\gev^{2}}\xspace
\newcommand{\mev}{\aunit{Me\kern -0.1em V}\xspace}
\newcommand{\kev}{\aunit{ke\kern -0.1em V}\xspace}
\newcommand{\ev}{\aunit{e\kern -0.1em V}\xspace}
\newcommand{\mevc}{\ensuremath{\aunit{Me\kern -0.1em V\!/}c}\xspace}
\newcommand{\gevc}{\ensuremath{\aunit{Ge\kern -0.1em V\!/}c}\xspace}
\newcommand{\mevcc}{\ensuremath{\aunit{Me\kern -0.1em V\!/}c^2}\xspace}
\newcommand{\gevcc}{\ensuremath{\aunit{Ge\kern -0.1em V\!/}c^2}\xspace}

\def\fb   {\ensuremath{\aunit{fb}}\xspace}
\def\invfb   {\ensuremath{\fb^{-1}}\xspace}

\def\deriv {\ensuremath{\mathrm{d}}}

\renewcommand{\gev}{\, \mathrm{GeV}}
\renewcommand{\tev}{\, \mathrm{TeV}}

\newcommand{\nn}{\nonumber}

\newcommand{\eq}[1]{\begin{equation} #1 \end{equation}}
\newcommand{\eqa}[1]{\begin{eqnarray} #1 \end{eqnarray}}
\renewcommand{\C}[1]{{\cal C}_{#1}}
\renewcommand{\Cp}[1]{{\cal C}_{#1'}}

\newcommand{\re}{{\rm Re}}
\newcommand{\im}{{\rm Im}}

\def \azeL{{\ensuremath{A_0^L}}\xspace}
\def \azeR{{\ensuremath{A_0^R}}\xspace}

\def \apaL{{\ensuremath{A_\parallel^L}}\xspace}
\def \apaR{{\ensuremath{A_\parallel^R}}\xspace}

\def \apeL{{\ensuremath{A_\perp^L}}\xspace}
\def \apeR{{\ensuremath{A_\perp^R}}\xspace}

\def \re{\text{Re}}
\def \im{\text{Im}}

\usepackage{mciteplus}
\usepackage{multirow}
\usepackage{booktabs}

\begin{document}

\renewcommand{\thefootnote}{\fnsymbol{footnote}}
\setcounter{footnote}{1}

\renewcommand{\thefootnote}{\arabic{footnote}}
\setcounter{footnote}{0}
\maketitle
\flushbottom

\cleardoublepage

\pagestyle{plain} 
\setcounter{page}{1}
\pagenumbering{arabic}


\section{Introduction and Motivation}

Recent years have witnessed rising interest in the $B$-flavour anomalies as potential hints of New Physics (NP).  On the one side quantitatively, due to the observation of an increasing number of observables deviating from their Standard Model (SM) predictions; and on the other side, qualitatively, via an enhancement of the statistical significance of the NP hypotheses in $b \to s \ell^+\ell^-$ global analyses.  Recent analyses~\cite{Alguero:2019ptt,quim_moriond} (see also \cite{Geng:2021nhg,Altmannshofer:2021qrr,Ciuchini:2020gvn,Hurth:2020ehu}), show that some NP hypotheses exhibit a pull with respect to the SM of more than 7$\sigma$ and point to a NP contribution that is dominantly left-handed with a vector (or axial-vector) coupling to muons that breaks Lepton Flavour Universality. Solutions with additional small NP contributions from right-handed currents or  Lepton Flavour Universal (LFU) NP contributions~\cite{Alguero:2018nvb}  are also compatible with the data. Measurements of $B^{0} \to K^{*0}(\to \Kp\pim)\ell^+\ell^-$ decays with the $\Kp\pim$ system in an P-wave configuration  give rise to several of the anomalies observed and an improved understanding of these decays is essential to distinguish between the SM and possible NP scenarios.
The LHCb collaboration has observed the presence of a large $\Kp\pim$ S-wave component in \BdKpill decays~\cite{LHCb:2016ykl, LHCb:2020lmf}. However, the lack of reliable $B\to\Kp\pim$ S-wave form factors means that the physics potential of this component remains untapped.

In this paper we present the potential of \BdKpill transitions to search for physics beyond the SM, considering both P- and S-wave contributions to the $\Kp\pim$ system. For other works studying the impact of the S-wave contribution we refer the reader to Refs.~\cite{Becirevic:2012dp,Matias:2012qz,Das:2014sra,Das:2015pna,Blake:2012mb,Meissner:2013hya,Meissner:2013pba,Shi:2015kha}. Key to this work is the identification of the symmetries 
of the five dimensional decay rate that underpins the complete set of \BdKpill observables and the relations between them.
In particular, we identify new observables related to the interference between the S- and P-wave amplitudes of the $\Kp\pim$ system, and use the symmetry relations to investigate the potential of S-wave observables as precision probes of NP. 
We work under the hypothesis of no scalar or tensor NP contributions in our study of the symmetries.
In addition, we present a new and robust way to extract information on both NP scenarios and non-perturbative hadronic contributions by studying the common position in dilepton mass squared at which a subset of P- and S-wave observables cross zero.

Using pseudoexperiments that account for both background and detector effects, we make the first study of the capability of the LHCb experiment to extract the complete set of P- and S-wave observables from a single fit to the five-dimensional differential decay rate of \BdKpill decays. We also investigate the potential of combinations of the new S-wave observables to separate between relevant NP scenarios, in light of the current anomalies, for both current and future data sets. The complexities of both experimental and theoretical techniques to study \BdKpill transitions lend themselves to systemic errors. We therefore use the symmetry relations to devise stringent and model-independent cross-checks of the validity of both experiment and theory methodologies.

The paper is organised as follows. In section~\ref{sec:structure decay}, we discuss the structure of the differential angular distribution including P- and S-wave contributions. In the case of P-wave observables with massive leptons, we study the sensitivity of previously identified observables to new scalar and pseudoscalar contributions. In the case of the S wave, we define new observables. In section~\ref{sec3} 
we first perform an analysis of the degrees of freedom required to fully describe the angular distribution, identify the symmetries of the angular distributions and derive a set of  relations between P- and S-wave observables that are a consequence of the transformation symmetries of the angular distribution. These relations offer control tests for both experimental and theoretical analyses. Significantly given the lack of knowledge of S-wave form factors, these relations also enable predictions for some combinations of S-wave observables in terms of P-wave observables. In section~\ref{sec:bounds}, the relations are used to obtain the first bounds on the complete set of S-wave observables and the potential to observe NP with some of these observables is discussed. 
In section~\ref{sec:zeroes}, a set of P- and S-wave observables that share a zero at the same position in dilepton invariant mass is highlighted and the resulting information on both NP scenarios and on hadronic effects is discussed. The experimental prospects for determining all of the P- and S-wave observables discussed, in both massless and massive cases, are presented in section~\ref{sec:experiment}. Finally, a summary and conclusion are presented in section~\ref{sec:summary}.

\section{Structure of the differential decay rate: P and S waves}\label{sec:structure decay}
The differential decay rate of the four-body transition $B\to K\pi \ell^+\ell^-$ receives contributions from the amplitude of the P-wave decay $B\to K^*(\to K\pi) \ell^+\ell^-$, as well as from the amplitude of the S-wave decay $B\to K_0^*(\to K\pi) \ell^+\ell^-$, with $K_0^*$ being a broad scalar resonance. The rate can then be decomposed into:
\begin{equation}
\label{eq:pdftotal}
\frac{d^5\Gamma}{dq^2\,dm_{K\pi}^2\,d\Omega}\,=\,  \frac{d^5\Gamma_{P}}{dq^2\,dm_{K\pi}^2\,d\Omega} + \frac{d^5\Gamma_{S}}{dq^2\,dm_{K\pi}^2\,d\Omega} \end{equation}
where $d\Omega=d\ctl{}d\ctk{}d\phi$ and $\Gamma_P$ contains the pure P-wave contribution and $\Gamma_S$ contains the contributions from pure S-wave exchange, as well as from S-P interference. Here, $q^2$ denotes the square of the invariant mass of the lepton pair and $m_{K\pi}$ the invariant mass of the $K\pi$ system. The angles $\theta_\ell$, $\theta_K$ describe the relative directions of flight of the final-state particles, while $\phi$ is the angle between the dilepton and the dimeson plane (see Ref.~\cite{Egede:2010zc} for definitions). 
The differential rate for a \Bzb decay to a final state in the P-wave configuration is
\begin{align}
    \frac{d^5\Gamma_{P}}{dq^2\,dm_{K\pi}^2\,d\Omega} = \frac{9}{32\pi}\big[ & J_{1s}\sin^{2}\theta_{K} + J_{1c}\cos^{2}\theta_{K} + J_{2s}\sin^{2}\theta_{K}\cos2\theta_{\ell} 
    \nonumber \\
    + &J_{2c}\cos^{2}\theta_{K}\cos2\theta_{\ell} 
    + J_{3}\sin^{2}\theta_{K}\sin^{2}\theta_{\ell}\cos2\phi \nonumber \\ +&J_{4}\sin2\theta_{K}\sin2\theta_{\ell}\cos\phi 
    +  J_{5}\sin2\theta_{K}\sin\theta_{\ell}\cos\phi \nonumber \\ +&J_{6s}\sin^{2}\theta_{K}\ctl 
    +  J_{6c}\cos^{2}\theta_{K}\ctl \nonumber \\ +&J_{7}\sin2\theta_{K}\sin\theta_{\ell}\sin\phi + J_{8}\sin2\theta_{K}\sin2\theta_{\ell}\sin\phi \nonumber \\
    + & J_{9}\sin^{2}\theta_{K}\sin^{2}\theta_{\ell}\sin2\phi\big] \times |BW_{P}(\mkpi)|^{2} ,
    \label{eq:pdfpwaveraw}
\end{align}
    
\noindent with a similar form for the \Bz rate. The \mkpi dependence, denoted by $BW_{P}(\mkpi)$, can be modelled by a relativistic Breit-Wigner amplitude describing the $\Kstarz$ resonance, including the apposite angular momentum and phase-space factors. The Breit-Wigner amplitude is normalised such that the integral of the modulus squared of the amplitude over the \mkpi region of the analysis is one. For the exact form of the Breit-Wigner functions $BW_i(m_{K\pi})$ we refer the reader to Ref.~\cite{Becirevic:2012dp}.

The differential rate of the S-wave final state configuration is
\begin{equation}
\begin{split}
\frac{d^5\Gamma_{S}}{dq^2\,dm_{K\pi}^2\,d\Omega} =
+ \frac{1}{4\pi} & \left[ (\tilde{J}_{1a}^{c}  + \tilde{J}_{2a}^{c}\cos2\theta_{\ell}) |BW_S(\mkpi)|^2 \right. \\
& + \tilde{J}^{c}_{1b} \cos\theta_{K} 
 + \tilde{J}^{c}_{2b}  \cos2\theta_{\ell} \cos\theta_{K}  \\
& + \tilde{J}_{4}  \sin 2 \theta_{l} \sin\theta_{K} \cos\phi  
 + \tilde{J}_{5}  \sin\theta_{l} \sin\theta_{K}\cos\phi   \\
& + \tilde{J}_{7}  \sin\theta_{l} \sin\theta_{K}\sin\phi 
 \left. + \tilde{J}_{8}  \sin 2\theta_{l} \sin\theta_{K}\sin\phi \right] \, .
\end{split}
\label{eq:pdfswaveraw}
\end{equation}
The coefficients $J_i$, $\tilde J_{1a}^{c}$ and $\tilde{J}_{2a}^{c}$ are functions of  $q^2$. Those for the interference, $\tilde{J}_{1b}^{c}$, $\tilde{J}_{2b}^{c}$ and $\tilde{J}_{4-8}$ depend on both \qsq and \mkpi.
The \mkpi amplitude for the S wave, $BW_{S}(\mkpi)$ may be described with the LASS parameterisation~\cite{ASTON1988493,Rui:2017hks}. Similarly to the P wave, the S-wave \mkpi-amplitude is normalised such that the integral of the modulus squared of the amplitude over the analysed \mkpi range is one. 

If not explicitly stated otherwise, we will not consider the presence of scalar or tensor contributions in the following (this implies, in particular, that $J_{6c}$ in Eq.(\ref{eq:pdfpwaveraw}) is taken to be zero). The decays $B\to K^*\ell^+\ell^-$ and $B\to K_0^*\ell^+\ell^-$  are described by seven complex amplitudes $A_{\|,\bot,0}^{L,R}$, $A_t$ and three complex amplitudes $A_0^{\prime L,R}$, $A_t^\prime$, respectively, where the upper index $L,R$ refers to the chirality of the outgoing lepton current, while in the case of the P-wave the lower index $\|,\bot,0$ indicates the transversity amplitude of the $K^*$-meson.

Since the distribution is summed over the spins of the leptons,  the observables $J_i$ and $\tilde J_i$ are described in terms of spin-summed squared amplitudes of the form $A_i^{L*}A_j^{L}\pm A_i^{R*}A_j^{R}$. This structure suggests that the amplitudes can be arranged in a set of two-component complex vectors:
\begin{equation}
\label{eq:nvecs}
n_\|=\binom{A_\|^L }{A_\|^{R*} }\ ,\quad\!
n_\bot=\binom{A_\bot^L }{-A_\bot^{R*} }\ ,\quad\!
n_0=\binom{A_0^L }{A_0^{R*} }\ ,\quad\!
n_S=\binom{A_0^{\prime L} }{A_0^{\prime R*} },\quad\!  
n_S^\prime=\binom{A_0^{\prime L} }{-A_0^{\prime R*} } .
\end{equation}
Two vectors are needed to parametrize the $L$ and $R$ components of the $A_0^\prime$ amplitude, and the $A_t$ and $A_t^\prime$ amplitudes are not expressed in terms of two-complex vectors. Except for the lepton mass terms that mix the $L$ and $R$ components and include the $A_{t}$ (or $A_t^\prime$) amplitudes, one can express the coefficients of the distribution in terms of these vectors.  The expression for the coefficients in the P-wave terms can be found in Ref.~\cite{Egede:2010zc} and~\cite{Matias:2012xw}. For the S-wave terms we find 
\begin{eqnarray}
{\tilde J_{1a}^c}&=&\frac{3}{8} \left[|A_0'^L|^2 +| A_0'^R|^2+(1-\beta^2)\left(|A'_{t}|^2+2{\rm Re}\left[A'^{L}_{0}A'^{R*}_{0}\right]\right)\right] \, ,\hfill\quad \quad \quad \quad \quad \quad \quad \quad \quad \quad \quad \nn \\
{\tilde J_{2a}^c}&=&-\frac{3}{8}\beta^2\left(|A_0'^L|^2 +| A_0'^R|^2\right)=-\frac{3}{8}\beta^2|n_S|^2 .
\end{eqnarray}
Similarly for the P-S (real) interference terms
\begin{eqnarray}
{\tilde J_{1b}^c}&=& \frac{3}{4}\sqrt{3}   {\rm Re} \left[\left(
A_0'^L A_0^{L*} + A_0'^R A_0^{R*}+(1-\beta^2)\left(A'^{L}_{0}A^{R*}_{0}+A^{L}_{0}A'^{R*}_{0}+A'_{t}A^{*}_{t}\right)\right)BW_S BW_P^*
  \right] \nn \\
 &=& {\tilde J}_{1b}^{c, \, r}  {\rm Re}(BW_S BW_P^*)-{\tilde J}_{1b}^{c, \, i}  {\rm Im}(BW_S BW_P^*) 
   \nn \\
{\tilde J_{2b}^c}&=&-\frac{3}{4}\sqrt{3}\beta^2 {\rm Re}\left[\left(A'^{L}_{0}A^{L*}_{0}+A'^{R}_{0}A^{R*}_{0}\right)BW_S BW_P^*\right] \nn \\
   &=& {\tilde J}_{2b}^{c, \, r}  {\rm Re}(BW_S BW_P^*)-{\tilde J}_{2b}^{c, \, i}  {\rm Im}(BW_S BW_P^*) 
   \nn 
   \end{eqnarray}
   \begin{eqnarray}\label{eq:swave_new_1}
{\tilde J_4}&=& \frac{3}{4}\sqrt{\frac{3}{2}} \beta^2 {\rm Re} \left[(
A_0'^L A_\|^{L*} + A_0'^R A_\|^{R*})BW_S BW_P^*
\right]\hfill\quad \quad \quad \quad \quad \quad \quad \quad \quad \quad \quad \quad \quad \quad \nn \\ &=& \tilde{J}^{r}_{4} {\rm Re}(BW_S BW_P^*) - \tilde{J}^{i}_{4} {\rm Im}(BW_S BW_P^*) 
\nn \\
{\tilde J_5}&=&
\frac{3}{2}\sqrt{\frac{3}{2}} \beta {\rm Re} \left[(
A_0'^L A_\perp^{L*} - A_0'^R A_\perp^{R*})BW_S BW_P^*
\right] \nn \\
&=& \tilde{J}^{r}_{5} {\rm Re}(BW_S BW_P^*) - \tilde{J}^{i}_{5} {\rm Im}(BW_S BW_P^*)
\end{eqnarray}

\noindent and finally for the P-S (imaginary) interference terms
\begin{eqnarray}
\label{eq:swave_new_2}
{\tilde J_7}&=& \frac{3}{2}\sqrt{\frac{3}{2}}  \beta {\rm Im} \left[(
A_0'^L A_\|^{L*} - A_0'^R A_\|^{R*})BW_S BW_P^*
\right]\hfill\quad \quad \quad \quad \quad \quad \quad \quad \quad \quad \quad \quad \quad \quad \nn \\
&=& \tilde{J}^{r}_{7} {\rm Im}(BW_S BW_P^*) + \tilde{J}^{i}_{7} {\rm Re}(BW_S BW_P^*)
\nn \\ 
{\tilde J_8}&=&\frac{3}{4}\sqrt{\frac{3}{2}} \beta^2  {\rm Im} \left[(
A_0'^L A_\perp^{L*} + A_0'^R A_\perp^{R*})BW_S BW_P^*
  \right]\nn \\
 &=& \tilde{J}^{r}_{8} {\rm Im}(BW_S BW_P^*) + \tilde{J}^{i}_{8} {\rm Re}(BW_S BW_P^*) ,
\end{eqnarray}
where $\beta=\sqrt{1-4 m_\ell^2/q^2}$
and the superscript indices $r$ and $i$ (here and for the rest of the paper) refer to the real and imaginary parts of the bilinears, respectively.

The study of the S-wave observables presented in this paper is the first to consider the complete set of observables that arise when the decay rate is written differentially with respect to $m_{K\pi}$. As a consequence, the interference between the S-P-wave $m_{K\pi}$ lineshapes projects out additional bilinear combinations of S- and P-wave amplitudes, giving rise to the 12 new observables ${\tilde J}_{i}^{r,\, i}$ given in Eqs.(\ref{eq:swave_new_1}) and~(\ref{eq:swave_new_2}). Previous studies, such as those of Ref.~\cite{Hofer:2015kka}, only considered the differential decay rate integrated over $m_{K\pi}$. In this case one obtains the six well-known S-P interference observables ${\tilde J}_{i}$ that can be described by a single two-dimensional S-wave amplitude vector $n_S$, without the need for $n_S^\prime$.

\subsection{P-wave massive observables}\label{sec21}

The so-called optimized observables are designed to reduce form factor uncertainties. The set of such observables that describes the P-wave $K\pi$ system has been discussed at length in a series of papers~\cite{Matias:2012xw,Descotes-Genon:2013vna,Descotes-Genon:2015uva}. However, due to  improvements in experimental precision, there is increasingly  sensitivity to observables 
that are suppressed by factors of the lepton mass. 
For the optimized observables, $P_i$, the impact that lepton masses have in the very low \qsq region via the kinematical prefactor $\beta$ is well known. 

Our interest here is to explore two further optimized observables $M_1$ and $M_2$, introduced in Ref.\cite{Matias:2012xw}, that can be neglected in the massless limit.
These observables are defined in terms of the coefficients of the distribution as follows\footnote{In order to make the comparison with experimental prospects easier, in this work we have slightly changed the definition of $M_{1,2}$ by removing the constant terms appearing in Ref.~\cite{Matias:2012xw}.}:
\begin{equation}
    M_1=\frac{ J_{1s}}{ 3 J_{2s}}      \qquad 
    M_2=-\frac{ J_{1c}}{J_{2c}}.
\end{equation}
For this specific type of observable it makes sense to explore the impact from NP scalar and pseudoscalar contributions. Therefore, we will relax in this section the hypothesis of no scalar or pseudoscalar contributions.

Even considering a large set of NP scenarios, the observable $M_1$ is found to be practically insensitive to NP and is not analysed further. By contrast, $M_2$ can potentially provide information on scalar and/or pseudoscalar NP scenarios. In order to explore reasonable values of (pseudo)scalar contributions, we constrain the range for the coefficients ${\cal C}_{P,S}$ by considering only those values allowed by the experimental measurement of ${\cal B}(B_s\to\mu\mu)$. Thus, we write the following ratio~\cite{Fleischer:2017ltw}, which is used to define the $1\sigma$ region from ${\cal B}^{\rm exp}(B_s\to\mu\mu)$:

\begin{equation}
    R_{B_s\to\mu\mu}=\frac{{\cal B}^{\rm exp}(B_s\to\mu\mu)}{{\cal B}^{\rm SM}(B_s\to\mu\mu)} = |S|^2+|P|^2 \, ,
\end{equation}
where the quantities $S,P$\footnote{Not to be confused with the P- and S-wave components of the decay, this $S,~P$ refer to  Scalar and Pseudoscalar  NP contributions entering $R_{B_s\to\mu\mu}$. The latter includes the SM axial-vector contribution.} contain the different NP contributions and are given by:
\begin{align}
    S&=\sqrt{1-4\frac{m^2_{\mu}}{m^2_{B_s}}}\frac{m^2_{B_s}}{2m_bm_{\mu}}\left(\frac{{\cal C}_{S}-{\cal C}_{S'}}{{\cal C}^{\rm SM}_{10\mu}}\right) \, , \label{eq:S}\\
    P&=\frac{{\cal C}^{\rm SM}_{10\mu}+{\cal C}^{\rm NP}_{10\mu}-{\cal C}_{10'\mu}}{{\cal C}^{\rm SM}_{10\mu}}+\frac{m^2_{B_s}}{2m_bm_{\mu}}\left(\frac{{\cal C}_{P}-{\cal C}_{P'}}{{\cal C}^{\rm SM}_{10\mu}}\right). \label{eq:S-P}
\end{align}

\begin{figure}[htp]
    \centering
    \includegraphics[width = 0.6\textwidth]{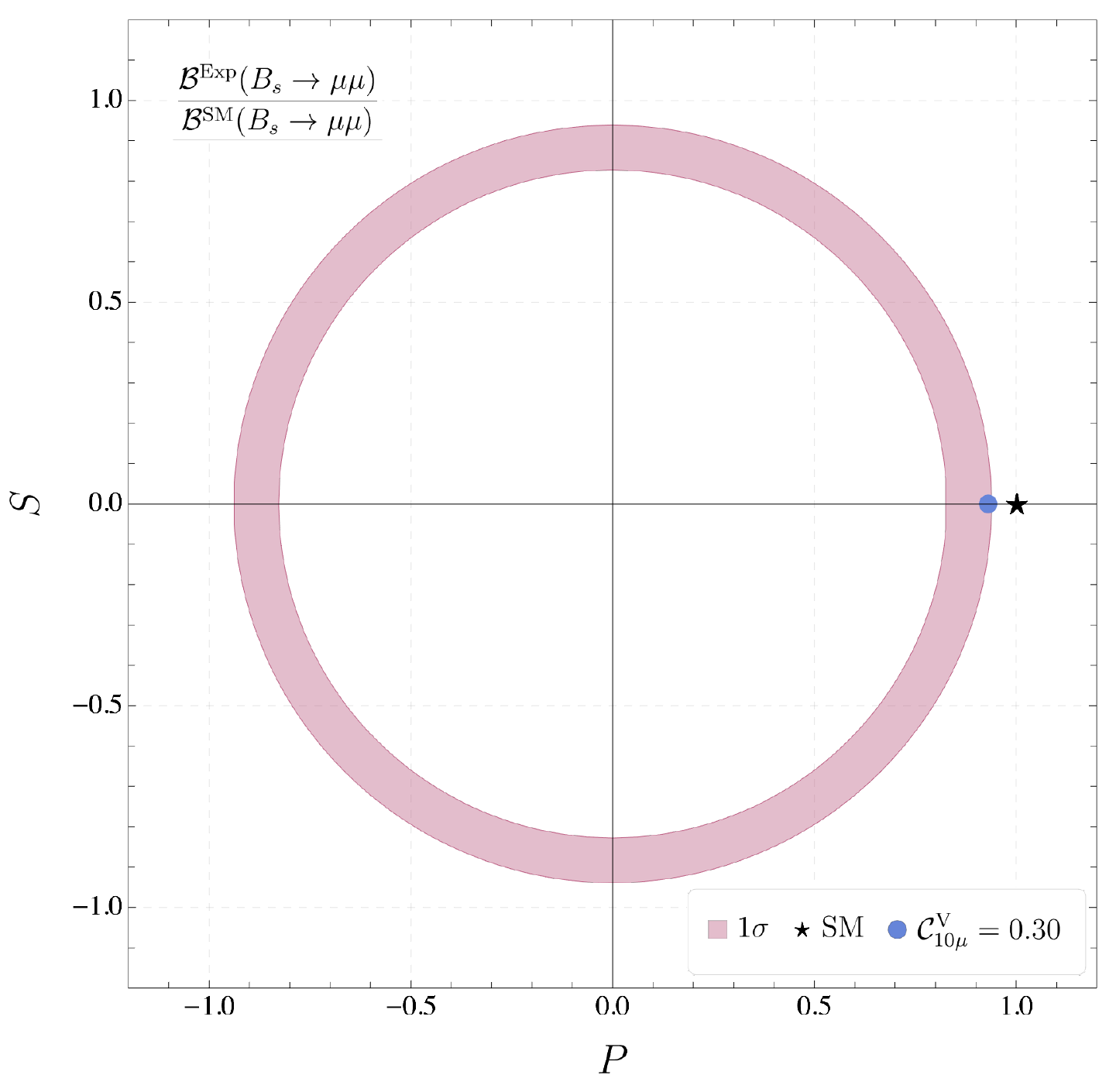}
    \caption{Region of allowed values for $S,P$ that fulfill the condition $|R^{\rm SM}_{B_s\to\mu\mu}-R^{\rm NP}_{B_s\to\mu\mu}| \leq 0.10$. In order to illustrate the sensitivity of this observable to NP contributions, we display its value in the SM (black star) and in one of the favoured scenario from Ref.~\cite{quim_moriond} (blue dot): \{${\cal C}_{9\mu}^{\rm V}=-{\cal C}_{10\mu}^{\rm V}=-0.30, {\cal C}_9^{\rm U}=-0.92$\}. Only the dependence on ${\cal C}_{10\mu}$ is displayed in the plot. The tiny difference of this scenario with the SM illustrates that $M_2$ is an observable with low sensitivity to the preferred scenarios of present global fits. For this reason we explore its sensitivity under other types of NP, namely scalars and pseudoscalars. }
    \label{fig:SP_constraint}
\end{figure}

Fig.~\ref{fig:SP_constraint} shows the allowed region for $S$ and $P$ once the latest experimental value for ${\cal B}(B_s\to\mu\mu)=(2.85\pm 0.34)$~\cite{diegoMartinezPrivateCommunication} is included, corresponding to $R_{B_s\to\mu\mu}=(0.78\pm0.10)$. In this analysis we have not allowed for the presence of right-handed currents. 

We perform an analysis of the behaviour of $M_2$ under different hypotheses for (pseudo)scalar NP contributions that are compatible with Fig.~\ref{fig:SP_constraint}. The case $S=0, P=1$ corresponds to the SM, as can be seen from Eqs.~\eqref{eq:S} and~\eqref{eq:S-P}. We consider  three other possible scenarios, corresponding to maximal values of $S,P$:

\begin{itemize}
    \item[i)] $S=\pm0.94, P=0$,
    \item[ii)] $S=P=0.66$,
    \item[iii)] $S=0, P=-0.94$.
\end{itemize}

\begin{figure}[htp]
    \centering
    \includegraphics[width = 0.8\textwidth]{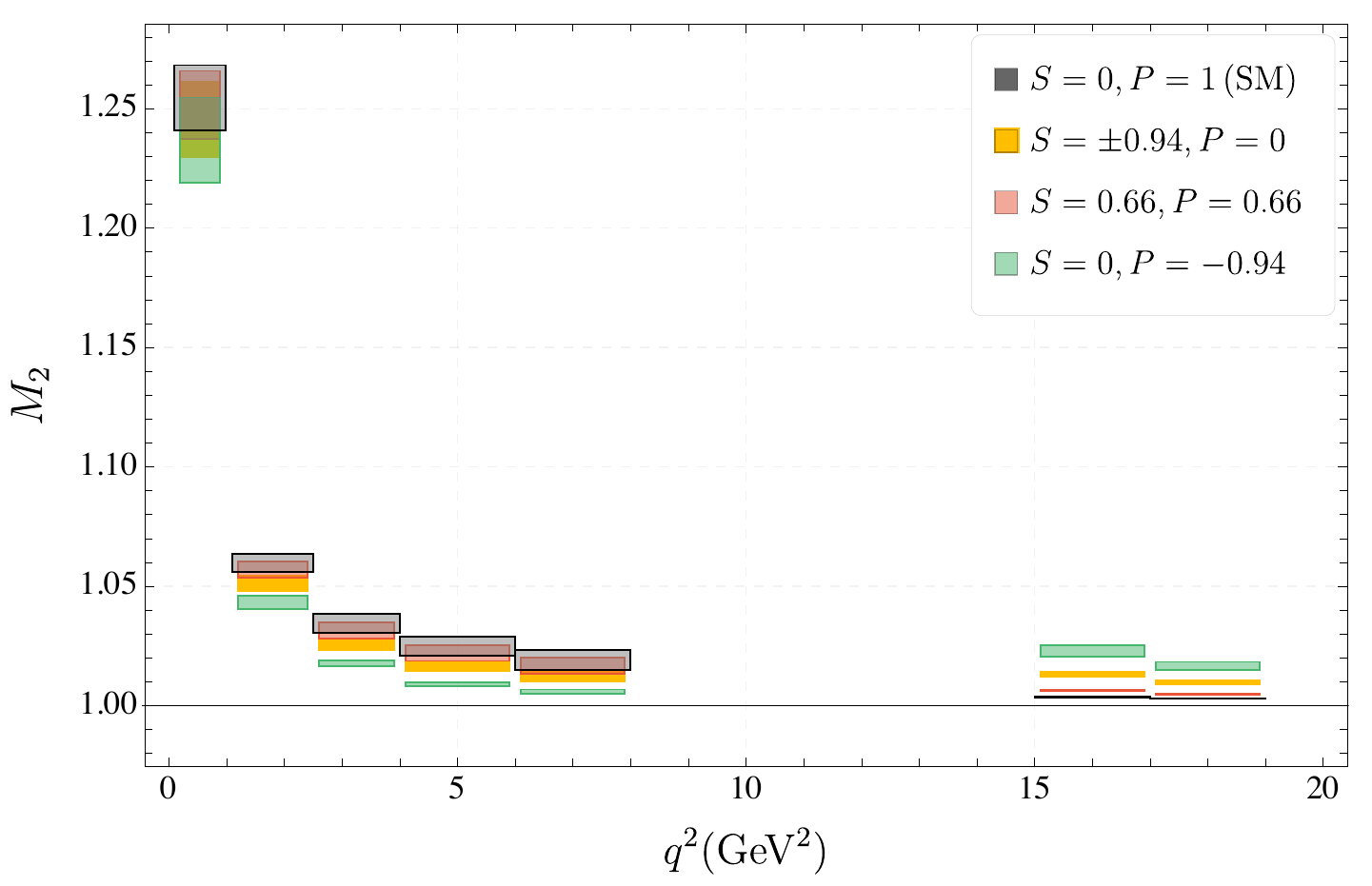}
    \caption{Binned theoretical predictions for $M_2$ in the SM and in selected NP scenarios including pseudoscalar and scalar contributions.}
    \label{fig:M2_SP}
\end{figure}

These three benchmark cases are: i) only a scalar contribution (with two possible signs) and no pseudoscalar NP, ii) both $S$ and $P$ contributions present and equal in magnitude and iii) the opposite sign of the SM case with a negative pseudoscalar contribution. 
Fig.~\ref{fig:M2_SP} shows the theoretical prediction of the large- and low-recoil bins of $M_2$ in the four scenarios mentioned above.

It is evident from Fig.~\ref{fig:M2_SP} that the rather small sensitivity of $M_2$ to (pseudo)scalar contributions  makes it difficult to get a significant distinction between the different scenarios. This is especially the case in the very low \qsq region, where the uncertainties associated with the theoretical prediction of this observable are larger. Only for the $S=0,P=-0.94$ scenario in the large-recoil region is a  clean separation between hypotheses possible, given suitably high precision measurements. The situation is somewhat better in the low-recoil \qsq region,  where the theoretical errors are smaller but an  even higher experimental resolution will be required. The experimental prospects for such a separation of NP hypotheses is outlined in section~\ref{sec:experiment:massive:scalar}.

\subsection{Definition of S-wave observables: massless and massive case}\label{sobservables}

In this section we define the list of S-wave observables that can be constructed using the coefficients of the distribution. They follow from the previous section including P and S waves in the massless case but also taking into account lepton mass terms. The S-wave observables that were mostly treated  as nuisance parameters thus far will become an interesting target for future  experimental analyses. 

Our goal  here will be  to define the S-wave observables but it is beyond the scope of this paper to provide SM predictions and enter into a discussion of the form factors or other hadronic uncertainties. Our first interest  is to determine how many of the observables  are genuinely independent. The question of the number of degrees of freedom is critical for the stability of experimental fits and  is discussed further in section~\ref{sec3}.

As discussed in section~\ref{sec:structure decay}, the new $S$-$P$ interference observables defined in Eqs.(\ref{eq:swave_new_1}) and(\ref{eq:swave_new_2}), can be defined in terms of the vectors in Eq.(\ref{eq:nvecs}) as follows:
\eqa{ {S_{S1}^r}&\!=\!&{ -\frac{3}{4}}\sqrt{3}{\frac{1}{\Gamma^\prime}} \beta^2 {\rm Re} (n_0^\dagger\, n_S )+ CP, \qquad \qquad
{ S_{S1}^i} = { -\frac{3}{4}}\sqrt{3}{\frac{1}{\Gamma^\prime}} \beta^2 {\rm Im} (n_0^\dagger\, n_S^\prime )+ CP,
\nn \\
{S_{S2}^r}&\!=\!&{ \frac{3}{4}}\sqrt{\frac{3}{2}}{\frac{1}{\Gamma^\prime}} \beta^2 {\rm Re} (n_\|^\dagger\, n_S   )+ CP, \qquad \qquad
\,\, {S_{S2}^i}= { \frac{3}{4}} \sqrt{\frac{3}{2}}{\frac{1}{\Gamma^\prime}} \beta^2 {\rm Im} (n_\|^\dagger n_S^\prime   )+ CP,
\nn \\
{S_{S3}^r}&\!=\!&{ \frac{3}{2}}  \sqrt{\frac{3}{2}}{\frac{1}{\Gamma^\prime}} \beta {\rm Re} (n_\perp^\dagger\, n_S  )+ CP,\qquad \qquad
\,\,\,{S_{S3}^i}=
{ \frac{3}{2}} \sqrt{\frac{3}{2}}{\frac{1}{\Gamma^\prime}} \beta {\rm Im} ( n_{\perp}^\dagger n_S^\prime )+ CP,
\nn \\
{S_{S4}^r}&\!=\!&{ \frac{3}{2}}  \sqrt{\frac{3}{2}}{\frac{1}{\Gamma^\prime}} \beta {\rm Re} ( n_\|^{\dagger} n_S^\prime )+ CP, \qquad \qquad
\,\,\,\,\, {S_{S4}^i}={ \frac{3}{2}} \sqrt{\frac{3}{2}}{\frac{1}{\Gamma^\prime}}\beta {\rm Im} ( n_\|^{\dagger} n_S )+ CP
\nn \\
{S_{S5}^r}&\!=\!&{ \frac{3}{4}}\sqrt{\frac{3}{2}}{\frac{1}{\Gamma^\prime}} \beta^2 {\rm Re} ( n_\perp^{\dagger} n_S^\prime)+ CP\qquad \qquad
\,\,\,\,{S_{S5}^i}= { \frac{3}{4}} \sqrt{\frac{3}{2}}{\frac{1}{\Gamma^\prime}} \beta^2 {\rm Im} (n_\perp^{\dagger} n_S )+ CP\label{sset}
}
where
\begin{eqnarray}
\Gamma^\prime&=&\Gamma_P^\prime+\Gamma_S^\prime + { CP}\nn \\
\Gamma_P^\prime&=&\frac{3}{4}(2 J_{1s}+J_{1c}) -\frac{1}{4} (2 J_{2s}+J_{2c})+ { CP} \nn \\
\Gamma_S^\prime&=&2 \tilde{J}^c_{1a} - \frac{2}{3} \tilde{J}^c_{2a} + { CP} .
\end{eqnarray}
Here the prime stands for the differential distribution. Note that once we include lepton mass terms, $F_S$ should be extracted from ${\tilde J_{2a}^c}$  and not from the combination with ${\tilde J_{1a}^c}$ such that:
\begin{equation}\label{fseq}
F_S=\frac{|n_S^\dagger n_S|}{\Gamma^\prime}
=-\frac{8}{3\beta^2} \frac{\tilde J_{2a}^c}{\Gamma^\prime}.
\end{equation}
In order not to overload excessively the notation  it should be understood that from
 Eq.(\ref{fseq}) to Eq.(\ref{defs})
each explicit $J$ or $\tilde{J}$ is accompanied by its $CP$-conjugate partner. In the case that $B^0$ and $\bar{B}^0$ decays were experimentally separated, a set of $CP$-asymmetries corresponding to each $J$ and $\tilde{J}$ observable would also become accessible  (see section~\ref{sec:experiment:setup}).

In terms of these observables, the angular distribution in the massless limit (taking $\beta \to 1$ in Eq.\eqref{sset}) is given:
\begin{equation}
\begin{split}
\left.\frac{1}{\deriv(\Gamma+\bar{\Gamma})/\deriv q^2}\,\frac{\deriv^4(\Gamma+\bar{\Gamma})}{\deriv\qsq\,\deriv\vec{\Omega}}\right|_{{\rm S}+{\rm P}}~&=~ 
(1-F_{\rm S})  |BW_P|^2 \left.\frac{1}{\deriv(\Gamma+\bar{\Gamma})/\deriv q^2}\,\frac{\deriv^4(\Gamma+\bar{\Gamma})}{\deriv\qsq\,\deriv\vec{\Omega}}\right|_{\rm P} +  \\
 &\!\!\!\! \!\!\!\! +{ \frac{1}{4\pi}}  \left[ { \frac{3}{4}} F_{\rm S} |BW_S|^2 \sin^{2}\theta_{\ell} \right. \\
&\!\!\!\! \!\!\!\!{-2} [S^r_{S1} {\rm Re}(BW_S BW_P^*) - S^i_{S1} {\rm Im}(BW_S BW_P^*)] \sin^{2}\theta_{\ell} \cos\theta_{K} \\
& \!\!\!\! \!\!\!\! + [S^r_{S2} {\rm Re}(BW_S BW_P^*) - S^i_{S2} {\rm Im}(BW_S BW_P^*)] \sin 2 \theta_{\ell} \sin\theta_{K} \cos\phi  \\
&\!\!\!\! \!\!\!\! + [S^r_{S3} {\rm Re}(BW_S BW_P^*) - S^i_{S3} {\rm Im}(BW_S BW_P^*)] \sin\theta_{\ell} \sin\theta_{K}\cos\phi  \\
& \!\!\!\! \!\!\!\! + [S^r_{S4} {\rm Im}(BW_S BW_P^*) + S^i_{S4} {\rm Re}(BW_S BW_P^*)] \sin\theta_{\ell} \sin\theta_{K}\sin\phi \\
&\!\!\!\! \!\!\!\! \left. + [S^r_{S5} {\rm Im}(BW_S BW_P^*) + S^i_{S5} {\rm Re}(BW_S BW_P^*)] \sin 2\theta_{\ell} \sin\theta_{K}\sin\phi \right].
\end{split}\label{eq:pdfswave}
\end{equation}
The corresponding angular distribution in the massive case can be obtained from Eq.(\ref{eq:pdfswaveraw}) using optimized S-wave observables and mass terms defined by:
\begin{align}
    M_{3}' = \frac{-\beta^{2}\tilde{J}_{1a}^{c} - \tilde{J_{2a}^{c}}}{\tilde{J}_{2a}^{c}},
\end{align}
together with the extra S-P interference massive optimized terms:
\begin{align}
    M_{4}' &= \frac{-\beta^{2}\tilde{J}_{1b}^{c,r} - \tilde{J}_{2b}^{c,r}}{\sqrt{J_{2c}\tilde{J_{2a}^{c}}}}, \nn \\
    M_{5}' &= \frac{-\beta^{2}\tilde{J}_{1b}^{c,i} - \tilde{J}_{2b}^{c,i}}{\sqrt{J_{2c}\tilde{J_{2a}^{c}} }}.
\end{align}
Then the  massive distribution becomes:
\begin{equation}
\begin{split}
\left.\frac{1}{\deriv(\Gamma+\bar{\Gamma})/\deriv q^2}\,\frac{\deriv^4(\Gamma+\bar{\Gamma})}{\deriv\qsq\,\deriv\vec{\Omega}}\right|_{{\rm S}+{\rm P}}~&=~ 
(1-F_S^\prime)  |BW_P|^2 \left.\frac{1}{\deriv(\Gamma+\bar{\Gamma})/\deriv q^2}\,\frac{\deriv^4(\Gamma+\bar{\Gamma})}{\deriv\qsq\,\deriv\vec{\Omega}}\right|_{\rm P} +   \\
 &\!\!\!\! \!\!\!\! +{ \frac{1}{4\pi}}  \left[ 
 \left( \frac{3}{8} F_S (1+M_3^\prime) -\frac{3}{8}\beta^2 F_S \cos 2 \theta_l \right) |BW_S|^2 \right. \\ 
 &\!\!\!\! \!\!\!\!+\left( - \frac{1}{\beta^2} (S_{S1}^r + M_4^\prime N_L) {\rm Re}(BW_S BW_P^*)\right. \\
 &\!\!\!\! \!\!\!\! \left. + \frac{1}{\beta^2} (S_{S1}^i + M_5^\prime N_L) {\rm Im}(BW_S BW_P^*) \right) \cos\theta_K \\
&\!\!\!\! \!\!\!\!+ [S^r_{S1} {\rm Re}(BW_S BW_P^*) - S^i_{S1} {\rm Im}(BW_S BW_P^*)] \cos {2}\theta_{l} \cos\theta_{K} \\
& \!\!\!\! \!\!\!\! + [S^r_{S2} {\rm Re}(BW_S BW_P^*) - S^i_{S2} {\rm Im}(BW_S BW_P^*)] \sin 2 \theta_{l} \sin\theta_{K} \cos\phi  \\
&\!\!\!\! \!\!\!\! + [S^r_{S3} {\rm Re}(BW_S BW_P^*) - S^i_{S3} {\rm Im}(BW_S BW_P^*)] \sin\theta_{l} \sin\theta_{K}\cos\phi  \\
& \!\!\!\! \!\!\!\! + [S^r_{S4} {\rm Im}(BW_S BW_P^*) + S^i_{S4} {\rm Re}(BW_S BW_P^*)] \sin\theta_{l} \sin\theta_{K}\sin\phi \\
&\!\!\!\! \!\!\!\! \left. + [S^r_{S5} {\rm Im}(BW_S BW_P^*) + S^i_{S5} {\rm Re}(BW_S BW_P^*)] \sin 2\theta_{l} \sin\theta_{K}\sin\phi \right].
\end{split}
\label{eq:pdfswavemass}
\end{equation}
We define
\begin{eqnarray}
N_{L}&=&\sqrt{J_{2c} \tilde{J}^c_{2a}}=\frac{1}{2} \sqrt{\frac{3}{2}} \beta^2\Gamma^\prime \sqrt{(1-F_S^\prime) F_S F_{L}} \nonumber \\
N_{T}&=&\sqrt{-J_{2s} \tilde{J}^c_{2a}}=\frac{1}{4} \sqrt{\frac{3}{2}} \beta^2\Gamma^\prime \sqrt{(1-F_S^\prime) F_S F_{T}}
\end{eqnarray}
and 
\begin{equation} \label{defFsprime}
F_S^\prime=\frac{\Gamma_S^\prime}{\Gamma^\prime}=F_S-\epsilon_S   \quad \quad \epsilon_S=\frac{1}{4} F_S (1-\beta^2 -3 M_3^\prime) .
\end{equation}
Notice that in the massless limit ($M_i^{(\prime)} \to 0$, $\beta \to 1$) Eq.(\ref{eq:pdfswavemass}) reduces to Eq.(\ref{eq:pdfswave}).

Finally, in order to write the whole distribution with massive terms and optimized observables, the substitution:
\begin{equation}
S_{S1}^{r/i} \to PS_{1}^{r/i} \frac{N_L}{{\Gamma'}}  \quad S_{S2-S5}^{r/i} \to PS_{2-5}^{r/i} \frac{N_T}{{\Gamma'}}
\end{equation}
is needed, where the optimised observables for the interference terms in all \qsq bins are
\begin{align}
    PS_{1}^{r/i} = \frac{\tilde{J}_{2b}^{c,r/i}}{\sqrt{J_{2c}\tilde{J_{2a}^{c}} }}, \quad  
    PS_{2-5}^{r/i} = \frac{\tilde{J}_{4-8}^{r/i}}{\sqrt{-J_{2s}\tilde{J_{2a}^{c}} }}\, . \label{swaveopt}
\end{align}

\noindent Using the expressions\footnote{One may add to this list another observable, related to the presence of scalars, associated with the coefficient $J_{6c}$. Given that in the present paper we only allow for scalars when analyzing the observable $M_2$, we direct the reader to Ref.\cite{Matias:2012xw}, where this case is discussed.} 
\begin{eqnarray} 
 {J_{2s}}&=&\frac{1}{4} N_1, \quad
{ J_{2c}}=-N_2,  \quad
  J_3  = \frac{1}{2}  { P_1}   N_1,
\quad
{J_4} = \frac{1}{2}   { P_4'}  N_3, 
\quad 
 { J_5} =   { P_5'}  N_3, \nonumber \\
 { J_{6s}} &=& 2  { P_2}  N_1,
 \quad
{J_7} =-{ P_6'} N_3,
    \quad
 {J_8} =-\frac{1}{2}  { P_8'}  N_3,
    \quad
   {J_9} = -   {P_3}  N_1,  \label{defs}    
    \end{eqnarray}
where $N_{1,2}=\beta^2 F_{T,L} \Gamma_P^\prime$, $N_3=\beta^2\sqrt{F_T F_L} \Gamma_P^\prime$ (and the addition of the CP conjugate in $\Gamma_P^\prime$ is implicit) and including the definitions of $M_{1,2}$, one finds:

 \begin{eqnarray}
 \label{distfull}
 \hspace*{-0.2cm}\left.\frac{1}{\deriv(\Gamma+\bar{\Gamma})/\deriv q^2}\,\frac{\deriv^4(\Gamma+\bar{\Gamma})}{\deriv\qsq\,\deriv\vec{\Omega}}\right|_{{\rm P}} 
 =&&
\nn\\[1.5mm]
&&\hspace{-5.3cm}\frac9{32\pi} \bigg[
\frac{3}{4} {\hat{F}}_T M_1 \sin^2\theta_K + {\hat{F}}_L M_2 \cos^2\theta_K  
+ (\frac{1}{4} {\hat{F}}_T \sin^2\theta_K - {\hat{F}}_L \cos^2\theta_K) \cos 2\theta_l\nn\\[1.5mm]
&&\hspace{-5.3cm}+ \frac{1}{2} P_1 {\hat{F}}_T \sin^2\theta_K \sin^2\theta_l \cos 2\phi + \sqrt{\hat{F}_T \hat{F}_L} \left(\frac{1}{2} P_4'  \sin 2\theta_K \sin 2\theta_l \cos\phi  + P_5'  \sin 2\theta_K \sin\theta_l \cos\phi  \right)\nn\\[1.5mm]
&&\hspace{-5.3cm}+ 2 P_2 \hat{F}_T \sin^2\theta_K   \cos\theta_l    
- \sqrt{\hat{F}_T \hat{F}_L} \left( P_6'  \sin 2\theta_K \sin\theta_l \sin\phi  +\frac{1}{2} P_8^\prime  \sin 2\theta_K \sin 2\theta_l \sin\phi \right) \nn\\[1.5mm]
&&\hspace{-3.5cm}- P_3 \hat{F}_T \sin^2\theta_K \sin^2\theta_l \sin 2\phi \bigg]\, .
\end{eqnarray}
where a global pre-factor $\beta^2$ has been absorbed inside the re-definition ${\hat {F}_{T,L}}=\beta^2 F_{T,L}$.

\section{Symmetries of the distribution}\label{sec3}

In this section we present the explicit form of the symmetry transformations of the amplitudes  that leave  the full distribution (including P and S wave) invariant, and obtain explicitly the relations among the observables.  The massless and the massive cases are discussed separately.

The number of symmetries of the distribution are determined by performing an infinitesimal transformation $\vec{A}^\prime=\vec{A}+ \vec{\delta}$,
where $\vec{A}$ is a vector collecting the real and imaginary parts of all  the amplitudes entering the distribution (the vector $\vec{A}$ depends on whether the massless or massive hypothesis is taken), and the condition to be a symmetry is that the vector $\vec{\delta}$ is perpendicular to the hyperplane spanned by the set of gradient vectors:
\begin{equation} \label{symmetry}
      \forall i \in J_i,\tilde{J}_i : \vec{\nabla}_i \perp  \vec{\delta} \, .
\end{equation}
The gradients are defined then by the derivatives of the coefficients with respect to the real and imaginary parts of all the amplitudes.
The difference between the dimension of the hyperplane that the gradient vectors span if they are all independent (equal to the number of coefficients of the distribution) and the dimension of the hyperplane that they effectively span tells us the number of relations among the coefficients that exist. By relations we will refer only to non-trivial relations. We will discuss these relations in the following subsections. 
For completeness, we first find explicitly the form of the continuous symmetries.

In Ref.~\cite{Matias:2012xw}, the massless and massive symmetries were discussed for the P wave. There it was found that, in the massless case, four symmetries (two phase transformations for the left and right components and two ``angle rotations") leave the P-wave part of the distribution invariant. Alternatively, using the vectors $n_i$ we can implement the four symmetry transformations by means of a $2 \times 2$ unitary matrix, i.e,  $n_i^\prime=U n_i$ with $i=\perp,\|,0$. However, the inclusion of the S wave that requires two different vectors $n_S$ and $n_S^\prime$ breaks two of the symmetries\footnote{This is easily shown by simply transforming the sum  $n_S+n_S^\prime$} and only the two independent phase transformations survive, i.e.,
\begin{equation} \label{masslesssym}
    {A_i^{L}} \to e^{i \phi_L} A_i^{L}, \quad \quad    {A_0^{\prime L}} \to e^{i \phi_L} A_0^{\prime L}, \quad \quad 
       {A_i^{R}} \to e^{i \phi_R} A_i^{R}, \quad \quad 
       {A_0^{\prime R}} \to e^{i \phi_R} A_0^{\prime R}
\end{equation}
with $i=0,\perp,\|$. 

The massive case is relatively similar and again only two phase transformations survive. However, the existence of interference terms between left and right components fixes $\phi_L=\phi_R=\phi$, but this is compensated by the independent transformation of the extra amplitudes
$A_{t}^{(\prime)}$:
\begin{eqnarray}\label{massivesym}
    & A_i^L \to e^{i\phi} A_i^L, 
    \quad \quad    
        & A_i^R \to e^{i\phi} A_i^R, \nonumber \\
    &{A_0^{\prime L}} \to e^{i \phi} A_0^{\prime L},
  \quad \quad 
      &{A_0^{\prime R}} \to e^{i \phi} A_0^{\prime R}
    \nonumber \\
   &  A_t \to e^{i\varphi} A_t, \quad \quad  &A_t^\prime \to e^{i\varphi} A_t^\prime
\end{eqnarray}
with $i=0,\perp,\|$\footnote{Another example of the convenience of using symmetries but in the semileptonic charged-current $b\to c \ell\nu$ transition can be found in Ref.~\cite{Alguero:2020ukk}.} .

\subsection{Counting degrees of freedom: massive and massless cases}
One important question is how many degrees of freedom there are or, in other words, how many  observables in the set discussed in section~\ref{sobservables} are independent.
The number of independent observables to fully describe the distribution depends on whether massless or massive leptons are considered. We again work under the hypothesis that there are no scalar contributions but pseudoscalar ones are allowed in the massive case.

The number of observables that can be constructed out of the complex amplitudes is given by:
\begin{equation}
    n_{obs}=2 n_A - n_{sym}\, .
\end{equation}
Each symmetry transformation of the amplitudes that leaves the distribution invariant reduces the number of independent observables.

In the following, we determine the number of relations for the massless and massive case and consequently the number of independent observables required to have a full description of the corresponding distribution.

\bigskip
\noindent {\bf Massless case:}
\bigskip

Assuming the absence of scalars, we have  11 coefficients for the P-wave and 14 coefficients for the S-wave distribution.
Under the approximation of negligible lepton masses,  there are two trivial relations for the P-wave coefficients:
\begin{equation}
     J_{1s}=3 J_{2s}    \quad \quad J_{1c}=-J_{2c}
\end{equation}
and three trivial relations for the S-wave coefficients:
\begin{equation}
 \tilde{J}_{1a}^c=-\tilde{J}_{2a}^c \quad \quad \tilde{J}_{1b}^{c \, r}=-\tilde{J}_{2b}^{c \,r} \quad \quad \tilde{J}_{1b}^{c \, i}=-\tilde{J}_{2b}^{c \,i} \, ,
 \end{equation}
 reducing the number of coefficients to $n_c=20$.
 The vector $\vec{A}$ in the massless case is given by:
     \begin{eqnarray}
  \label{eq:Avector}
  \vec{A} & = & 
      \left(\re(\apeL),\im(\apeL),\re(\apaL),\im(\apaL),\re(\azeL),\im(\azeL), \re(\azeL^\prime),\im(\azeL^\prime),
      \phantom{)}\right. \nonumber \\ 
      & &\left. \phantom{(}  \re(\apeR),\im(\apeR),\re(\apaR),\im(\apaR),\re(\azeR),\im(\azeR), \re(\azeR^\prime),\im(\azeR^\prime)
      \right)
\end{eqnarray} 
Using Eq.(\ref{symmetry}), we find that the  dimension of the space spanned by the gradient vectors (given by the rank of the matrix $M_{ij}=\nabla_i X_j$ with $X=J,\tilde J$ and $i$ being the elements of $\vec{A}$ in Eq.(\ref{eq:Avector})) is $n_{rank}=14$. This rank gives the number of independent observables $n_{obs}$.
According to the discussion above, the number of relations fulfills:
\begin{equation} \label{eq:nrel}
    n_{rel}=n_c - n_{rank} \, .
\end{equation}
Therefore for the massless case $n_{rel}=6$.
There is one well-known relation among the coefficients for the P wave (see Ref.~\cite{Egede:2010zc,Matias:2014jua}) and five, previously unknown, relations for the S wave.
 An independent cross check of the rank of the matrix is provided by the fact that the number of degrees of freedom counting amplitudes minus symmetries, or coefficients minus relations should agree. This implies the equation:
 \begin{equation}
     2 n_A - n_{sym}=n_{rank}=n_c -n_{rel}\, .
 \end{equation}
 The number of complex amplitudes $n_A=8$ and the number of symmetries of the full distribution (P and S wave) is $n_{sym}=2$ (see Eq.(\ref{masslesssym})).
 
 The set of 14 independent observables consists of 8 (9 coefficients minus one relation) independent 
 observables for the P wave and 6 (11 coefficients minus 5 relations) independent observables for the S wave.
 This implies that in the massless case the basis of 20 observables,
 \begin{eqnarray}\label{eq:massless_obs_list}
{\cal O}_{m_\ell=0}&=&\{\Gamma^\prime, F_L, P_1, P_2, P_3, P_4^\prime, P_5^\prime, P_6^\prime, P_8^\prime,\hfill \nonumber \\ &&\quad \quad \quad \quad F_S, { S}_{S1}^r, S_{S2}^r, S_{S3}^r, S_{S4}^r, S_{S5}^r, { S}_{S1}^i, S_{S2}^i, S_{S3}^i, S_{S4}^i, S_{S5}^i \}, 
\end{eqnarray}
 has some redundancy. Among these 20 observables there are 6 relations leading to only 14 independent observables.
  The set of 6 massless relations can be obtained from the 6 massive expressions given below, after taking the massless limit. Notice that the seventh relation, given in the appendix, is exactly zero in the massless limit.
 
 \bigskip
 \noindent {\bf Massive case:}
 \bigskip{}

 The counting in this case, following the same steps as in the massless case, goes as follows. Our starting point is the same number of coefficients 11 (14) for the P wave (S wave), but now there are no trivial relations, i.e., $n_c=25$. Here the vector $\vec{A}$ is:
      \begin{eqnarray}
  \label{eq:Avectormassive}
  \vec{A}  & = & 
      \left(\re(\apeL),\im(\apeL),\re(\apaL),\im(\apaL),\re(\azeL),\im(\azeL), \re(\azeL^\prime),\im(\azeL^\prime),\re{A_t},\im{A_t},
      \phantom{)}\right. \nonumber \\ 
      & &\left. \phantom{(}  \re(\apeR),\im(\apeR),\re(\apaR),\im(\apaR),\re(\azeR),\im(\azeR), \re(\azeR^\prime),\im(\azeR^\prime),\re{A_t^\prime},\im{A_t^\prime} \nonumber
      \right).
\end{eqnarray} 
 Notice that pseudoscalar contributions are included in the amplitude $A_t$. Evaluating the rank of the corresponding matrix $M_{ij}$, one finds $n_{rank}=18$, indicating that in the massive case the number of independent observables is $n_{obs}=18$.
 Following Eq.(\ref{eq:nrel}), one immediately finds that the number of relations should be 7. These relations are discussed and presented in the next subsection.
 
 As in the previous case, we can repeat the counting using the amplitudes that build the observables. The number of complex amplitudes is $n_A=10$ with the same number of symmetries $n_{sym}=2$ (see Eq.(\ref{massivesym})) as in the massless case, such that we confirm that there are 18 independent observables.
 
 The set of 18 independent observables in the massive case consists of 10 (11 coefficients minus one relation) independent observables for the P wave and 8 (14 coefficients minus 6 relations) independent observables for the S wave.  The corresponding basis of 25 observables is:
 \begin{eqnarray}\label{eq:massive_obs_list}
{\cal O}_{m_\ell\neq 0}&\!\!=\!\!&\{\Gamma^\prime, F_L, M_1, M_2, P_1, P_2, P_3, P_4^\prime, P_5^\prime, P_6^\prime, P_8^\prime,\hfill \nonumber \\ &&\quad \quad \quad \quad F_S, M_3^\prime, M_4^\prime, M_5^\prime, { S}_{S1}^r, S_{S2}^r, S_{S3}^r, S_{S4}^r, S_{S5}^r, { S}_{S1}^i, S_{S2}^i, S_{S3}^i, S_{S4}^i, S_{S5}^i \}\,. \quad\quad
\end{eqnarray}
Therefore, among this set of 25 observables there are 7 relations and only 18 observables are independent.

\subsection{P-wave and S-wave symmetry relations among observables}
\label{sec:SymRel}
In this subsection we present for the first time the full set of symmetry relations of the P and S wave in the massive case. These complete the previous partial results given in Refs.~\cite{Egede:2010zc,Matias:2012xw,Matias:2014jua,Hofer:2015kka}. It is helpful to express the observables $J_i$ and $\tilde J_i$  in terms of scalar products $n_i^\dagger n_j$, as shown in Eq.\eqref{sset}.
{ All the relations found in this section are functions of $J_i$ and $\tilde{J}_i$ and an equivalent set of relations in terms of the $CP$-conjugate partners $\bar{J}_i$ and $\bar{\tilde{J}}_i$ can be written. However, the observables are functions of the coefficients and their CP partners. This means that when writing one of these relations in terms of observables the substitution $J_j \to a P_i$ is strictly speaking $J_j \to a (P_i+P_i^{CP})/2$ (with $a$ being some normalization factor). The observable $P_i^{CP}$ is the $CP$ asymmetry associated with the observable $P_i$, defined in Ref.~\cite{Descotes-Genon:2013vna,Matias:2014jua}, and similarly for $\tilde{J}_i$.
For the following analysis and  for simplicity, we will neglect the  $CP$ asymmetries for both the P and S wave. This is a very good approximation, given that such asymmetries are tiny both in the SM and in presence of NP models that do not have  large NP phases.
}

Following the strategy in Ref.~\cite{Hofer:2015kka}, we exploit the fact that a couple of  $n_i$ vectors (with $i=\perp,\|,0,S$ or $i=\perp,\|,0,S^\prime$)  span the space of complex 2-component vectors. We therefore express the other vectors as linear combinations of these vectors. For instance,
\eq{\label{eq:decompbasic}
n_i=a_in_\|+b_in_\bot,\quad i=0,S. 
}
Contracting with the vectors $n_\|$ and $n_\bot$, we obtain a system of linear equations~\cite{Hofer:2015kka}
\begin{eqnarray}
   n_\|^\dagger n_i&=&a_i|n_\||^2+b_i (n_\|^{\dagger} n_\bot), \nonumber\\
  n_\bot^\dagger n_i&=&a_i (n_\bot^{\dagger} n_\|) + b_i |n_\bot|^2,
\end{eqnarray}
which can be solved for $a_i,b_i$:
\begin{eqnarray}
  a_i=\frac{|n_\bot|^2(n_\|^\dagger n_i)-(n_\|^\dagger n_\bot)(n_\bot^\dagger n_i)}{|n_\||^2|n_\bot|^2-|n_\bot^\dagger n_\||^2},\qquad
  b_i=\frac{|n_\||^2(n_\bot^\dagger n_i)-(n_\bot^\dagger n_\|)(n_\|^\dagger n_i)}{|n_\||^2|n_\bot|^2-|n_\bot^\dagger n_\||^2}.
\end{eqnarray}
Using the decomposition of $n_0,n_S$ in terms of $n_\|,n_\bot$ (Eq.\eqref{eq:decompbasic}) to calculate the scalar products $|n_0|^2,|n_S|^2,n_0^\dagger n_S$, the first three relations are obtained. We leave the expressions explicitly in terms of $J_i$ to let the reader choose between different bases or conventions to write the P-wave observables.\bigskip

\noindent I. From $i=0$ in Eq.\eqref{eq:decompbasic} one finds 
$|n_0|^2=a_0(n_0^\dagger n_\|)+b_0(n_0^\dagger n_\bot)$ yielding the first relation:
\eqa{ 0=\!&+\!\!&J_{2c} (16 J_{2s}^2 - 4 J_3^2 - \beta^2 J_{6s}^2 - 4 J_9^2)
 + 
 2 (J_3 (4 J_4^2 + \beta^2 (- J_5^2 + J_7^2) - 4 J_8^2)
 \\
 \!&+\!\!&
    2 J_{2s} (4 J_4^2 + \beta^2 (J_5^2 + J_7^2) + 4 J_8^2) - 
    2 (\beta^2(J_4 J_5 J_{6s} + J_{6s} J_7 J_8 + J_5 J_7 J_9) - 4 J_4 J_8 J_9))\,. \nonumber}

This first relation was found in the massless case in Ref.~\cite{Egede:2010zc} and in the massive case in Ref.~\cite{Descotes-Genon:2013vna} and its consequences discussed in Ref.~\cite{Matias:2014jua} once re-expressed in terms of optimized observables:
\begin{eqnarray}  \label{relation_old}{ P_2}= \frac{( { P_4^\prime} { P_5^\prime} + \delta_{1})}{2 k_1} + \frac{1}{2 k_1\beta }\sqrt{(-1+{ P_1} + { P_4^{\prime 2}})(-1-{ P_1} + \beta^2 { P_5^{\prime 2}}) +\delta_{2} + \delta_3 { P_1} +\delta_4 { P_1}^2  }\,\,\,\quad
\,  \end{eqnarray}
 where the parameters $k_1$ and $\delta_i$ (with $i=1,...4$) are defined in Ref.~\cite{Matias:2014jua}.
\bigskip

\noindent II. Similarly for $i=S$ in Eq.(\ref{eq:decompbasic})
one finds $|n_S|^2=a_S(n_S^\dagger n_\|)+b_S(n_S^\dagger n_\bot)$ and this translates to:
\eqa{0=&&\!\!\!
-{ \frac{27}{16}} \beta^4 {F_S}  J_{6s}^2 +{ \Gamma^{\prime}}[- 8 (2 J_{2s} + J_3) S_{S2}^{r \, 2} - 16 J_9 S_{S2}^r S_{S5}^i + 
8 (-2 J_{2s} + J_3) S_{S5}^{i \,2} ]
\nonumber \\
&&+ 
2 \beta^2 (  \frac{27}{8}  {F_S}  (4 J_{2s}^2 - J_{3}^2 - J_{9}^2) 
+ { \Gamma^{\prime }} [(-2 J_{2s} + 
      J_3) S_{S3}^{r\, 2} + 2 J_9 S_{S3}^{r} S_{S4}^{i} \nonumber \\
    &&  - (2 J_{2s} + J_{3}) S_{S4}^{i \,2} + 
   2 J_{6s} (S_{S2}^{r} S_{S3}^{r} + S_{S4}^{i} S_{S5}^{i})]) , \label{secondrel}
}
once expressed in terms of S-wave observables.
\bigskip

\noindent III. Finally, the scalar product $n_0^\dagger n_S$ leads to the third relation:
\eqa{0=&&\!\!\! { 2\,}[-16 J_{2s}^2  + 4 J_3^2  + \beta^2 J_{6s}^2  + 4 J_9^2 ]{ S_{S1}^{r}}+ 
4  [ \beta^2 J_5 J_{6s}  - 4 J_8 J_9 -8 J_{2s} J_4 -4 J_3 J_4 ] S_{S2}^{r}
 \nonumber \\ 
 &&
 +4 \beta^2 [  J_4 J_{6s}  
 +   J_7 J_9 -2  J_{2s} J_5  +   J_3 J_5 ] S_{S3}^{r} + 4 \beta^2 [  J_{6s} J_8  +   J_5 J_9
\nonumber \\ 
 && -2 J_{2s} J_7 -  J_3 J_7 ] S_{S4}^{i}   +4 [ \beta^2 J_{6s} J_7  - 4 J_4 J_9 -8 J_{2s} J_8+ 4 J_3 J_8 ] S_{S5}^{i} .\label{thirdrel} }

\noindent
Eq.(\ref{secondrel}) and Eq.(\ref{thirdrel}) are the generalizations of the massless limit ($\beta\to 1$) expressions found in Ref.~\cite{Hofer:2015kka}.

Following the same methodology but using instead the vector $n_S^\prime$ yields three new relations. Expressing $n_S^\prime$ in terms of $n_\perp$ and $n_\|$:
\eq{\label{eq:decomp}
n_S^\prime=a_S^\prime n_\|+b_S^\prime n_\bot,\quad 
}
and contracting with $n_\|$ and $n_\bot$ we get a system of linear equations
\begin{eqnarray}
   n_\perp^{\dagger} n_S^\prime &=&a_S^\prime  (n_\perp^{\dagger} n_\|)   +b_S^\prime  |n_\perp|^2     
   , \nonumber\\
   n_\|^{\dagger} n_S^\prime  &=&a_S^\prime |n_\||^2 + b_S^\prime (n_\|^\dagger n_\perp).
\end{eqnarray}
We can determine $a_S^\prime$ and $b_S^\prime$:
\begin{eqnarray}
  a_S^\prime=\frac{(n_\|^{\dagger} n_S^\prime) |n_\perp|^2    -     (n_\perp^{\dagger} n_S^\prime)   (n_\|^\dagger n_\perp)
}{   |n_\||^2    |n_\perp|^2     -   |n_\perp^{\dagger} n_\||^2 }    
  ,\qquad
  b_S^\prime=\frac{(n_\|^{\dagger} n_S^\prime)   (n_\perp^{\dagger} n_\|)  -    (n_\perp^{\dagger} n_S^\prime)    |n_\||^2}{  
  |n_\perp^{\dagger} n_\||^2  - |n_\||^2 |n_\perp|^2   } .
\end{eqnarray}
Using the properties of the vector $n_S^\prime$ we then obtain the following three relations:
\bigskip

\noindent IV. From the equality of the modulus of both vectors $n_S$ and $n_S^\prime$ one obtains  \eqa{|n_S^\prime|^2=|n_S|^2=a_S^\prime (n_S^{\prime\dagger} n_\|) + b_S^{\prime} (n_S^{\prime\dagger} n_\bot)\,,} 
which implies the following relation:
\eqa{    
0=\!&&\!\!+{ \frac{27}{16}} \beta^2 {{F_S}} (16 J_{2s}^2 - 4 J_3^2 - \beta^2 J_{6s}^2 - 4 J_9^2) - 
 2 { \Gamma^{\prime}} [-2 (\beta^2 J_{6s} S_{S2}^{i} S_{S3}^{i} - \beta^2 J_9 S_{S3}^{i} S_{S4}^{r}  \nonumber \\ &&
      + 4 J_9 S_{S2}^{i} S_{S5}^{r} + \beta^2 J_{6s} S_{S4}^{r} S_{S5}^{r}) + 
    4 S_{S2}^{i \, 2} (J_3+ 2 J_{2s}) + \beta^2 S_{S3}^{i \, 2}  (2 J_{2s}-J_3) 
  \nonumber \\ &&  
    + \beta^2   S_{S4}^{r \,2} (J_3+2 J_{2s}) + 4 S_{S5}^{r \,2}(2 J_{2s}-J_3)] .
    }

\noindent V. Above we focus on relations constructed from the real part of the product of vectors. The imaginary parts provide additional new relations:
\eq{ {\rm Im}[n_0^\dagger n_S^\prime]={\rm Im}[a_S^\prime (n_0^\dagger n_\|) + b_S^\prime (n_0^\dagger n_\perp)]\,,
}
which leads to:
\eqa{
 0=\!\!&&\!\!2\, [ -16 J_{2s}^2  + 4 J_3^2  + \beta^2 J_{6s}^2  +4 J_9^2 ] { S_{S1}^{i}}+ 
 [4 \beta^2 J_5 J_{6s}  - 16 J_8 J_9-16 J_3 J_4-32 J_{2s} J_4 ] S_{S2}^{i}\nn \\ &&+4 \beta^2 [ J_4 J_{6s}   
 +J_7 J_9+  J_3 J_5 -2 J_{2s} J_5] S_{S3}^{i}+
 4 \beta^2 [-  J_{6s} J_8  -  J_5 J_9  +  J_3 J_7 +2 J_{2s} J_7 
 ] S_{S4}^{r} \nn \\ &&+
  [-4 \beta^2 J_{6s} J_7  + 16 J_4 J_9    
  -16  J_3  J_8 +32 J_{2s} J_8]  S_{S5}^{r} .
\quad \quad    }

\noindent 
VI. Finally, combining the vectors $n_S$ and $n_S^\prime$ one finds:
\eq{ {\rm Im}[n_S^\dagger n_S^\prime]=0=a_S^\prime (n_S^\dagger n_\|) + b_S^\prime (n_S^\dagger n_\perp)\,,} 
which corresponds to
\eqa{ 
0=\!&&\! \! \beta^2 J_{6s}[ - S_{S2}^{r} S_{S3}^{i} -  S_{S2}^{i} S_{S3}^{r} +  S_{S4}^{r} S_{S5}^{i}  
+  S_{S4}^{i} S_{S5}^{r}]
+ J_9 [- \beta^2 S_{S3}^{i} S_{S4}^{i} + 
 \beta^2 S_{S3}^{r} S_{S4}^{r} + 4  S_{S2}^{i} S_{S5}^{i}
\nonumber \\
&&
- 4  S_{S2}^{r} S_{S5}^{r}]  + 
 2 J_{2s} [4 S_{S2}^{i} S_{S2}^{r} 
  + \beta^2 (S_{S3}^{i} S_{S3}^{r} - S_{S4}^{i} S_{S4}^{r})
 - 4 S_{S5}^{i} S_{S5}^{r}] 
 +J_3 [4 S_{S2}^{i} S_{S2}^{r} 
\nn \\ && 
 - \beta^2( S_{S3}^{i} S_{S3}^{r} + S_{S4}^{i} S_{S4}^{r})
 + 4 S_{S5}^{i} S_{S5}^{r}].
 }

These six relations are common to the massive and massless case, and they reduce to the massless case when taking the limit $\beta \to 1$.  There is a very long seventh relation that applies only in the massive case, i.e. it is zero in the limit of massless leptons. For this reason and given that it is difficult to extract information from such a long relation, we refrain from writing it explicitly and, instead, provide only the main steps to obtain this relation in the Appendix.

\begin{figure}[h!] \begin{center}
    \includegraphics[width = 0.49\textwidth]{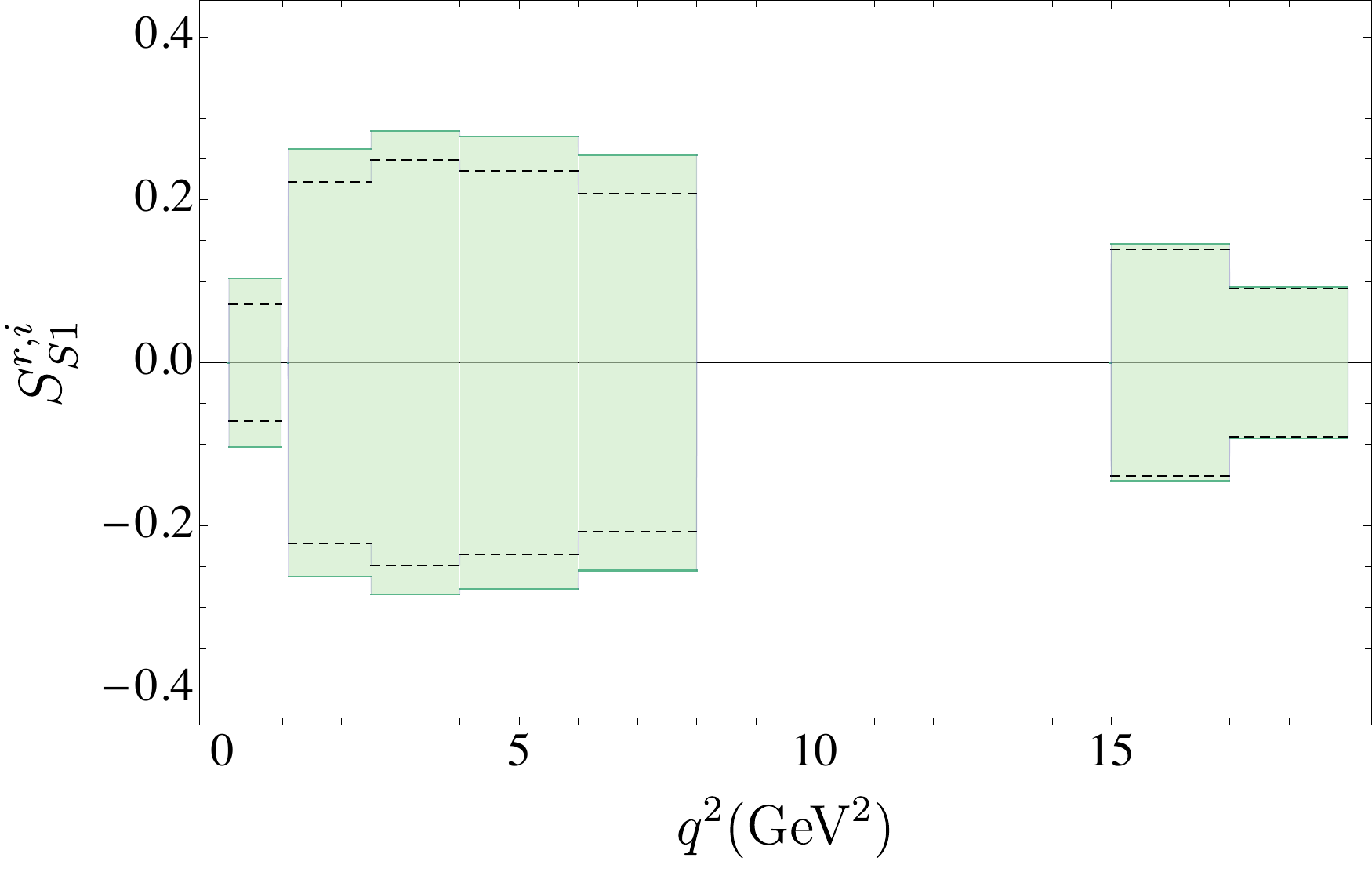}
    \includegraphics[width = 0.49\textwidth]{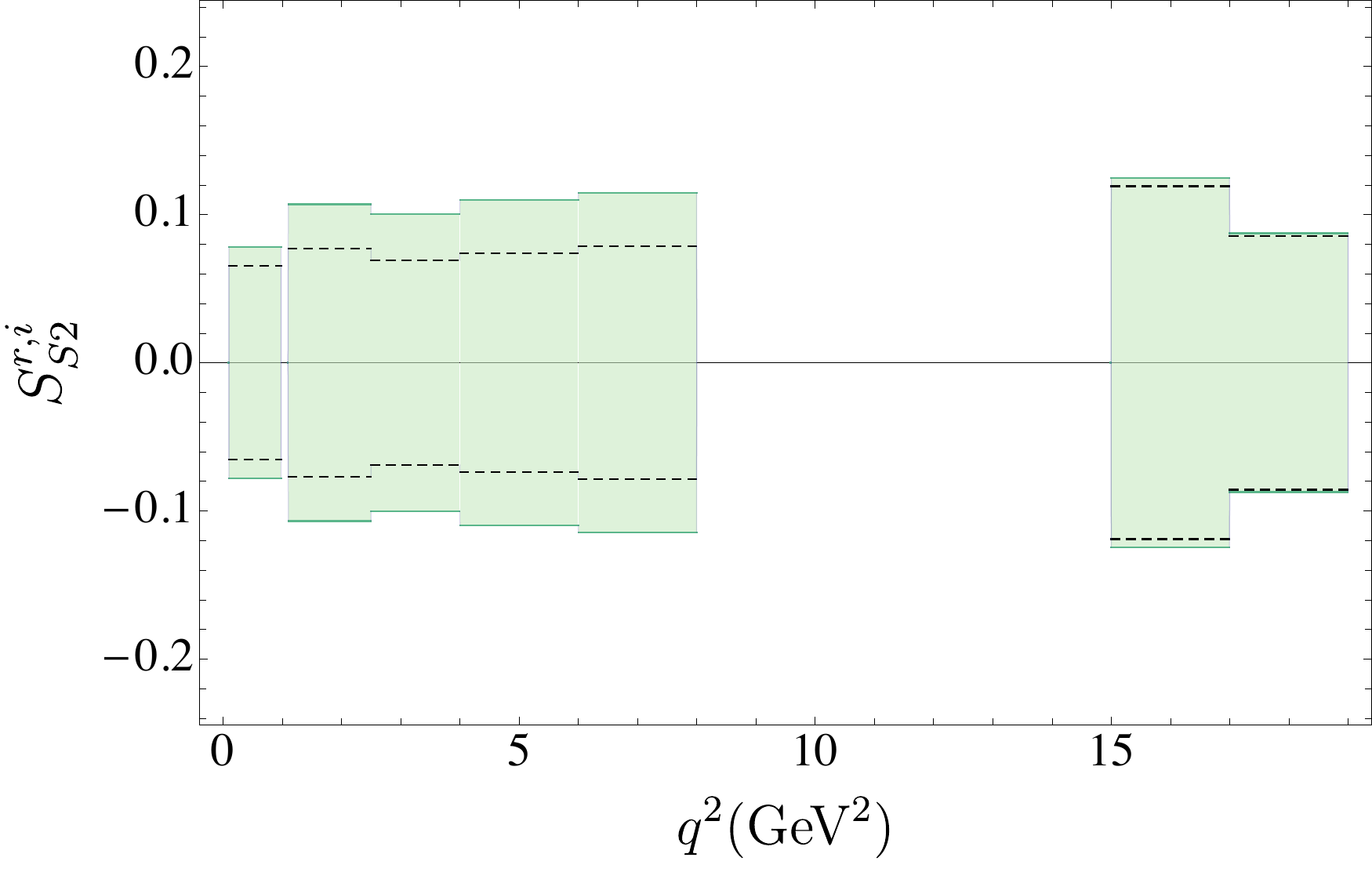}
        \includegraphics[width = 0.49\textwidth]{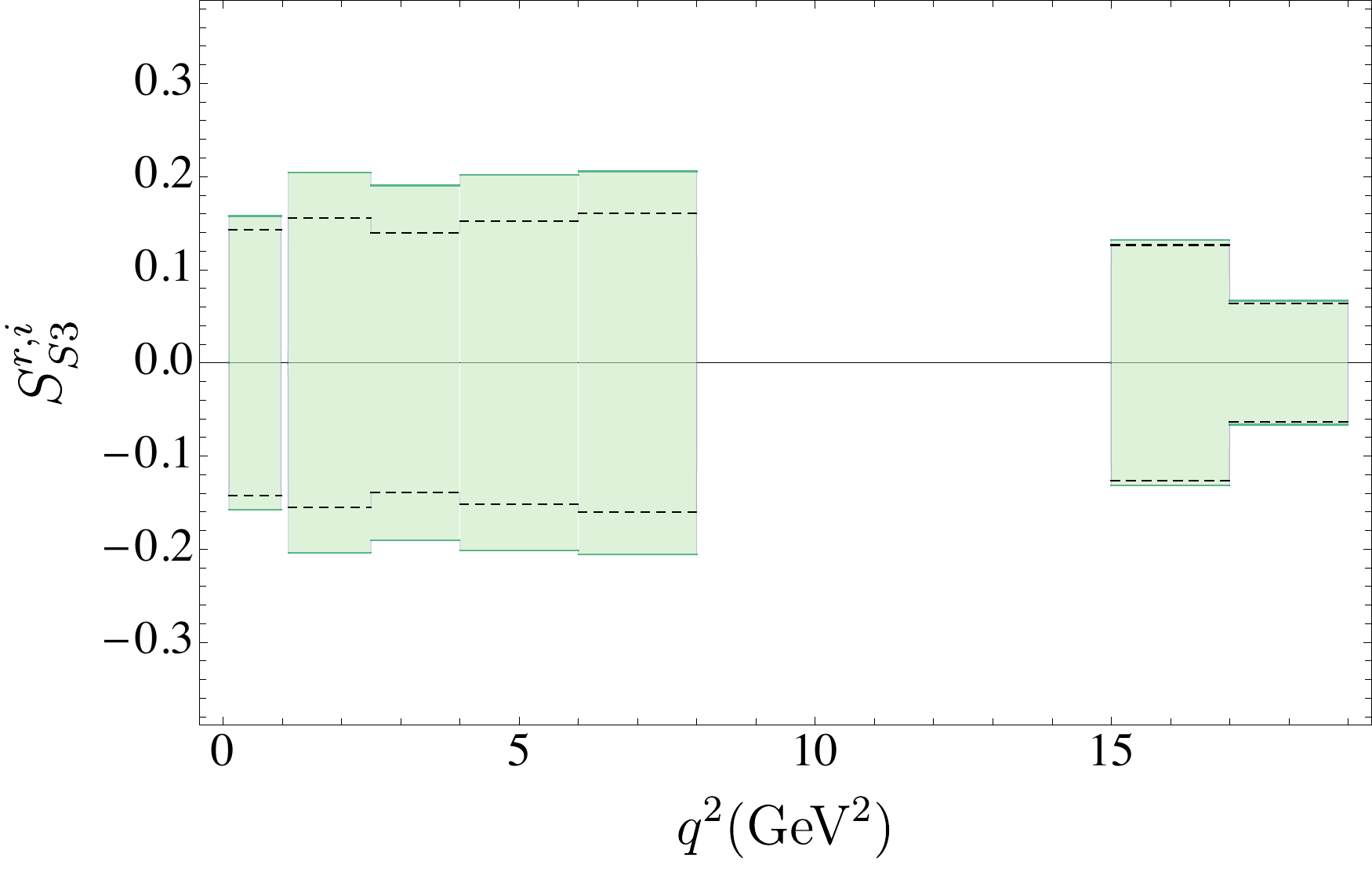}    
        \includegraphics[width = 0.49\textwidth]{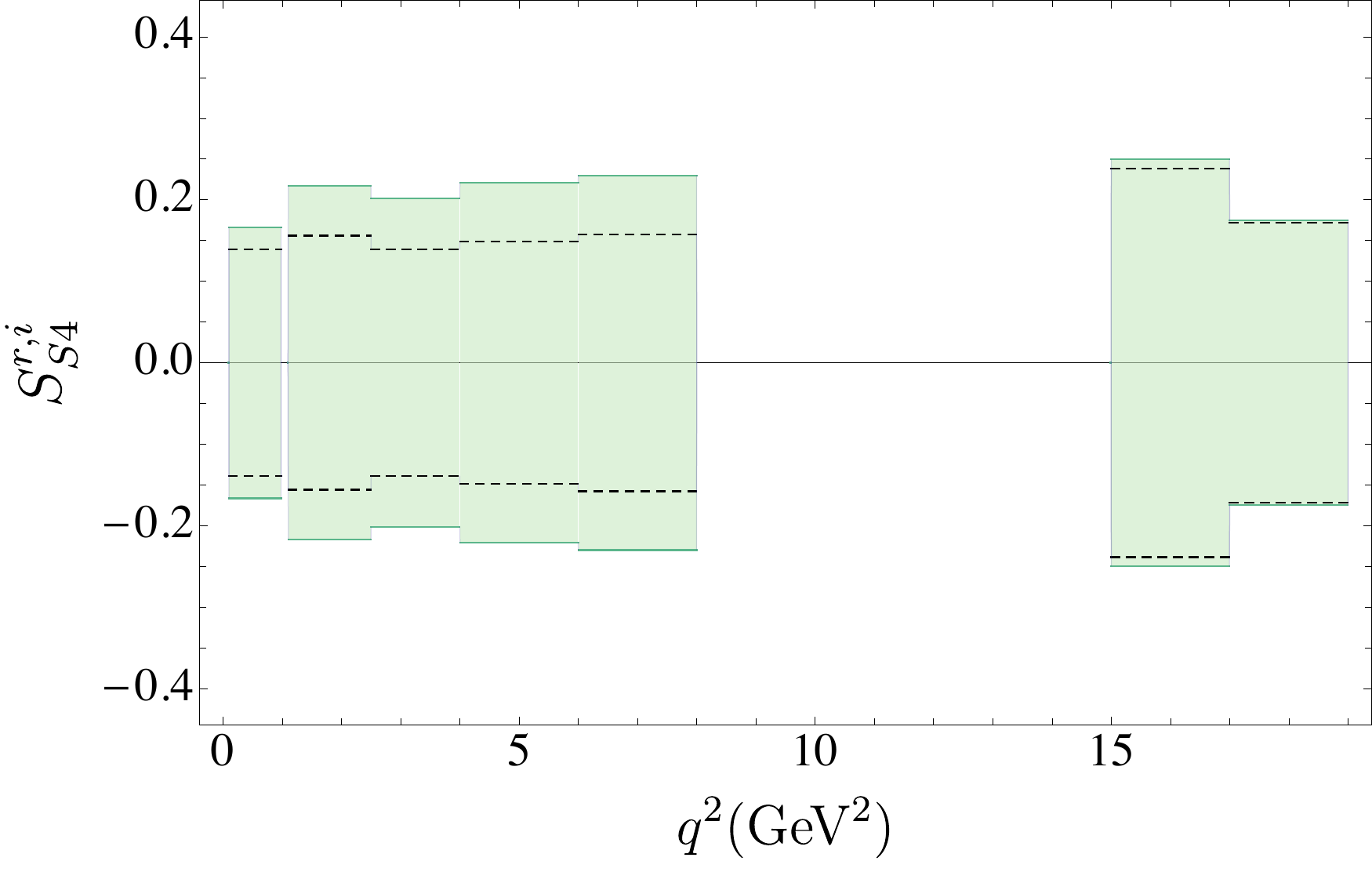}    
           \includegraphics[width = 0.49\textwidth]{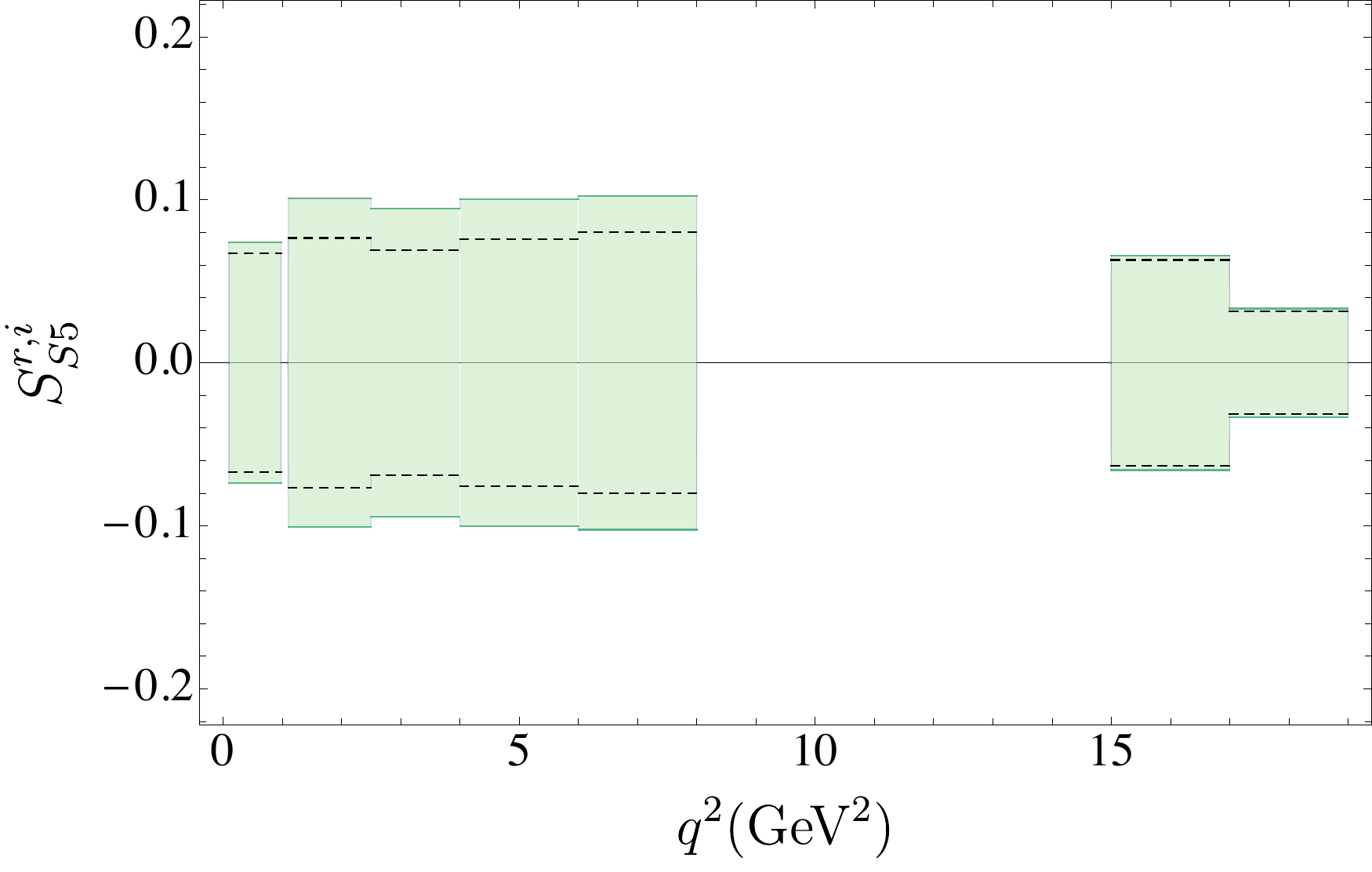} \end{center}
                   \caption{Bounds for $S^{r,i}_{Si}$ binned observables with $i=1,2,3,4,5$. The dashed line corresponds to the central value of the bound in the SM, while the green regions include the uncertainty of the observables that define the bound in the SM.    } 
    \label{fig:p_basis_summary}
\end{figure}

\begin{figure}[h!]
              \includegraphics[width = 0.49\textwidth]{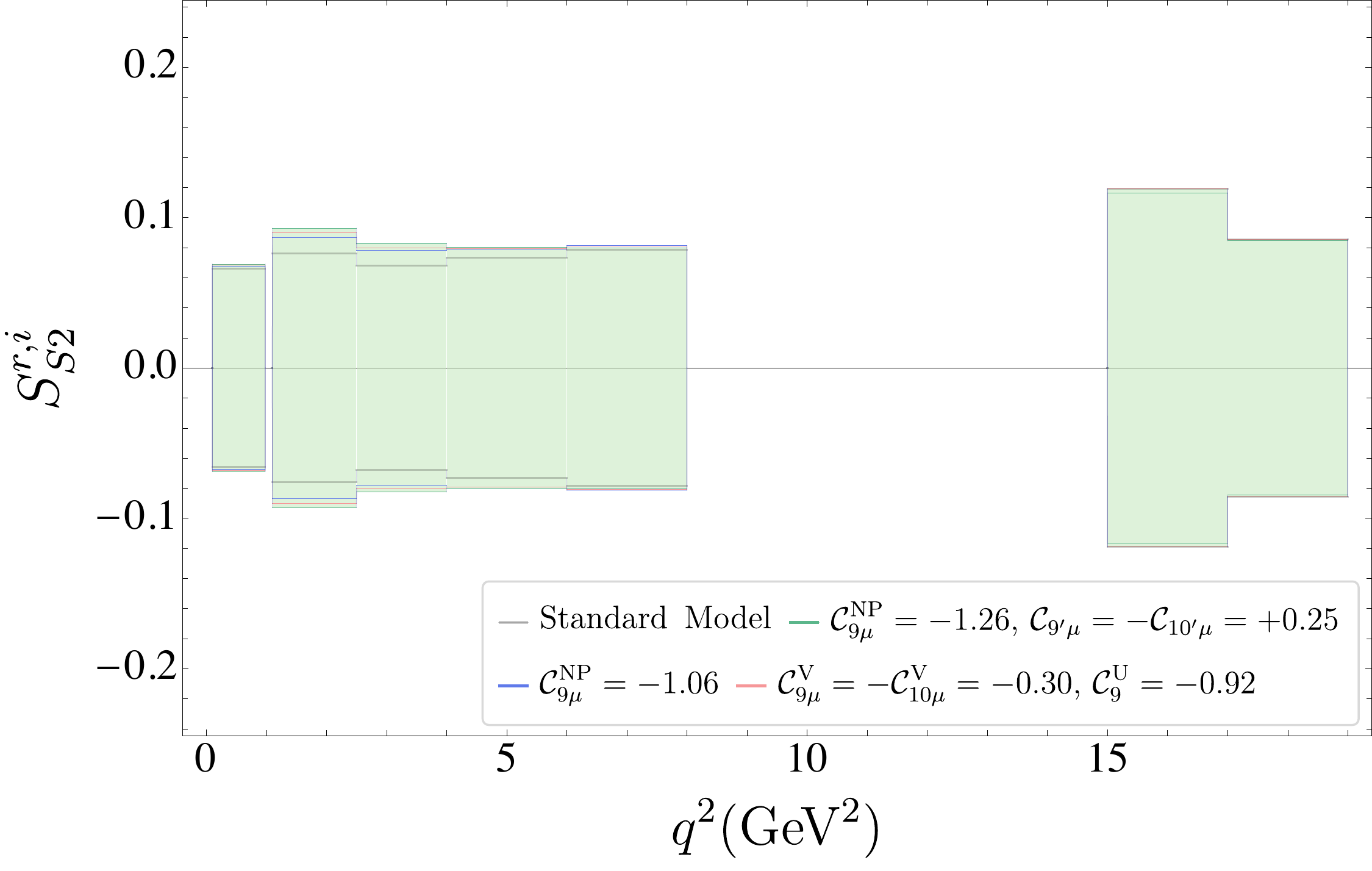}
        \includegraphics[width = 0.49\textwidth]{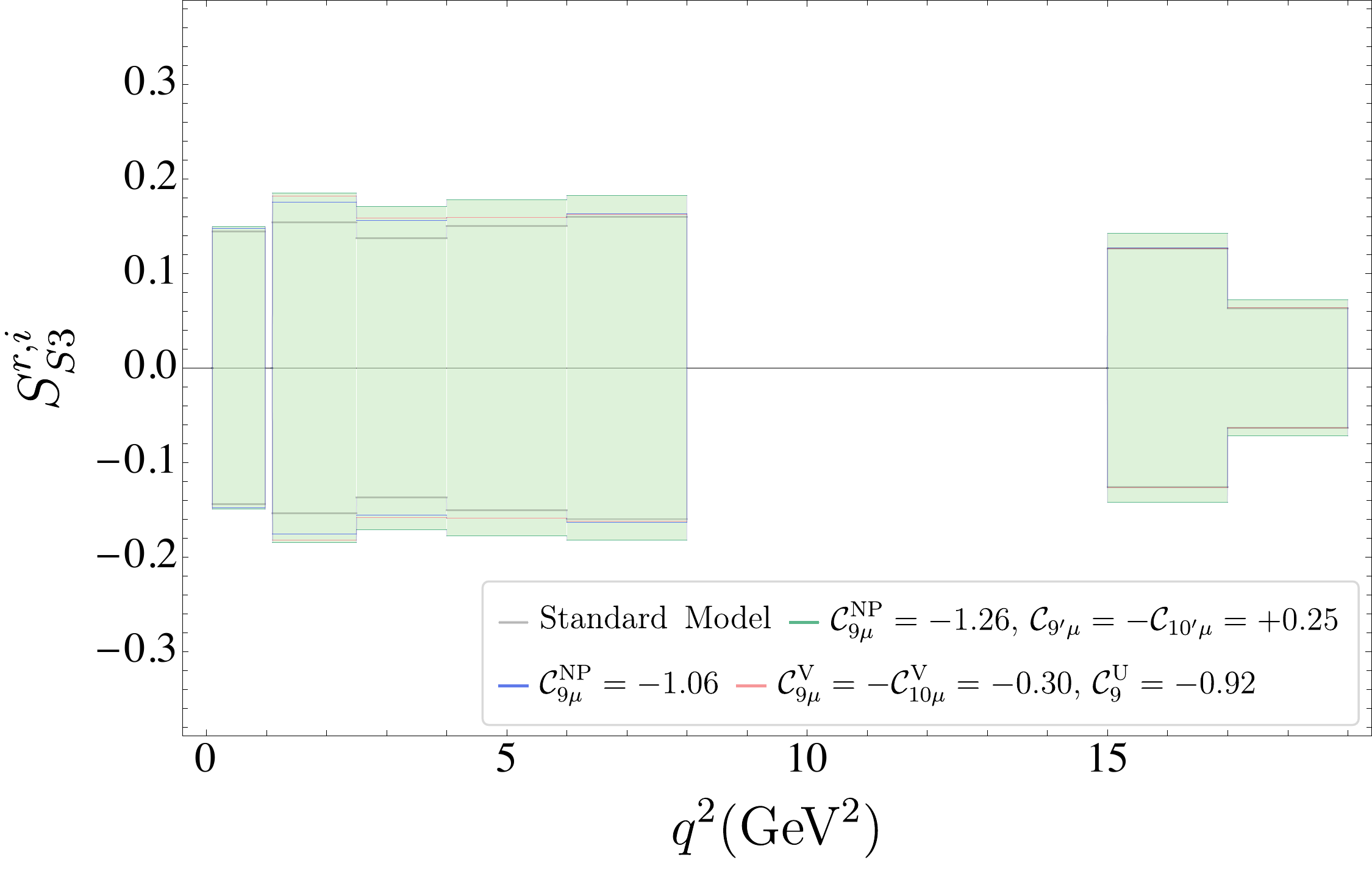}               
    \caption{Illustration of the sensitivity of the central value of the bound to the preferred NP scenarios for two observables $S^{r,i}_{S2}$ and $S^{r,i}_{S3}$. We have checked explicitly that the variation of the bound in the most significant NP scenarios amounts to at most a 20-25\% enhancement. }     \label{fig:p_basis_summary_NP}
\end{figure}

\section{Bounds on  S-wave observables and \texorpdfstring{$W_{1,2}$}{W1,2}  observables}\label{sec:bounds}

Following the strategy of Ref.~\cite{Hofer:2015kka}, the relations found in the previous section enable bounds to be placed on the $S^r_{Si}$ observables and the newly defined $S^i_{Si}$ observables.
For instance,  solving for $S_{S2}^r$ and imposing a real solution in relation II gives:
\begin{eqnarray}
0\leq\Delta(S_{S2}^r)=\!&-&\!\beta^2 x (S_{S3}^{r})^2 - 4 x (S_{S5}^{i})^{2} - \beta^2 (2 P_3 S_{S3}^r + (1+ P_1) S_{S4}^i - 4P_2 S_{S5}^i)^2 \nonumber \\
&+&{ \frac{27}{16}} \beta^4 x F_S (1-F_S^\prime)  F_T (1+ P_1)\,,   \label{dets2r}
\end{eqnarray}
where $x=1-P_1^2-4\beta^2 P_2^2 - 4 P_3^2 \geq 0$ (see Ref.\cite{Hofer:2015kka}). The first three terms are negative definite and each of them separately has to be smaller  than the last positive definite term. In a similar way but solving for $S_{S3}^r$ and imposing a real solution one finds:
\begin{eqnarray}
0\!\leq\!\Delta(S_{S3}^r)=\!&-&\!\beta^2 x (S_{S4}^{i})^2 - 4 x (S_{S2}^{r})^{2} - 4 (2 P_3 S_{S2}^r - (1- P_1) S_{S5}^i + \beta^2 P_2 S_{S4}^i)^2\nonumber \\
&+& { \frac{27}{16}} \beta^4 x F_S (1-F_S^\prime) F_T (1- P_1)\,. \label{dets3r}
\end{eqnarray}
\noindent This implies the following constraints for $S_{S2,3}^r$:
\begin{equation}
|S_{S2}^r| \leq \beta^2 { \frac{3}{4}} \sqrt{\frac{3}{4} F_S (1-F_S^\prime) F_T (1-P_1)} \quad 
|S_{S3}^r| \leq \beta { \frac{3}{4}} \sqrt{3 F_S (1-F_S^\prime) F_T (1+P_1)}, \quad 
\end{equation}
and for $S_{S4,5}^i$:
\begin{equation}
|S_{S4}^i| \leq  \beta { \frac{3}{4}} \sqrt{3 F_S (1-F_S^\prime) F_T (1-P_1)} \quad
|S_{S5}^i| \leq \beta^2 { \frac{3}{4}} \sqrt{\frac{3}{4} F_S (1-F_S^\prime) F_T (1+P_1)}\,.
\end{equation}
Similarly using relation IV, one finds
\begin{eqnarray}
0\leq\Delta(S_{S4}^r)=\!&-&\!\beta^2 x (S_{S3}^{i})^2 - 4 x (S_{S5}^{r})^{2} - 4 ( 2 P_3 S_{S5}^r+(1+P_1) S_{S2}^i -\beta^2 P_2 S_{S3}^i)^2
\nonumber \\
&+&  { \frac{27}{16}} \beta^4 x F_S  { (1-F_S^\prime)} F_T (1+P_1)\,,  \label{dets4r} 
\end{eqnarray}
and
\begin{eqnarray}
0\leq\Delta(S_{S5}^r)=\!&-&\!\beta^2 x (S_{S4}^{r})^2 - 4 x (S_{S2}^{i})^{2} - \beta^2 ( 4 P_2 S_{S2}^i-(1-P_1) S_{S3}^i + 2 P_3 S_{S4}^r)^2\nonumber \\
&+& { \frac{27}{16}} \beta^4 x F_S { (1-F_S^\prime)} F_T (1-P_1)\,,\label{dets5r}
\end{eqnarray}
which leads to the following bounds:
\begin{equation}
|S_{S4}^r| \leq \beta { \frac{3}{4}} \sqrt{3 F_S (1-F_S^\prime) F_T (1-P_1)}\,, \quad 
|S_{S5}^r| \leq \beta^2 { \frac{3}{4}} \sqrt{\frac{3}{4} F_S (1-F_S^\prime) F_T (1+P_1)}\,, \quad
\end{equation}
and 
\begin{equation}
|S_{S2}^i| \leq  \beta^2 { \frac{3}{4}} \sqrt{\frac{3}{4} F_S (1-F_S^\prime) F_T (1-P_1)}\,, \quad
|S_{S3}^i| \leq  \beta { \frac{3}{4}} \sqrt{3 F_S (1-F_S^\prime) F_T (1+P_1)}\,.
\end{equation}
In summary, 
\begin{equation}
|S_{S2}^{r,i}| \leq  \beta^2 \frac{k_1}{2} ,  \quad |S_{S3}^{r,i}| \leq \beta k_2 , \quad |S_{S4}^{r,i}| \leq \beta k_1 , \quad |S_{S5}^{r,i}| \leq \beta^2 \frac{k_2}{2} ,
\label{boundsall}
\end{equation}
with $k_1={\frac{3}{4}}\sqrt{3 F_S (1-F_S^\prime) F_T (1-P_1)}$ and $k_2={\frac{3}{4} }\sqrt{3 F_S (1-F_S^\prime) F_T (1+P_1)}$. 
All the bounds above can alternatively be obtained using the Cauchy-Schwarz inequalities. For the observables $S_{S1}^{r,i}$ this is the only way to obtain the bounds. For instance, from $|n_0^\dagger n_S|^2 \leq |n_0|^2 |n_S|^2$ and a corresponding inequality with $n_S^\prime$ using the properties of the vectors Eq.(\ref{eq:nvecs}) one arrives at
\begin{equation}
    |S_{S1}^{r,i}| \leq \beta^2 \frac{3}{4} \sqrt{3} \sqrt{F_S (1-F_S^\prime) F_L}\,.
\end{equation}
All the bounds on the other observables can be re-derived using the four inequalities:
\begin{equation}
    |n_\|^\dagger n_S^{(\prime)}|^2 \leq |n_\||^2 |n_S|^2 \,, \quad \quad |n_\perp^\dagger n_S^{(\prime)}|^2 \leq |n_\perp|^2 |n_S|^2\,.
\end{equation}

We have computed explicitly the bounds of the $S^{r,i}_{Si}$ observables in the SM in Fig.~\ref{fig:p_basis_summary}. The relatively low sensitivity of the central value of the bound for $S^{r,i}_{S2,3}$ on the dominant NP scenarios is illustrated in Fig.~\ref{fig:p_basis_summary_NP}. We work under the approximation of substituting $q^2$ dependent observables by their binned equivalents, where we denote the latter using angular brackets. This introduces some uncertainty but, as shown in Ref.~\cite{Matias:2014jua}, this uncertainty is negligible, especially for slowly varying observables like those involved in the bounds. To compute the binned form of the bounds from Eq.\eqref{boundsall} we consider the theoretical prediction for the observables $\langle F_{L,T} \rangle, \langle P_1 \rangle$, taking into account the $1\sigma$ ranges of such observables. Therefore Fig.~\ref{fig:p_basis_summary} shows the maximum value allowed for such constraints. For $\langle F_S \rangle$ we extract the value from a reduced $m_{K\pi}$ resonance window, $0.795<m_{K\pi}<0.995 \, {\rm GeV}$. In Fig.~\ref{fig:p_basis_summary_NP} we evaluate $\langle F_{L,T} \rangle$ and $\langle P_1 \rangle$ in the corresponding NP scenarios, while taking the SM prediction for $\langle F_S \rangle $. The computation of $\langle F_S\rangle$ is the only place where we use S-wave form factors. If $\langle F_S\rangle$ is taken as an experimental input, then no S-wave form factors  are  required. Finally, notice that the bounds include a term $(1-F_S^\prime)$. However, in evaluating these bounds we have neglected a small lepton mass dependent term (see Eq.(\ref{defFsprime})) taking $F_S$ instead of $F_S^\prime$.

The third term in Eqs.\eqref{dets2r},\eqref{dets3r},\eqref{dets4r},\eqref{dets5r} should tend to zero when $x(q_1^2)\to 0$, in order not to violate the condition of a real solution. Indeed, if we repeat the same procedure using relation II but impose a real solution for $\Delta (S_{S4}^i)$ and  $\Delta (S_{S5}^i)$ and for relation IV impose a real solution for  $\Delta (S_{S2}^i)$ and  $\Delta (S_{S3}^i)$, we find respectively:
\begin{eqnarray}
\left((1+P_1) S_{S2}^r - \beta^2 P_2 S_{S3}^r - 2 P_3 S_{S5}^i \right)_{q_1^2}&=&0 \nn ,\\
\left(4 P_2 S_{S2}^r -(1-P_1) S_{S3}^r - 2 P_3 S_{S4}^i \right)_{q_1^2}&=&0 \nn ,\\
\left(4 P_2 S_{S5}^r - (1+P_1) S_{S4}^r + 2 P_3 S_{S3}^i \right)_{q_1^2}&=&0 \nn ,\\
\left((1-P_1) S_{S5}^r -\beta^2 P_2 S_{S4}^r + 2 P_3 S_{S2}^i \right)_{q_1^2}&=&0 .
\end{eqnarray}

Neglecting quadratically suppressed terms, $P_3 S_{Sj}^i \ll P_2 S_{Sj}^r$ with $j=2...5$, the previous equations can be combined to obtain:
\begin{eqnarray}
S_{S2}^r|_{q_1^2}&=&\left[\frac{\beta}{2} \sqrt{\frac{1-P_1}{1+P_1}} S_{S3}^r\right]_{q_1^2} \nn ,\\
S_{S5}^r|_{q_1^2}&=&\left[\frac{\beta}{2} \sqrt{\frac{1+P_1}{1-P_1}} S_{S4}^r\right]_{q_1^2},
\end{eqnarray}
and from $x(q_1^2)=0$, neglecting $P_3^2$, one finds at $q_1^2$ that $P_2=\sqrt{1-P_1^2}/(2\beta)$.

\begin{figure}[ht]
    \includegraphics[width=0.49\textwidth]{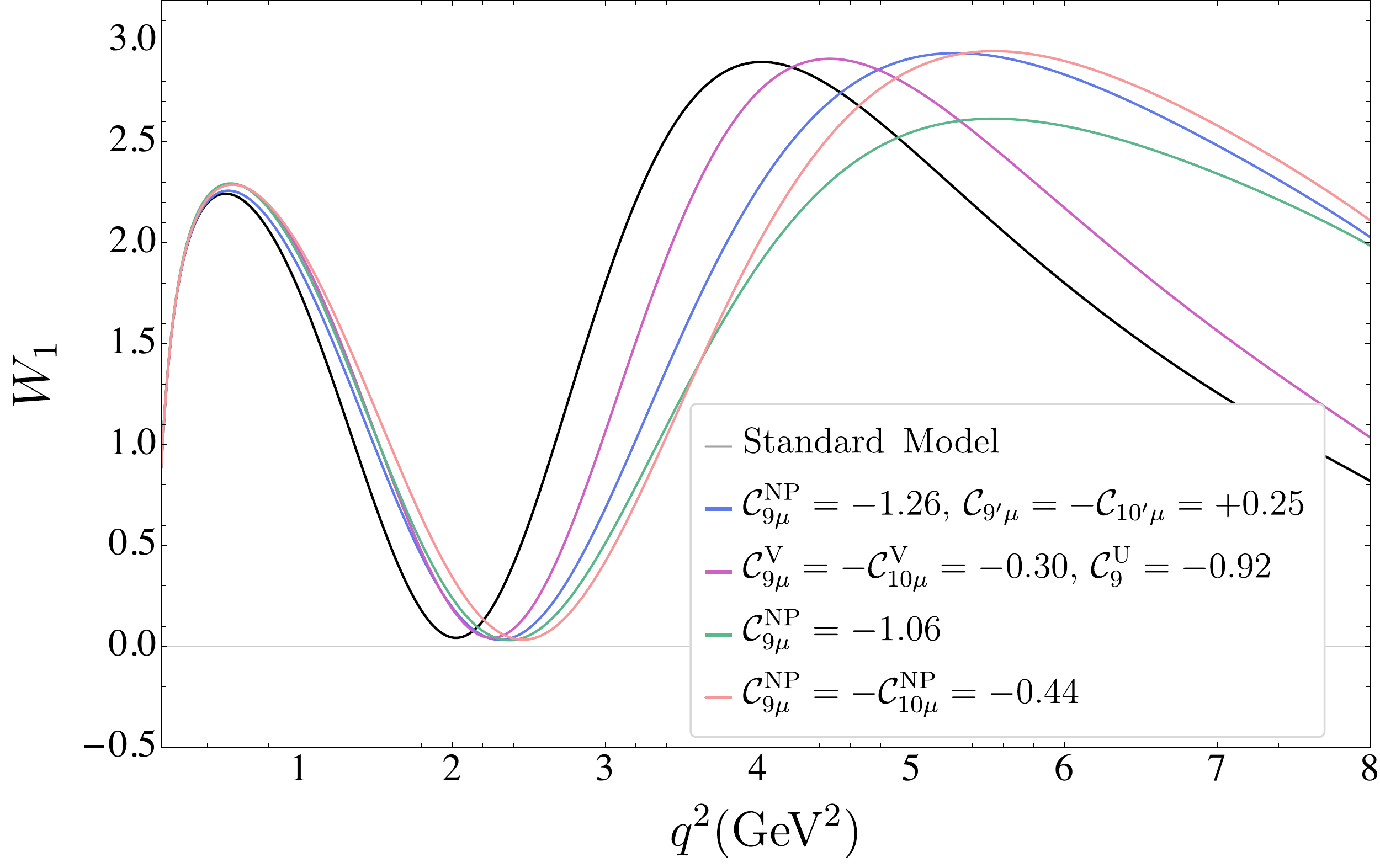}
    \includegraphics[width=0.49\textwidth]{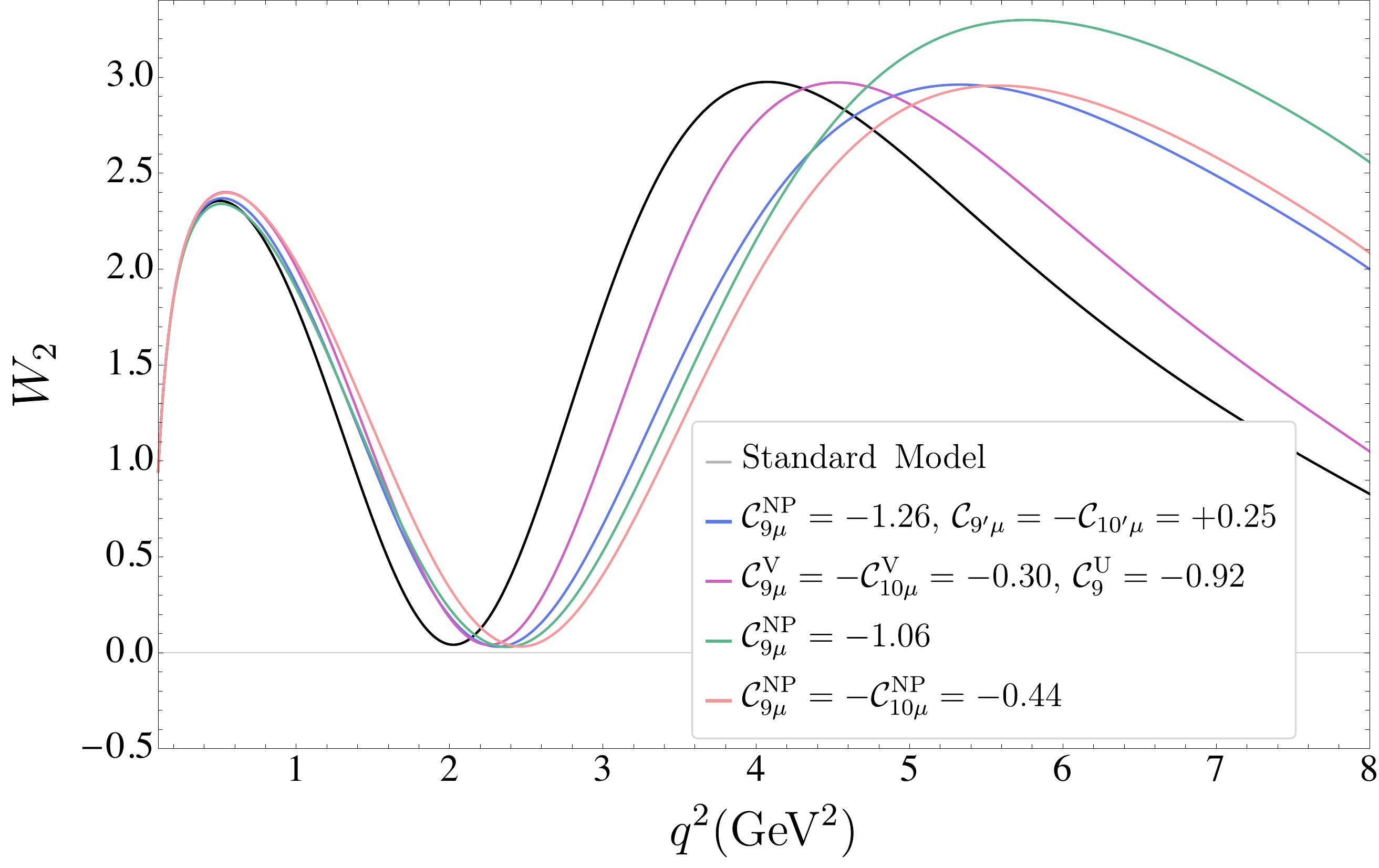}
   \includegraphics[width=\textwidth]{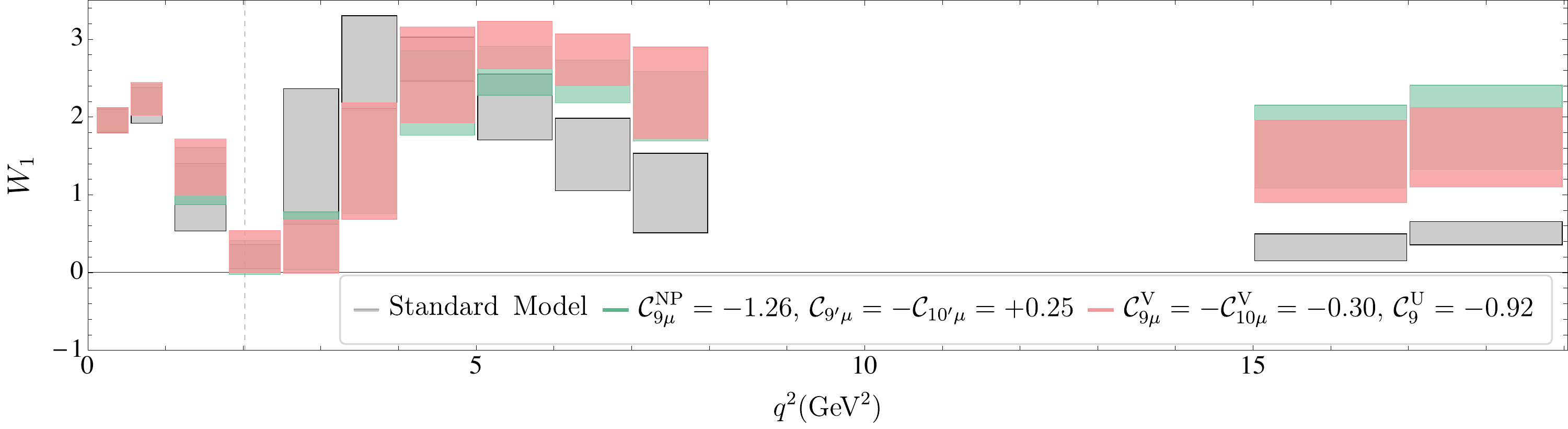}
   \includegraphics[width=\textwidth]{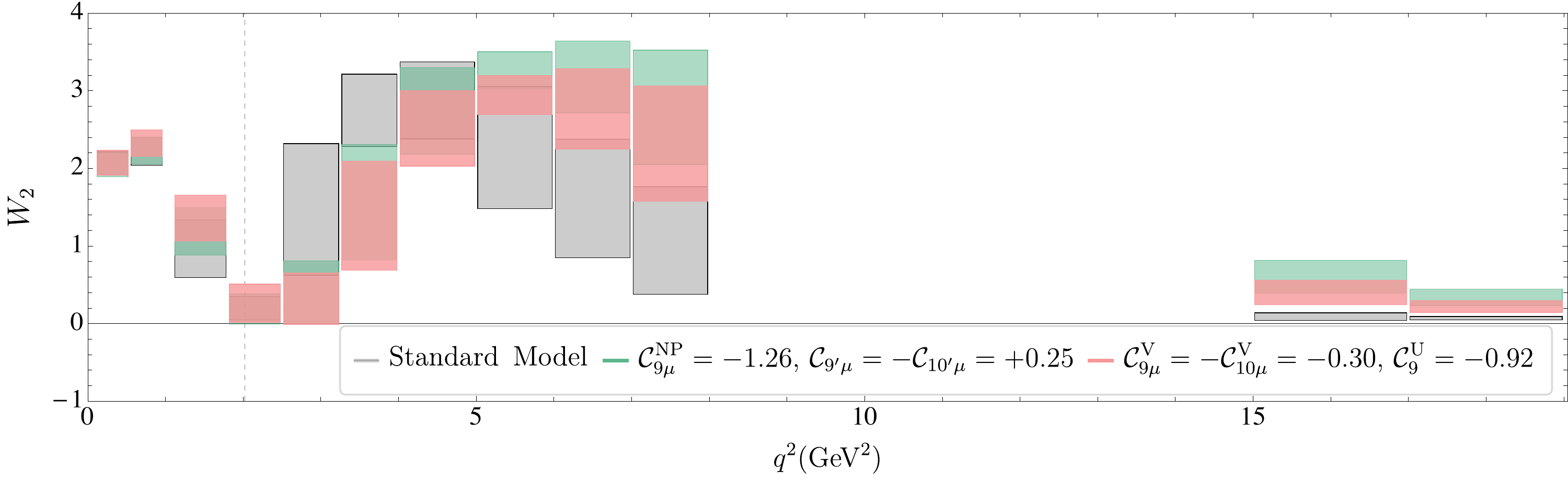}
  \caption{SM and NP predictions for the observables $W_1$ and $W_2$ as continuous functions of $q^2$ and binned in $q^2$.}\label{fig:w1w2}
\end{figure}

Another example of the information that can be extracted from the relations, neglecting quadratic terms of the type ${\cal O}(P_3 S_{Sj}^i, P_3^2)$, are the following expressions. These are valid for all $q^2$ and derive from relations II and IV, respectively. They can be tested as a cross-check of the experimental analyses:
\begin{equation} \label{eq54}
 W_1=(2\hat S_{S2}^r)^2 + p (\beta\hat S_{S3}^r )^2
 + q {\hat S_{S2}^r} {\hat S_{S3}^r} = 3 \beta^4 \frac{1}{1+P_1} x\,,
\end{equation}
\bigskip
where \begin{equation} \hat S_{Si}^r={ \frac{4}{3}}\frac{S_{Si}^r}{ \sqrt{(1-F_S^\prime)F_S F_T}}=\frac{\beta^2}{\sqrt{6}} PS_{i}^{r} ,\end{equation}
 with $i=2...5$,  $p= (1-P_1)/(1+P_1)$ and $q=-8 \beta^2 P_2/(1+P_1)$. Similarly,
\begin{equation} \label{eq55}
 W_2=(2\hat S_{S5}^r)^2 + p^\prime (\beta\hat S_{S4}^r )^2
 + q^\prime {\hat S_{S4}^r} {\hat S_{S5}^r} = 3 \beta^4 \frac{1}{1-P_1} x ,
\end{equation}
\bigskip
where  $p^\prime= (1+P_1)/(1-P_1)$ and $q^\prime=-8 \beta^2 P_2/(1-P_1)$. 

Eq.\eqref{eq54} is particularly interesting because at the zero of $\hat{S}_{S3}^r$ (or equivalently $PS_3^r$) one can predict the absolute value of $\hat{S}_{S2}^r$ (or $PS_2^r$) as a function of P-wave observables with no need to rely on any S-wave form factors. In the case of Eq.\eqref{eq55}, at the zero of $\hat{S}_{S4}^r$ one can  predict the absolute value of  $\hat{S}_{S5}^r$
at this particular value of $q^2$. These are valuable tests to compare with future predictions using calculations of the form factors.

Given that Eq.(\ref{eq54}) and Eq.(\ref{eq55}) are functions of P- and S-wave optimized observables (${P S_{i}^{r}}$ and $P_{1,2}$), $W_{1,2}$ are also optimized observables. We can compute SM and NP predictions for these two observables using the right hand side of Eq.(\ref{eq54}) and Eq.(\ref{eq55}), respectively. These relations then give access to the $n_{\perp,\|,0}$ components inside the new S-wave observables, cancelling the dependence on $n_{S}$ and $n_{S}^\prime$ and hence their predictions do not require the S-wave form factors. The $W_{1,2}$ observables  bring new information that can help to disentangle the SM from different NP scenarios, as illustrated in Fig.~\ref{fig:w1w2}. From $W_1$ in the region above 4 GeV$^2$, the SM and ${\cal C}^{\rm NP}_{9\mu}=-{\cal C}^{\rm NP}_{10\mu}$ are not distinguishable but all the other  scenarios shown can in principle be distinguished from the SM. The expected experimental precision for such measurements is detailed in section~\ref{sec:experiment}.

Finally we can use relation III, again neglecting all terms including quadratic products of observables sensitive to imaginary parts of bilinears ($P_3$, $P_{6^\prime,8^\prime}$ and $S_{S3,5}^i$), to find:
\begin{equation}
S_{S1}^r= { -\frac{1}{x} }\frac{F_L}{\sqrt{F_L F_T}} \left ( 2 (P_4^\prime (1+P_1)-2 \beta^2 P_2 P_5^\prime) S_{S2}^r + \beta^2 (P_5^\prime (1-P_1) - 2 P_2 P_4^\prime) S_{S3}^r \right)\,, 
\end{equation}
however, this does not give any additional experimental insight.

\section[Common zeroes of P- and S-wave observables]{Common zeroes of P- and S-wave observables}\label{sec:zeroes}

The optimized observable $P_2$ can be rewritten in terms of the $q^2$-dependent complex-vectors $n_\perp$ and $n_\|$ in the following way:
\begin{equation} \label{p2def}
P_2 = \frac{1}{2\beta} \left( 1- \frac{(n_\perp -n_\|)^\dagger (n_\perp-n_\|) + CP}{|n_\perp|^2+|n_\||^2+CP}\right) .
\end{equation}

In the absence of right-handed currents, the maximum of $P_2$, denoted $P_2^{max}$, occurs at a certain value of $q^2$, which we denote $q_1^2$. At the maximum, $P_2^{max}(q_1^2)\simeq 1/(2\beta)$. To a very good approximation, this maximum occurs when \begin{equation} n_\perp (q_0^2)\simeq n_\|(q_0^2)\,,
\end{equation}
where in principle a different $q^2$ is involved. This is because this expression is in fact four equations (two for the real and two for the imaginary part) and, moreover, they have to be combined with their $CP$ conjugated equivalents. Strictly speaking this would require that real and imaginary parts and left and right handed parts
have the zero at the same point in $q^2$, which is not the case. If we restrict ourselves to only ${\rm Re}(A^L_\bot(q_0^2))={\rm Re}(A^L_\|(q_0^2))$  the obtained position of the zero $q_0^2$ is in very good agreement
with the position of the maximum given by $q_1^2$,  as illustrated in Table~\ref{tab1}.

In the presence of right-handed currents the condition  $n_\perp(q_0^2)\simeq n_\|(q_0^2)$  can only be fulfilled  if a very concrete combination of Wilson coefficients is realized in Nature: \begin{equation}
    {\cal C}_{7^\prime}\simeq -\frac{{\cal C}_7^{\rm eff}}{{\cal C}_{10\mu}-{\cal C}_{9\mu}^{\rm eff}}({\cal C}_{10^\prime\mu}+{\cal C}_{9^\prime\mu})\label{rhcextra2} .\end{equation}
One of the NP scenarios that presently has the highest pull with respect to the SM,  
(${\cal C}_{7^\prime}=0,~{\cal C}^{\rm NP}_{9\mu}, {\cal C}_{9^\prime\mu}=-{\cal C}_{10^\prime\mu}$)
indeed fulfills this condition. From now on we will refer to this combination (Eq.(\ref{rhcextra2})) as condition$_R$.

In the SM, in the absence of right-handed currents, or in the presence  of right-handed currents that fulfill condition$_R$,  Table~\ref{tab1} illustrates that 
$q_0^2$ and $q_1^2$  are within 1\% of each other. This can be understood due to the small phases entering, but also because the equation:
\begin{equation} {\rm Re}(A^R_\bot(q_0^2))=-{\rm Re}(A^R_\|(q_0^2)) , \end{equation}
is exactly fulfilled in the large recoil limit in the absence of right handed currents, or if such currents are present but obey condition$_R$. 
Under these conditions, deviations from this relation then owe to departures from the large recoil limit. 
We can parametrize these tiny deviations and the effect of imaginary terms in the following form\footnote{
Besides the fact that we can compute $\delta$, $\epsilon_L$ and $\epsilon_R$, these quantities can be bounded experimentally using  Eq.\eqref{p2def} and Eq.\eqref{spurious} and  rewriting $P_2$ at the point of its maximum (again, for new physics scenarios with right-handed currents that satisfy  condition$_R$, or in the absence of right handed currents) as:
\begin{equation} \label{p2q0}
P_2(q_0^2) = \frac{1}{2\beta} \left( 1- N^2\frac{|\delta|^2+|\epsilon_L|^2+|\epsilon_R|^2 + CP}{|n_\perp|^2+|n_\||^2+CP}\right) .
\end{equation}
This implies that the tiny difference between $1/(2\beta)$ and the  maximum  imposes a  bound on each term $|\delta|$, $|\epsilon_L|$ and $|\epsilon_R|$ separately:
\begin{equation}
|\delta|^2, |\epsilon_L|^2, |\epsilon_R|^2\leq (1/(2\beta)-P_2^{\rm measured}(q_0^2)) F_T (d\Gamma/dq^2) /N^2 .
\end{equation}
However, measuring the difference $1/(2\beta)-P_2^{\rm measured}(q_0^2)$ would require an experimental precision that is presently unattainable.
}:

\begin{equation} \label{spurious}
\frac{1}{N}\left(n_\bot-n_\|\right)=\frac{1}{N}\binom{A_\bot^L(q_0^2)- A_\|^L(q_0^2)  }{-A_\bot^{R*}(q_0^2) -A_\|^{R*}(q_0^2)}=\binom{i \epsilon_L}{ \delta+i \epsilon_R} ,
\end{equation}
where $N$ is the normalization factor defined:
\begin{equation}
N = \sqrt{ \frac{G^2_{F}\alpha^2}{3\cdot 2^{10}\pi^5m^3_{B} }\lambda^2_ts\lambda^{1/2} \sqrt{1-4\frac{m_{\ell}^2}{s}}}  \, .
\end{equation}

\begin{table}[t]
    \centering
    \begin{tabular}{|c|c|c|c|}
    \hline 
     Hypotheses  & $q_0^2$ & $q_1^2$ & $q_2^2$    \\
    \hline 
    SM	&		2.03		&		2.02	&	2.02 \\
${\cal C}^{\rm NP}_{9\mu}$		&	2.31			&	2.30	&	2.30   \\
Hypothesis 5: $({\cal C}_{9\mu}^{\rm NP},{\cal C}_{9^\prime\mu}=-{\cal C}_{10^\prime\mu})$	&	2.37			&	2.37		&	2.36  \\
LFU Scenario 8: $({\cal C}_{9\mu}^{\rm V}=-{\cal C}_{10\mu}^{\rm V},{\cal C}_{9}^{\rm U})$	&	2.43			&	2.46	&	2.45    \\
Hypothesis 1: $({\cal C}_{9\mu}^{\rm NP}=-{\cal C}_{9^\prime\mu},{\cal C}_{10\mu}^{\rm NP}={\cal C}_{10^\prime\mu})$	&	2.45			&	2.34	&	2.18  \\
          \hline
    \end{tabular}
    \caption{Position of the zero evaluated from: a) ${\rm Re}(A_\bot(q_0^2))={\rm Re}(A_\|(q_0^2))$, b) position of $P_2^{\rm max}(q_1^2)$ and c) the exact position given by $X_2(q_2^2)$. This shows that only in the presence of right handed currents that do not fulfill condition$_R$, as in Hypothesis~1~\cite{quim_moriond}, do the zero points differ significantly from one another.
        }
    \label{tab1}
\end{table}

\begin{figure}[ht]
\begin{center}
  \includegraphics[width = 0.49\textwidth]{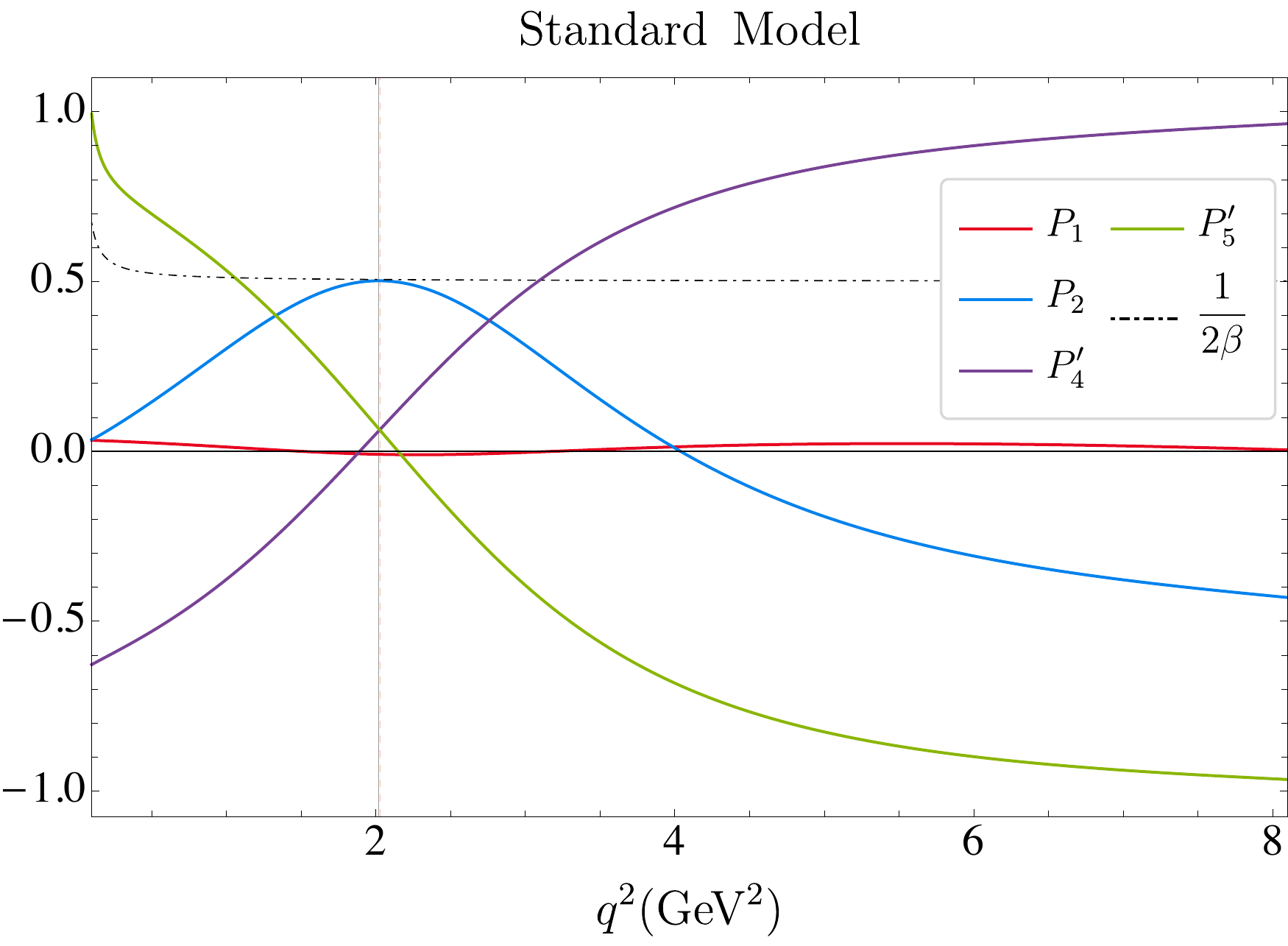} 
  \includegraphics[width = 0.49\textwidth]{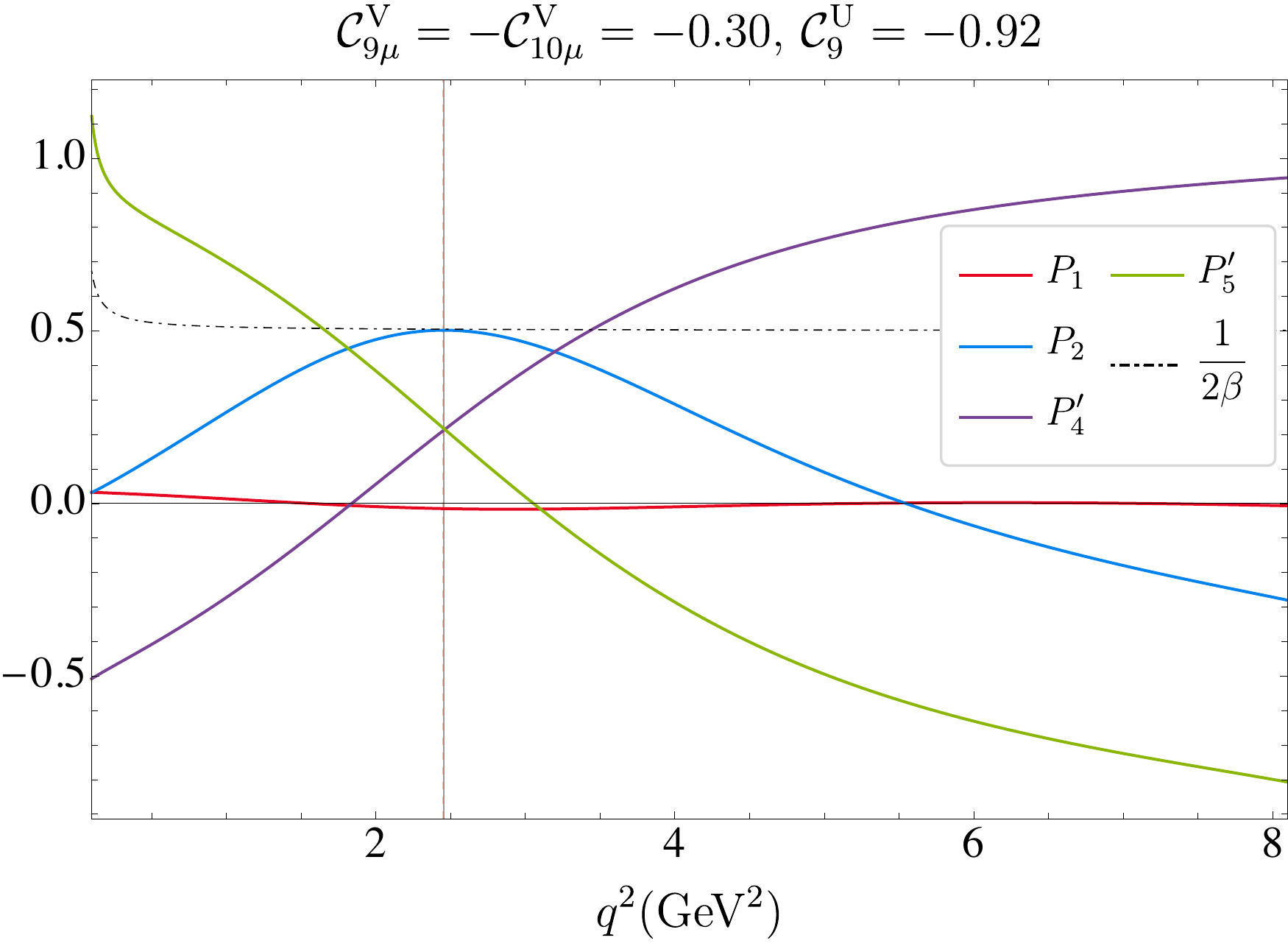} \\
   \includegraphics[width = 0.49\textwidth]{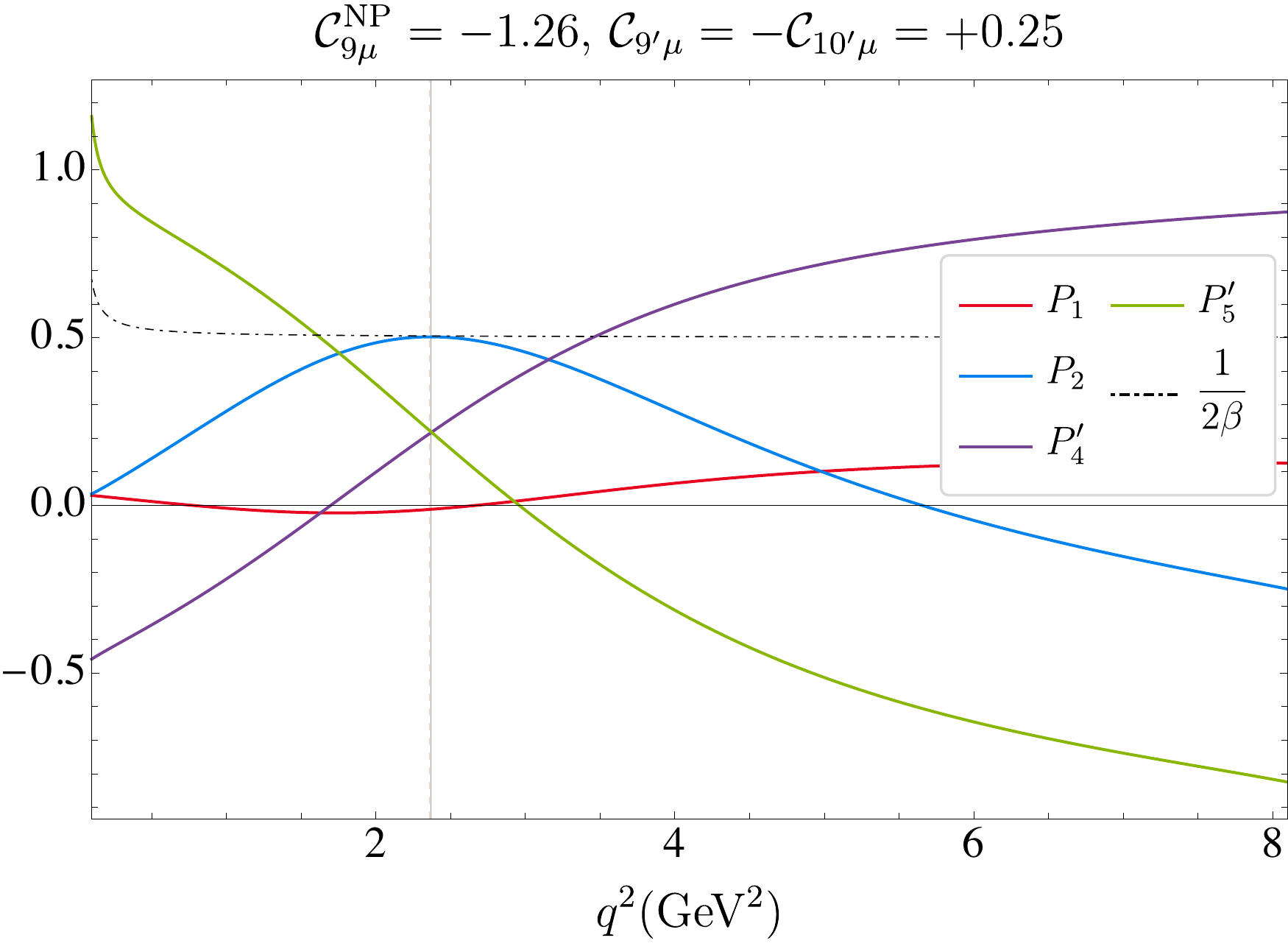} 
    \includegraphics[width = 0.49\textwidth]{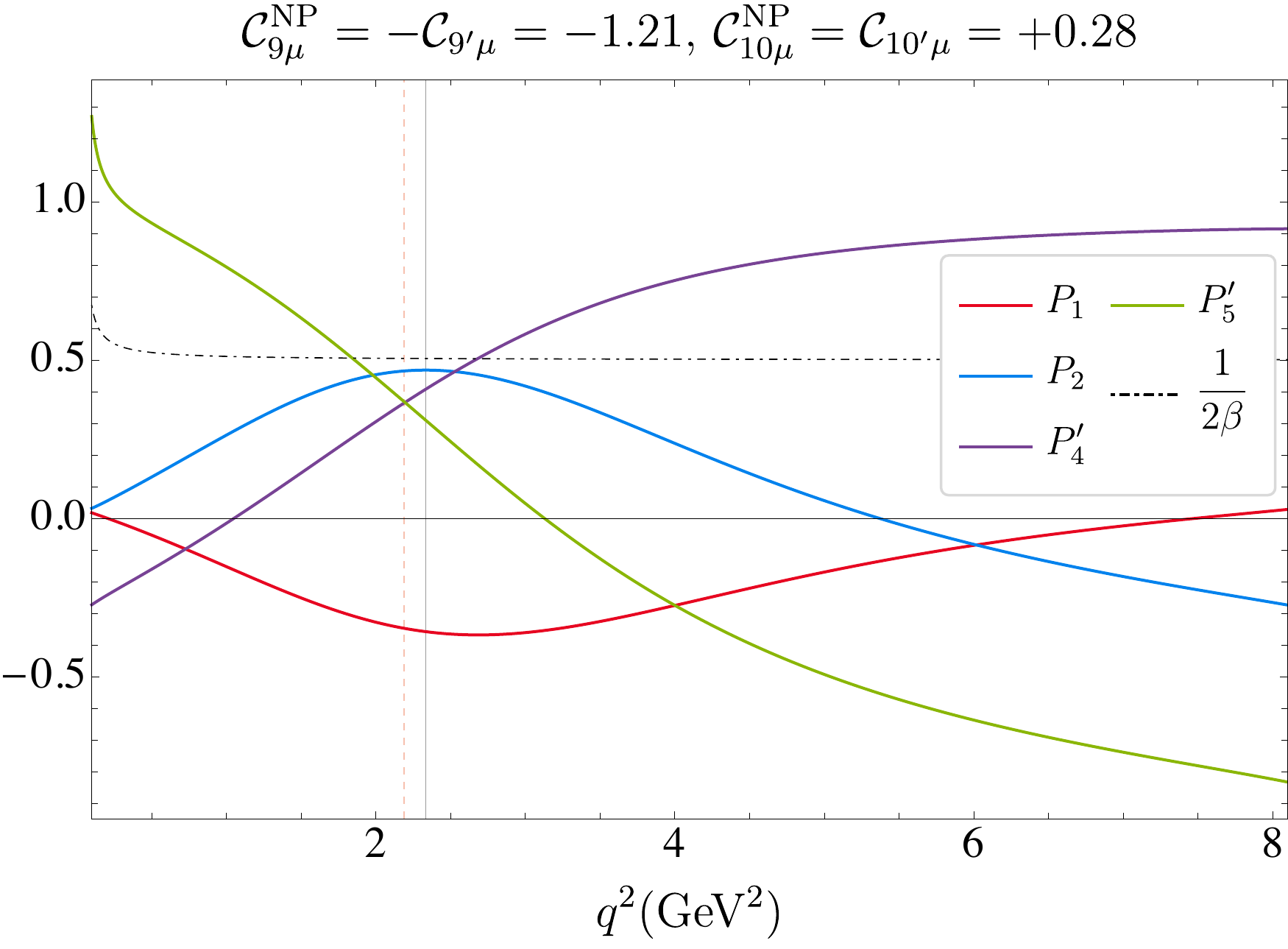} 
       \end{center}
 \caption{Predictions for different $P_i$ observables in (top left) the SM, (top right) Scenario 8, (bottom left)  Hypothesis~5 and (bottom right) Hypothesis~1.}\label{fig:rhcs1}
   \end{figure}

For new physics scenarios with right handed currents that satisfy  condition$_R$ or in the absence of right handed currents, a number of other observables are zero at the same point in $q^2$ at which $P_2$ is maximal. The relevant observables are formed from pairs of P- and S-wave angular observables: 
\begin{eqnarray}
X_1=P_2^{\rm max}(q_1^2), \quad 
X_2=\beta P_5^\prime-P_4^\prime, \quad
X_3=\beta S_{S4}^r- 2 S_{S5}^r, \quad
X_4=\beta S_{S3}^r - 2 S_{S2}^r. 
\end{eqnarray}
In Fig.~\ref{fig:rhcs1} the dependence of the position of the zero for several P-wave observables is shown for different NP scenarios. The observables $P_3$ and $P_{6,8}^\prime$ would also in principle give a further zero. However, given that they are numerically small over the entire low-q$^2$ region, they are difficult to determine experimentally. Moreover, the small contribution coming from the $\lambda_u=V_{ub} V^{*}_{us}$ piece of the Hamiltonian distorts  the position of the zero for such observables, which motivates their omission from the list above.  For the same reason,  the $S_{Si}^i $ observables and $P_1$ in the absence of right handed currents are also not included.

The $X_{2,3,4}$ observables then offer the possibility of looking at the compatibility of multiple zeros, rather than just the zero of single variables such as $A_{\rm FB}$. 
In the presence of sizeable right handed currents that do not fulfill condition$_R$, $P_2$ does not reach the maximal value $1/(2\beta)$, 
and a small difference between the $X_i$ observables should be observed. Misalignment between the zeroes of the $X_i$ observables could then help confirm a right handed current scenario, although another possible reason for a tiny misalignment is the presence of scalar or pseudoscalar contributions. 
The observable $X_1$ is not included in the list above because it is difficult to identify precisely the position of the maximum experimentally.  

\begin{figure}[ht]
\begin{center}
  \includegraphics[width = 0.49\textwidth]{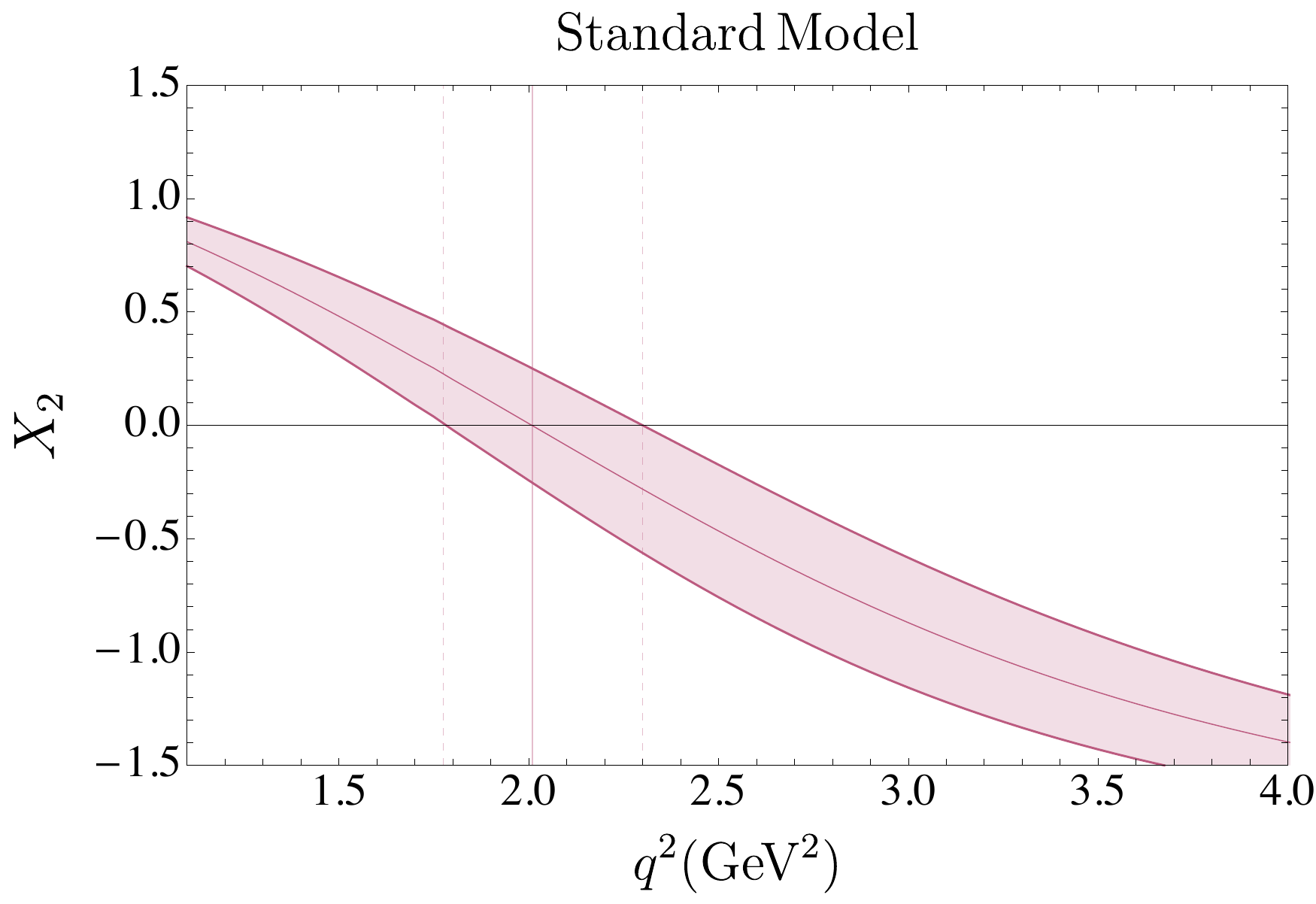} 
  \includegraphics[width = 0.49\textwidth]{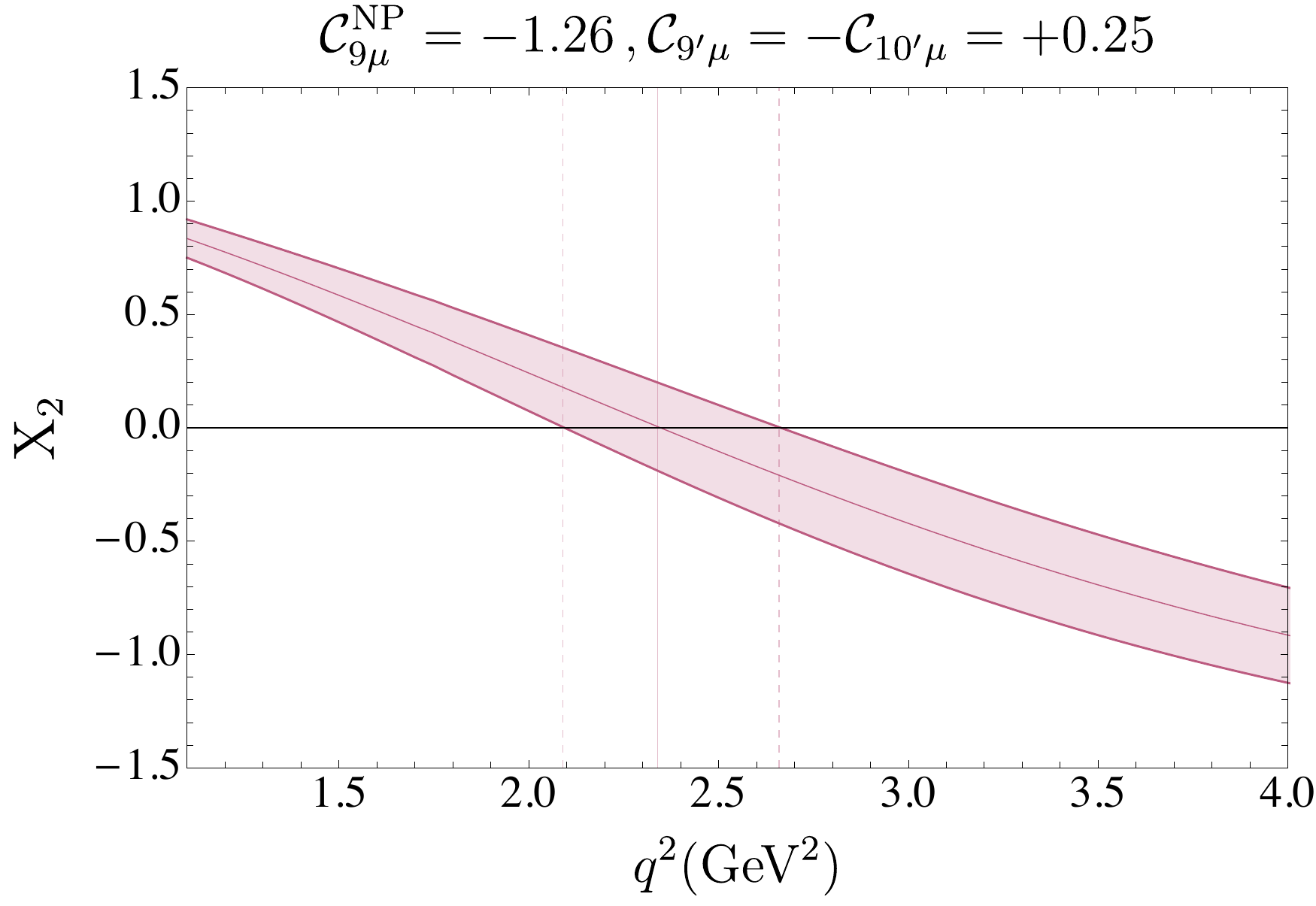}
  \includegraphics[width = 0.49\textwidth]{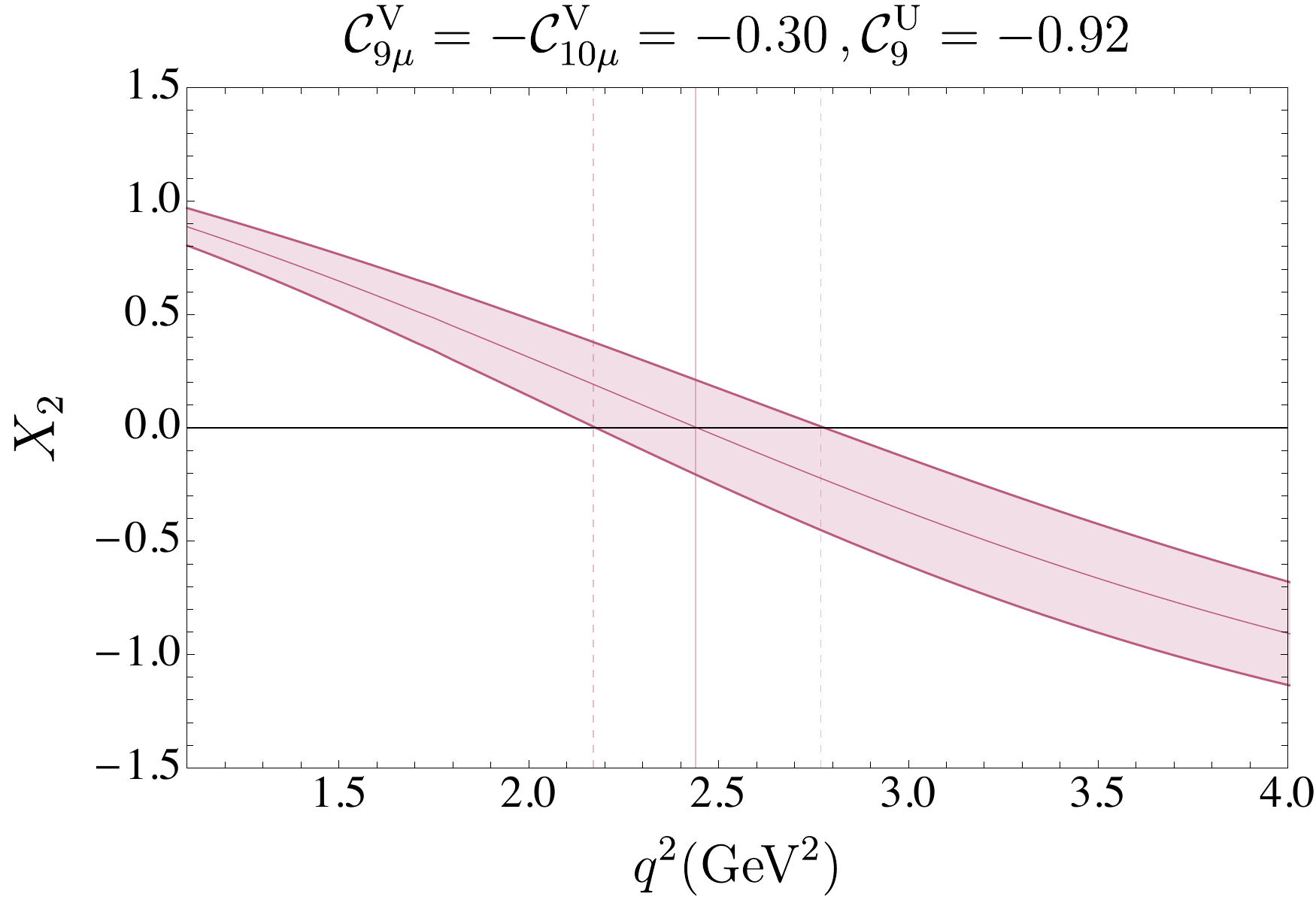}
       \end{center}
 \caption{$X_2$ predictions for the SM, Hypothesis V and Scenario 8, including theory uncertainties.}\label{fig:rhcs}
   \end{figure}

The point where $P_5^\prime$ and $P_4^\prime$ cross gives the zero of $X_2$, as shown in Fig.~\ref{fig:rhcs1}. Unfortunately, when including theory uncertainties using the KMPW computation~\cite{Khodjamirian:2010vf} of the form factors $V,A_{0,1,2},T_{1,2,3}$ and long-distance charm, the overlap between the zeroes of different NP scenarios is as shown  in Fig.~\ref{fig:rhcs}. This implies that further efforts are required to improve on the theoretical uncertainty of the observables.

We use the complete perpendicular and parallel amplitudes given by:

{\small
\begin{align}
\nonumber
A^{L,R}_{\perp} &= N\sqrt{2}\lambda^{1/2} \Bigg\{ \frac{2m_b}{s} \Bigg[ \left( {\cal C}^{\rm eff}_{7} + {\cal C}_{7'} + \frac{m_s}{m_b}{\cal C}^{\rm eff}_{7} \right)T_1 + \big( 1+r_1(s) \big) {\cal T}_{\perp} \Bigg] \\ 
&+ \left[ \left({\cal C}_{9}+{\cal C}_{9'} \right) \mp \left({\cal C}_{10} + {\cal C}_{10'}\right) + \big(1+r_1(s)\big)Y_t + \frac{\lambda_u}{\lambda_t}Y_u +c^{\rm long}_\perp(s) {s_{\perp}} \right]\frac{V}{m_{B}+m_{K^*}} \Bigg\}    \label{eq:Aperp}\\  
\nonumber
A^{L,R}_{\parallel} &= -N\sqrt{2}(m^2_{B}-m^{2}_{K^*}) \Bigg\{\frac{2m_b}{s} \left[\left( {\cal C}^{\rm eff}_{7} - {\cal C}_{7'} - \frac{m_s}{m_b}{\cal C}^{\rm eff}_{7} \right)T_2 + \big(1+r_2(s)\big){\cal T}_{\perp}\frac{\left( m^2_{B}-s \right)}{m^2_{B}}  \right]  \\
&+ \left[ \left({\cal C}_{9}-{\cal C}_{9'} \right) \mp \left({\cal C}_{10} - {\cal C}_{10'}\right) + \big(1+r_2(s)\big)Y_t + \frac{\lambda_u}{\lambda_t}Y_u + c^{\rm long}_{\|}(s) {s_{\parallel}} \right]\frac{A_1}{m_{B}-m_{K^*}} \Bigg\}     \label{eq:Apar}
\end{align}
}

\noindent and the longitudinal amplitude:

{\small
\begin{align}
\nonumber
A^{L,R}_{0} &= -\frac{N}{2m_{K^*}\sqrt{s}} \Bigg\{2m_b \Bigg[ \left(\left( {\cal C}^{\rm eff}_{7} - {\cal C}_{7'} - \frac{m_s}{m_b}{\cal C}^{\rm eff}_{7} \right) T_2 + \big(1+r_2(s)\big){\cal T}_{\perp}\frac{\left( m^2_{B}-s \right)}{m^2_B}\right)\times \nonumber  \\
&\times\left( m^2_B+3m^2_{K^*}-s \right)- \bigg(\left({\cal C}^{\rm eff}_{7} - {\cal C}_{7'} - \frac{m_s}{m_b}{\cal C}^{\rm eff}_{7}\right)T_3 + \big(1+r_3(s)\big){\cal T}_{\parallel} +{\cal T}_{\perp}\bigg)\frac{\lambda}{m^2_B-m^2_{K^*}} \Bigg] \nonumber \\ \nonumber
&+\bigg[\left({\cal C}_{9}-{\cal C}_{9'} \right) \mp \left({\cal C}_{10} - {\cal C}_{10'}\right) + \big(1+r_3(s)\big)Y_t + \frac{\lambda_u}{\lambda_t}Y_u +
c^{\rm long}_0(s) s_0
\bigg] \times \\
&\times \left[(m^2_B-m^2_{K^*}-s)(m_B+m_{K^*})A_1 - \frac{\lambda A_2}{m_B+m_{K^*}}\right]  \Bigg\} \, , \label{eq:a0}
\end{align}
}
%
\noindent where $\cal{T}_{\perp,\|}$ are defined in Ref.~\cite{Beneke:2001at}, $r_{1,2,3}(s)$ correspond to the different types of non-factorizable power corrections included in our analysis \cite{Descotes-Genon:2014uoa},  and  $c^{\rm long}_{\perp,\|,0}(s)$ is a parametrization of long distance charm contribution (see Refs.~\cite{Capdevila:2017ert,Descotes-Genon:2013wba} for the definition of the parameters):
\begin{eqnarray}\label{longdistance}
c^{\rm long}_{\perp,\|}(s)&=&\bigg( a^{c\bar{c}}_{\perp,\|} + b^{c\bar{c}}_{\perp,\|}(c^{c\bar{c}}_{\perp,\|}-s)s \bigg) \frac{1}{(c^{c\bar{c}}_{\perp,\|}-s)s} \nonumber \\
c^{\rm long}_{0}(s)&=&\left(a_{0}^{c\bar{c}}+b_{0}^{c\bar{c}}(c_{0}^{c\bar{c}}-s)(s+1) \right)\frac{s_{0}}{(c_{0}^{c\bar{c}}-s)(s+1)}\,.
\end{eqnarray}
Finally, the corresponding theoretical position of the zero is  the solution of the following implicit equation:

\begin{eqnarray}
\frac{q_0^2}{2 m_b}=\frac{{\cal C}_7^{\rm eff} \left[ T_1 (1+\frac{m_s}{m_b}) \lambda^{\frac{1}{2}}+T_2 (1-\frac{m_s}{m_b}) (m_B^2-m_{K^*}^2)\right] + {\cal T}_\perp \left[\lambda^{\frac{1}{2}}+(m_B^2-m_{K^*}^2) \frac{m_B^2-q_0^2}{m_B^2} \right]}{({\cal C}_{10}-{\cal C}_9^{\rm eff}(q_0^2)) (
\frac{\lambda^{\frac{1}{2}}}{m_B+m_{K^*}} V + (m_B+m_{K^*}) A_1)} ,\!\!\!\!\! \nonumber \\ \!\!\!\! \hspace*{-1cm}{ } \!\!\!\label{zeroq2}\,\,\hfill
\end{eqnarray}
where ${\cal C}_9^{\rm eff}(q^2)$ collects all pieces and, in order to simplify the expression,  we take all non-factorizable power corrections at their central values but keep long distance charm explicit inside ${\cal C}_9^{\rm eff}$ (taking $c^{\rm long}_\perp=c^{\rm long}_\|=c^{\rm long}$):
\begin{equation}
{\cal C}_{9 \perp,\|}^{\rm eff}(q^2)={\cal C}_{9}+ Y_t + \frac{\lambda_u}{\lambda_t}Y_u + c^{\rm long}_{\perp,\|}(s) {s_{\perp,\parallel}} .
\end{equation}

The form factors include soft form factors, $\alpha_s$  and power corrections and ${\cal T}_\perp$ also includes  the non-factorizable QCDF contribution.
Eq.(\ref{zeroq2}) offers an interesting combined test of form factors, Wilson coefficients and long-distance charm at a specific point in $q^2$.

\subsection{A closer look at the observable \texorpdfstring{$X_2$}{X2}: from New Physics to hadronic contributions}

In this section the properties of the observable  $X_2=\beta P_5^\prime-P_4^\prime$ are analyzed in detail, 
focusing on the $q^2$ bin where the zeroes fall both in the SM and in the NP scenarios considered~\cite{Alguero:2019ptt,quim_moriond}. While all the relevant observable information is already included inside global fits, analyzing particular observables like $X_2$ can provide guidance on how to disentangle NP effects in the longer term.
This observable has a simple structure in terms of Wilson coefficients when evaluated in the $q^2$ bin [1.8,2.5]:
\begin{equation}\label{P5mP4}
    \langle X_2\rangle_{[1.8,2.5]} \sim -0.14 + 0.22\, ( {\cal C}^{\rm NP}_{10\mu} - {\cal C}^{\rm NP}_{9\mu})+\epsilon\, ,
\end{equation}
\noindent
where $\epsilon$ in this equation refers to a tiny contribution that is non-zero only in the presence of right handed currents, in particular contributing to ${\cal C}_{9^{\prime}\mu}$, that can be cast as ${-0.02\,{\cal C}_{9^\prime\mu} (1+2 ({\cal C}_{9'\mu}-{\cal C}_{9\mu}^{\rm NP}))}$. As can be seen immediately from this equation, $\langle X_2\rangle^{\rm SM}_{[1.8,2.5]} \sim -0.14$. Independent of the details of the physics model, almost all NP scenarios with ${\cal C}_{9\mu}^{\rm NP} \neq 0$ yield  $ 0.88 < {\cal C}_{10\mu}^{\rm NP} - {\cal C}_{9\mu}^{\rm NP} < 1.26$, implying $ 0.05\leq \langle X_2\rangle_{[1.8,2.5]} \leq 0.14$. 
One relevant exception is Scenario~8, corresponding to $\langle X_2\rangle_{[1.8,2.5]} = 0.19$. This scenario contains a LFU contribution in ${\cal C}_{9}$, which would imply a contribution to the electronic mode too, $\langle X_{2e}\rangle_{[1.8,2.5]}=\langle \beta P_{5,e}^\prime-P_{4,e}^\prime\rangle_{[1.8,2.5]} \simeq 0.07$.  

\begin{figure}[t]
\begin{center}
 \includegraphics[width = 0.49\textwidth]{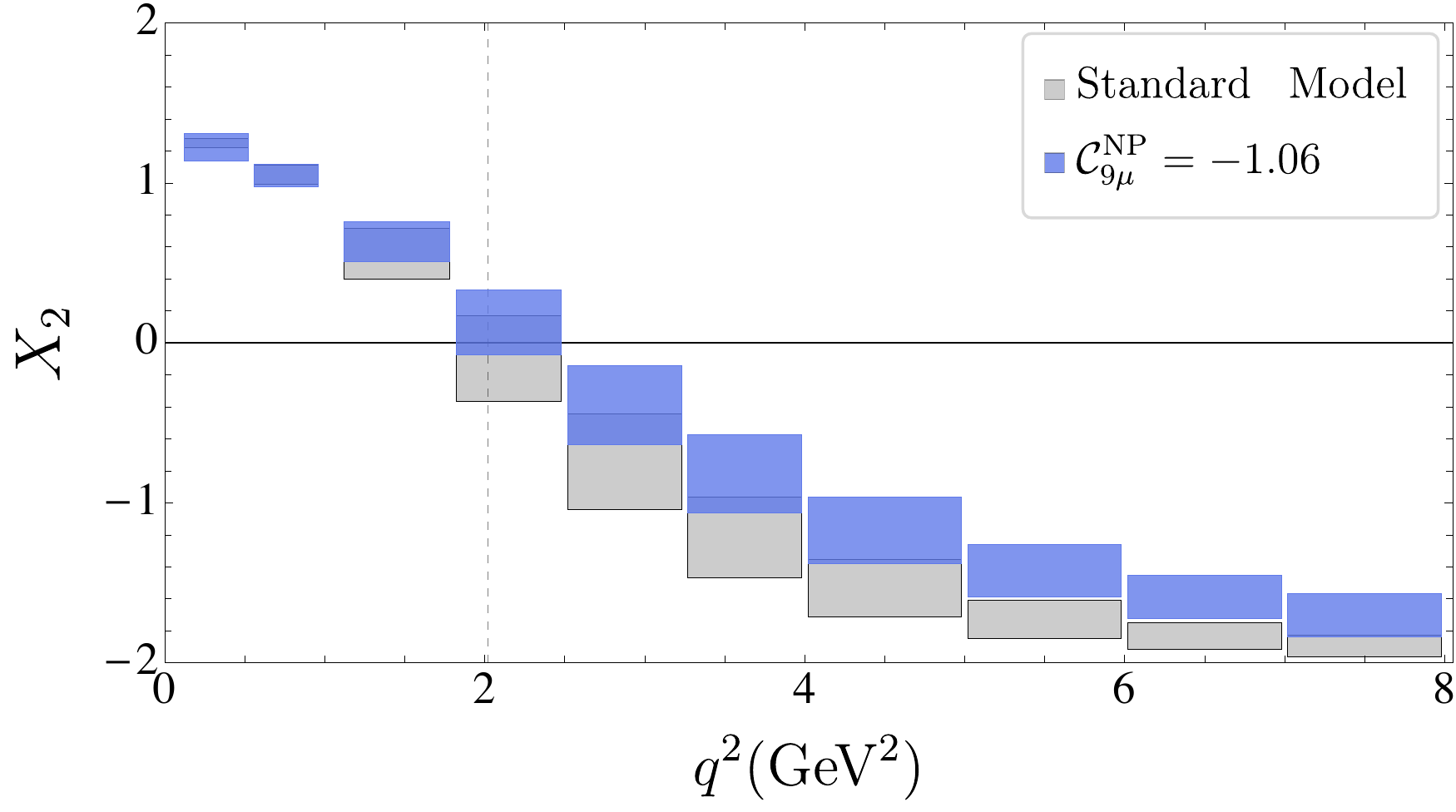} 
  \includegraphics[width = 0.49\textwidth]{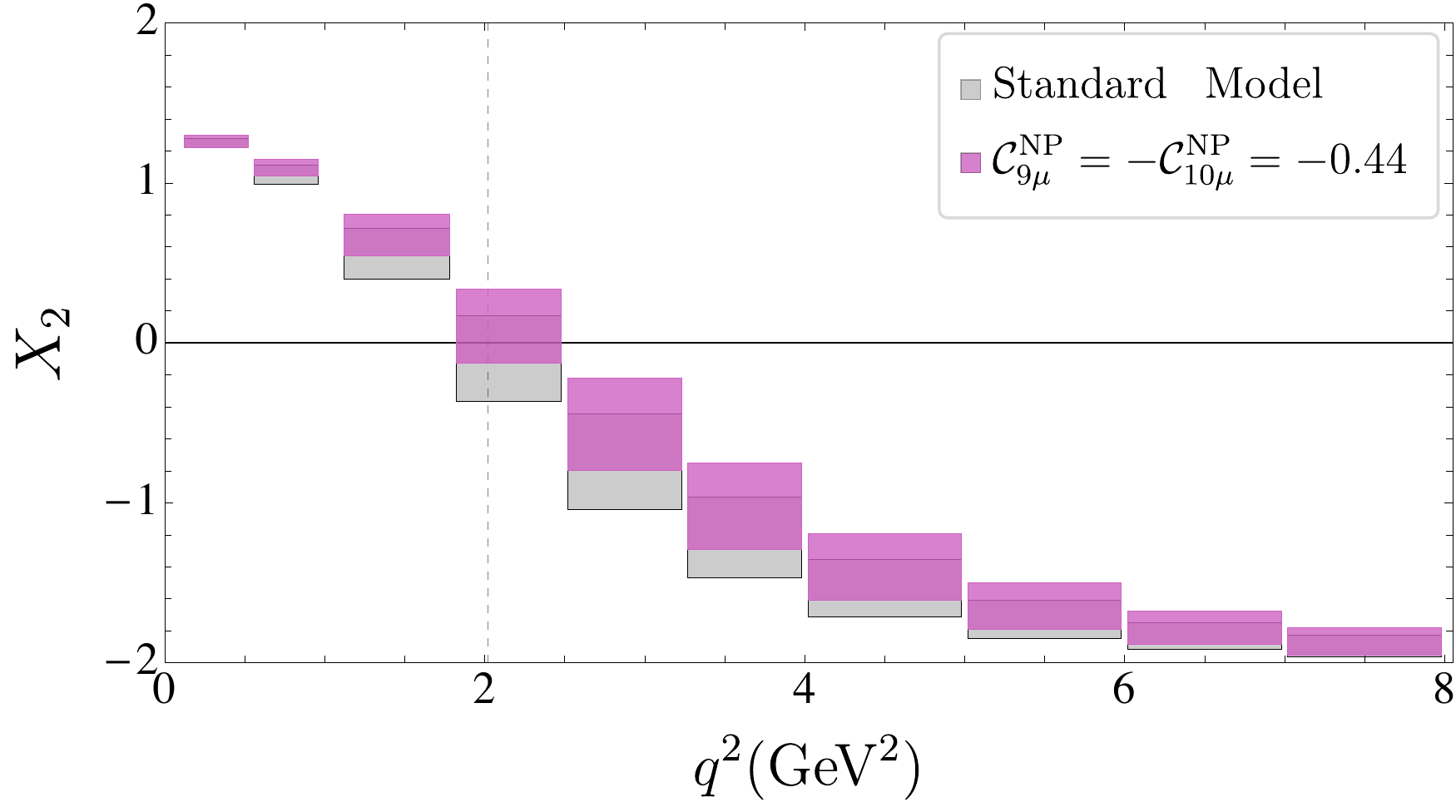}
  \includegraphics[width = 0.49\textwidth]{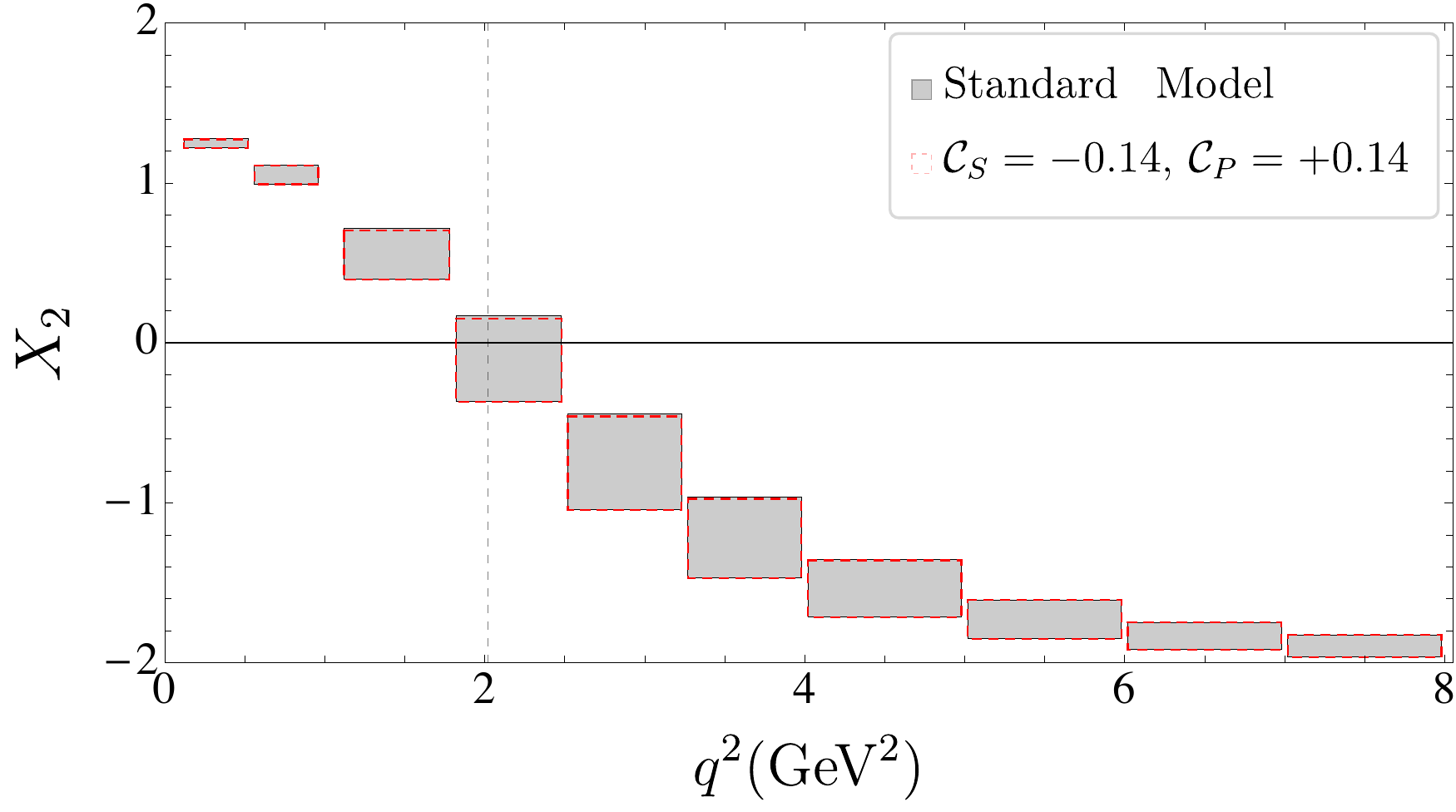}
  \includegraphics[width = 0.49\textwidth]{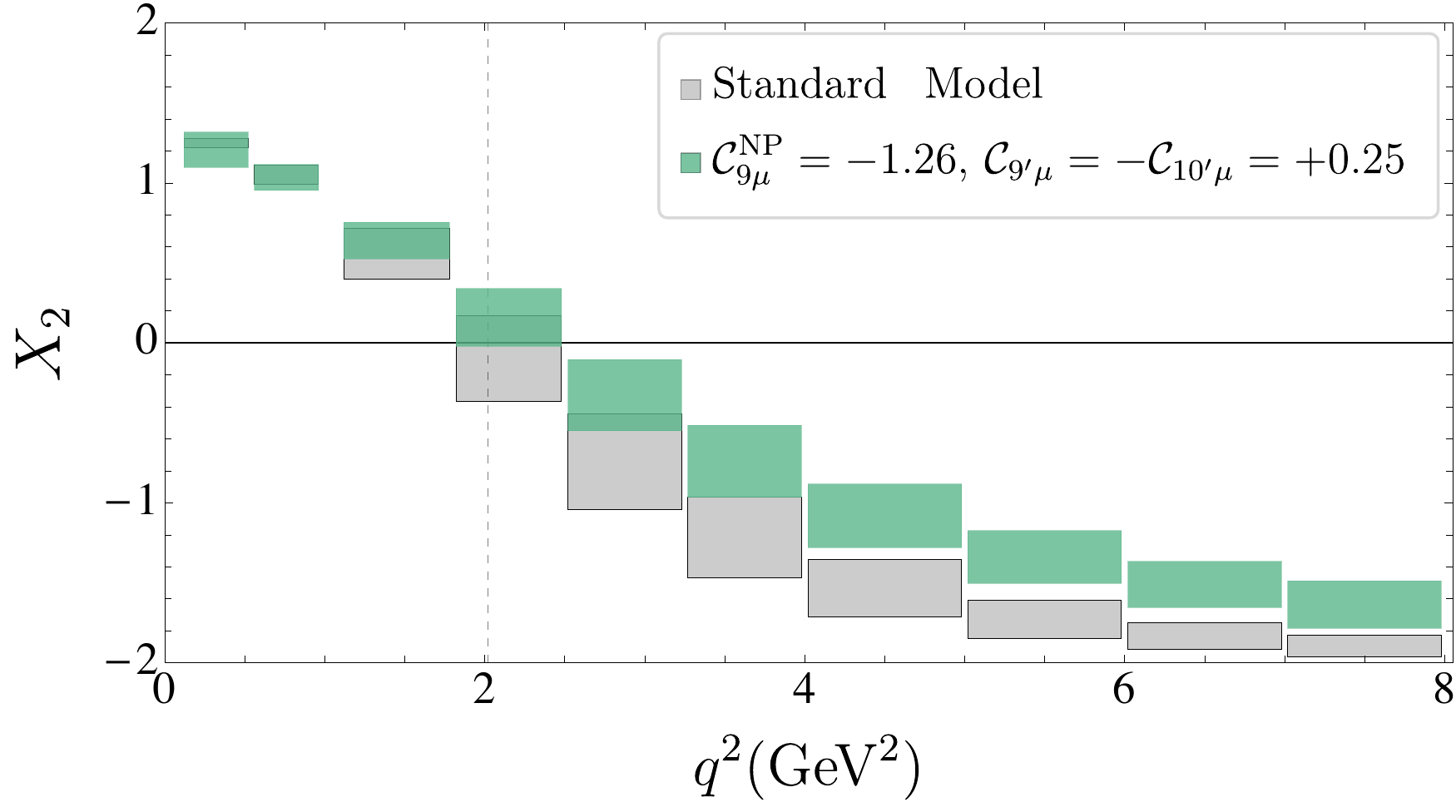}
    \includegraphics[width = 0.49\textwidth]{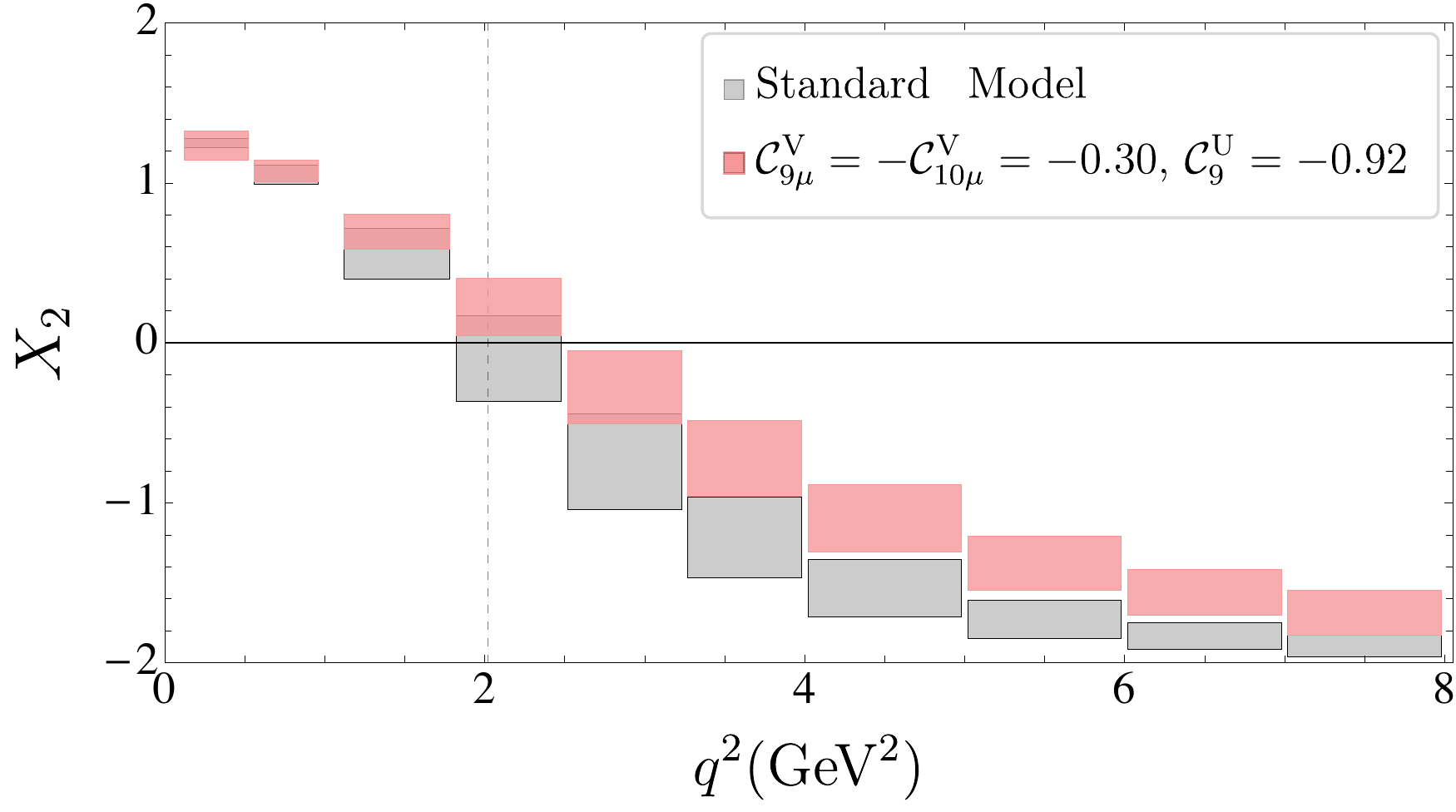}  \end{center}
 \caption{SM and NP predictions for $X_2$ binned in $q^2$.}\label{binnedX2}
  \end{figure}

In summary, given that $\langle X_2\rangle_{[1.8,2.5]}$ is predicted to be approximately $-0.1$ in the SM and up to $+0.2$ in some relevant NP scenarios, an experimental precision of $\pm 0.1$ would allow some of the NP scenarios to be disentangled from the SM. However, as shown above, with the present theoretical accuracy the theory predictions in $q^2$ bins yield a large overlap, preventing any clear discrimination. This is not surprising, because the deviation of $P_5^\prime$ in the  $[1.8,2.5]$ bin is not so large compared to the anomalies in the bins [4,6] or [6,8]. Moreover, given that $P_4^\prime$ is quite SM-like (see discussion below), it is expected that the largest deviation for this observable will occur in the [4,6] and [6,8] bins. This is confirmed in Fig.~\ref{binnedX2}.

Due to the stability of $X_2$ under most NP scenarios, it is essential  to improve on its theoretical uncertainties. In parallel 
we can explore the sensitivity that $P_5^\prime$ and $P_4^\prime$ may offer individually in the $[1.8,2.5]$ bin.
 For completeness, we provide the relevant expressions here: 
\begin{eqnarray}\label{P4P5}
    \langle P_4^\prime\rangle_{[1.8,2.5]} & \simeq & 0.13 -0.22 ({\cal C}^{\rm NP}_{10\mu} - {\cal C}_{10^\prime\mu}) + 0.03 ({\cal C}^{\rm NP}_{9\mu} - {\cal C}_{9^\prime\mu} )^2 \, , \\ \nonumber
     \langle P_5^\prime\rangle_{[1.8,2.5]} & \simeq & -0.01 +0.22 {\cal C}_{10^\prime\mu} - 0.26 {\cal C}^{\rm NP}_{9\mu} + 0.06 {\cal C}^{\rm NP}_{10\mu} {\cal C}_{10^\prime\mu} \, .
\end{eqnarray}

For the most prominent NP scenarios, we find the ranges  $0.04 < \langle P_4^\prime\rangle_{[1.8,2.5]} < 0.33$ and $0.10 < \langle P_5^\prime\rangle_{[1.8,2.5]} < 0.38$, with theory uncertainties of $\pm 0.20$ and $\pm 0.13$ 
for $\langle P_4^\prime\rangle_{[1.8,2.5]}$ and $\langle P_5^\prime\rangle_{[1.8,2.5]}$, 
respectively. These show that $\langle P_4^\prime\rangle_{[1.8,2.5]}$ exhibits a SM-like behaviour (in the absence of right handed currents) and gets a wider range only if right handed currents are rather large, as in Hypothesis~1 (see Table~\ref{tab1} and Refs.~\cite{Alguero:2019ptt,quim_moriond}).
This is different in the case  of $\langle P_5^\prime\rangle_{[1.8,2.5]}$, which exhibits an enhanced sensitivity to ${\cal C}_{9\mu}^{\rm NP}$ that drives the wider range. Moreover,  the current size of the theory uncertainty of $\langle P_4^\prime\rangle_{[1.8,2.5]}$  erases any possibility of discrimination between the SM and NP scenarios, but in the case of $\langle P_5^\prime\rangle_{[1.8,2.5]}$ the smaller size of the error leaves some discrimination power.

In order to discern hadronic contributions, the following strategy can be employed. The best fit point from a global fit that excludes $P_5^\prime$ and $P_4^\prime$ can be used to predict the NP contributions entering $\langle X_2 \rangle_{[1.8,2.5]}$, as well as $P_5^\prime$ and $P_4^\prime$ individually. These predictions can be constrasted with the experimental results in order to assess the SM contributions to $P_5^\prime$ and $P_4^\prime$. 
As noted above, the SM predicts $\langle X_2 \rangle_{[1.8,2.5]}=-0.14$, but  $\langle P_5^{\prime}\rangle_{[1.8,2.5]} = -0.01$ and $\langle P_4^{\prime}\rangle_{[1.8,2.5]} = 0.13$. Such values arise from a complex interplay between several SM sources, among them the hadronic form factors, ${\cal T}_{\perp}, {\cal T}_{\parallel}$ (these pieces encode, in particular, the non-factorizable power corrections), the value of the Wilson coefficients in the SM  but also  perturbative charm-loop contributions. 
Here we parametrize the remaining charm loop long-distance contributions in a manner that matches the non-perturbative computation from Ref.~\cite{Khodjamirian:2010vf}. In practice, when quoting long-distance charm loops we refer to Eq.\eqref{longdistance} for the transverse and perpendicular components and for the longitudinal one.

Using Eqs.(\ref{eq:Aperp}-\ref{eq:a0}) we can write the observables as follows\footnote{We neglect tiny contributions from ${\rm Im}{\cal T}_{\perp}$}:
{\small
\begin{eqnarray}\nonumber
\langle P_4^\prime\rangle^{{\rm SM}}_{[1.8,2.5]} & = & 0.35 + 10.63\, {\rm Re}{\cal T}_{\perp} +1.43\, {\rm Re}{\cal T}_{\parallel}+49.30\, ({\rm Re}{\cal T}_{\perp})^2 + 0.01 s_{\perp}-0.05 s_0    \, ,\\ 
\langle P_5^\prime\rangle^{\rm SM}_{[1.8,2.5]} & = & -0.34 - 11.71\, {\rm Re}{\cal T}_{\perp} +1.57\, {\rm Re} {\cal T}_{\parallel} -55.32\, {\rm Re}({\cal T}_{\perp})^2 -0.01 s_{\parallel}-0.05 s_0  \, ,\\\nonumber
\langle X_2\rangle^{\rm SM}_{[1.8,2.5]} & = & -0.68 - 22.33\, {\rm Re} {\cal T}_{\perp} +0.13 {\rm Re}{\cal T}_\|-104.62\, ({\rm Re} {\cal T}_\perp)^2
-0.01 s_{\parallel}-0.01 s_{\perp}\, ,
\end{eqnarray}
}
where in the SM in this particular bin one expects: ${\rm Re}{\cal T}_{\perp} \sim - 0.028$, ${\rm Re}{\cal T}_{\parallel} \sim + 0.025$, and in Refs.~\cite{Capdevila:2017bsm,Alguero:2019ptt,quim_moriond} $s_{\perp,\parallel,0}$ is taken as a nuisance parameter  allowed to vary in the range $s_i \in [-1,1]$ . The constant coefficients in these equations are intricate combinations of Wilson coefficients {and form factors} in the SM.

The first point to notice is that both $\langle P_4^\prime\rangle_{[1.8,2.5]}$ and $\langle P_5^\prime\rangle_{[1.8,2.5]}$ are dominated by 
${\rm Re}{\cal T}_{\perp}$ and the dominant long distance comes from $s_0$, in both cases with a very similar magnitude. Subleading contributions arise from ${\cal T}_{\parallel}$ and $s_{\perp,\parallel}$.  Secondly, $\langle X_2\rangle_{[1.8,2.5]}$ has a negligible sensitivity to ${\cal T}_{\parallel}$ and $s_0$, and the first long-distance piece enters via subleading contributions from $s_{\perp,\parallel}$. Thus this observable is basically dominated by ${\rm Re}{\cal T}_{\perp}$ and proves to be quite robust against long-distance charm loop contributions in this bin.

Finally, recalling the stability of $\langle X_2\rangle_{[1.8,2.5]}$ under different NP scenarios, we can parametrize this observable to a very good approximation as:
\begin{equation}
   \langle X_2\rangle_{[1.8,2.5]} =  -0.68 - 22.33 {\rm Re} {\cal T}_{\perp}
 -104.62\, ({\rm Re} {\cal T}_\perp)^2    
   + 0.22 ( {\cal C}^{\rm NP}_{10\mu} - {\cal C}^{\rm NP}_{9\mu} )\, ,
\end{equation}
\noindent
where the interplay between NP and the non-factorizable QCDF hadronic contributions is clearly encoded.
This implies that a measurement of $\langle X_2\rangle_{[1.8,2.5]}$ could provide an experimental constraint on ${\rm Re}{\cal T}_{\perp}$ in $[1.8,2.5]$, correlated with the NP scenario used, to be confronted with the SM prediction. The determination of ${\rm Re}{\cal T}_{\perp}$  can be seen as a non-trivial test of QCDF.
Notice also that, as discussed at the beginning of this section, ${\rm Re}{\cal T}_{\perp}$ has a significant impact on the position of the zero of $X_2$.
As soon as ${\cal T}_{\perp}$ is experimentally determined, the correlated measurement of the individual observables $\langle P_4^\prime\rangle_{[1.8,2.5]}$ and $\langle P_5^\prime\rangle_{[1.8,2.5]}$ will provide a handle on $s_{0}$, the dominant long-distance charm loop in this bin. The size of such effects should be clearly seen with the precision that should be attained during Run 4 of the LHC.

\section{Experimental prospects and precision}
\label{sec:experiment}

The angular observables in $\Bz\to\Kp\pim\mumu$ decays are usually extracted by means of a maximum likelihood fit of the decay rate in Eqs.\eqref{eq:pdftotal},~\eqref{eq:pdfpwaveraw} and~\eqref{eq:pdfswaveraw} to experimental data in bins of \qsq~\cite{LHCb_2016}. Practically, such a fit is achieved by the minimisation of a negative log-likelihood. Section~\ref{sec3} has elucidated that from the combination of the decay amplitudes not all of the angular observables are independent. In a fit to experimental data however, each observable is simply the coefficient of an angular term and is therefore an independent parameter. In the massless lepton case one can impose relations, for example the trivial $S_{1}^{c} = -S_{2}^{c}$, where the \CP-averaged observables are defined by $S_{i} = (J_{i} + \bar{J}_{i})/(\Gamma_{P} + \bar{\Gamma}_{P})$. However, these do not apply in the low \qsq region where the leptons should be treated as massive. 
In principle, the new relations of section~\ref{sec3} may also be used to reduce the number of free parameters in the fit. However they are too complex to be implemented in a minimisation procedure, as they cause discontinuities in the negative log-likelihood. Therefore the full bases of angular observables must be fitted in each of the massive and massless cases. The symmetry relations may instead be checked after the fit has been carried out in order to ensure that physically reasonable results have been obtained.

For a given data set there is no guarantee that all observables may be determined in a single maximum likelihood analysis. There may be large correlations between fit observables, apparent degeneracies due to the limited statistics, detector resolution effects and physical boundaries that distort the likelihood. Furthermore, the determination of the new interference observables that arise in the complete five-dimensional description of the decay can be distinguished only by making use of the \mkpi line shape, which has not been done experimentally before. All of these effects could impinge on the success of any new experimental analysis of the five-dimensional decay-rate. In order to study the stability of a fit to all the angular observables and to obtain an estimate of the experimental precision, a simple simulation was used to perform LHCb-like pseudo experiments. As the dominant effects on the experimental fits are statistical, rather than contingent on the experimental details, the results presented here will apply equally to future Belle~II analyses.

\subsection{Experimental setup}
\label{sec:experiment:setup}

Data sets are generated with the expected sample sizes collected by the LHCb collaboration at various points in time. The data the experiment currently has in hand, referred to as the Run~2 data set, is the combination of the Run~1 and Run~2 data with integrated luminosity of 9\invfb. Projections are made for future LHCb runs: Run~3 with 23\invfb, Run~4 with 50\invfb~\cite{LHCb-TDR-012}, and Run~5 representing the total data collected by the proposed Upgrade~II with 300\invfb~\cite{LHCb-PII-Physics}. The signal yields are extrapolated from those in Ref.~\cite{LHCb_2016}, scaling for the integrated luminosity, the $B$ production increase from Run~1 to Run~2, and an enlarged \mkpi window of $0.750<\mkpi<1.200\gev$ that is used to help determine the additional S-wave terms. The expected combinatorial background yields are similarly scaled. In a real analysis such a large \mkpi window would bring in extra partially-reconstructed backgrounds as well as more significant contributions from other P- and D- wave transitions, which would need to be accounted for in the systematic uncertainties. Furthermore, the exact form of the S-wave lineshape becomes more important in a wider window as the interference observables gain greater significance. Again, this would need to be accounted for in the systematic uncertainties which are beyond the scope of this paper.

The effect of the detector reconstruction and selection criteria is modelled using an angular acceptance function approximated to that in Ref.~\cite{LHCb-PAPER-2015-051}. For the \mkpi window considered, the acceptance is assumed to be constant with \mkpi, following Ref.~\cite{LHCb-PAPER-2016-012}.

The \qsq bins used are the same as those in Ref.~\cite{LHCb_2016}. An alternative configuration with each bin split in half is also trialled.
In contrast to previous experimental analyses, the fit is performed simultaneously with both $B$ flavours, in principle allowing the \CP-symmetric and \CP-asymmetric observables\footnote{The \CP-asymmetries are defined as $A_{i} = \frac{\bar{J}_{i} - J_{i}}{\Gamma_{P} + \bar{\Gamma}_{P}}$ for the P-wave observables and $AS_{Si}^{r/i}=\frac{\bar{\tilde{J}}_{i}^{r/i} - \tilde{J}_{i}^{r/i}}{\Gamma_{S} + \Gamma_{P} + \bar{\Gamma}_{S} + \bar{\Gamma}_{P}}$ for interference observables.} to be determined from a single fit. In order to fit the complete set of \CP asymmetries (including those for the S-wave and interference observables) one cannot solely rely on the angular description. The overall scale of the decay rate needs to be constrained with an extended term in the likelihood. The constraint may be the \CP asymmetry of the total decay rate, or the branching fraction of the average of the \Bz and \Bzb decays. The latter is preferred. It gives complementary information for use in global fits to the Wilson Coefficients that describe these decays and is thus of interest in its own right even when the \CP-asymmetries are not being extracted. An angular analysis is the only way to measure it in a model independent way, as the experimental efficiency can be corrected over all the kinematic variables of the $\Kp\pim\mumu$ system (within a \qsq bin these are the three angles and \mkpi).

Measuring absolute branching fractions is difficult due to systematic uncertainties that are hard to control. Instead the total P+S-wave rate relative to the mode $\Bz\to\jpsi\Kp\pim$ is taken, with the normalisation decay finishing in the same $\Kp\pim\mumu$ final state as the rare mode signal\footnote{Making the measurement relative to the $\Bz\to\jpsi\Kp\pim$ mode ensures that, to a large extent, nuisance production and detection asymmetries cancel. However, in an analysis of real data the normalisation mode is affected by contributions from exotic $\jpsi\pim$ states~\cite{LHCb-PAPER-2018-043}, both in terms of the signal yield and the angular distribution. Corrections to the fitted results will therefore need to be ascertained to produce the correct relative P-wave only rate and are beyond the scope of this paper.}. Using the measured S-wave fractions the P-wave relative branching fractions in each \qsq bin may be readily ascertained\footnote{Recent developments in calculations of $B^0\to K^+\pi^-$ form factors in a P-wave configuration~\cite{Descotes-Genon:2019bud} rely on a model to describe the lineshape of the $K^+\pi^-$ system. For a correct comparison between measurements and predictions of the branching fraction, any differences in the lineshape models used both in theory and experiment must be taken into account.}.

The SM values of the angular observables are used in the generation of the pseudo-data, except where stated. For the P-wave observables (and only for this experimental sensitivity study), the $B\to K^{*}$ form factors are taken from Ref.~\cite{Straub:2015ica} and rely on a combination of Light Cone Sum Rules and Lattice QCD calculations. For the S-wave observables, the $B\to K^{*}_{0}$ form factors are taken from Ref.~\cite{Meissner:2013hya}. For all observables the non-local charm contribution is taken from Ref.~\cite{Blake:2017fyh}, with the longitudinal and S-wave phase difference for all $J^{PC}=1^{--}$ dimuon resonances relative to the rare mode set to zero. The exact choice of these parameters has no impact on the conclusions of this study. The stability of the fit and the experimental precision on the P-wave observables is largely independent of the details of the model. Background events are simulated using a representative PDF constructed as the product of second order polynomials for each angular fit variable and an exponential function in \mB.

As is customary, the observable $F_L$ is used, here defined by $S_{2c} = -\beta^{2}F_L$ such that in the limit of no \CP violation $F_L$ is exactly representative of the longitudinal polarisation fraction divided by the total P-wave rate. The \CP asymmetry observable $AF_{L}$ is defined by $(\bar{J}_{2c} - J_{2c})/(\Gamma_{P} + \bar{\Gamma}_{P}) = -\beta^{2}AF_{L}$. Furthermore the forward-backward asymmetry is used as a fit parameter, with the customary definition
\begin{align}
    A_{\rm FB} = \frac{3}{4}S_{6s} + \frac{3}{8}S_{6c}. 
\end{align}

Similarly for the S-wave contribution the observable $F_S$ is used, defined by $\tilde{S}_{2a}^{c} = (\tilde{J}
_{2a}^c + \bar{\tilde{J}}_{2a}^{c})/\Gamma^\prime=-\frac{3}{8}\beta^{2}F_S$. Again, $F_S$ is a direct representation of the relevant transversity amplitude ($|n_{S}|^{2}$), divided by the total P- and S-wave rate. In the massless-lepton limit the integral of the S-wave component is therefore $F_S$ (the S-wave fraction). The corresponding \CP asymmetry observable, $AF_{S}$ is defined by $(\bar{\tilde{J}}
_{2a}^c - \tilde{J}_{2a}^{c})/\Gamma^\prime=-\frac{3}{8}\beta^{2}AF_{S}$.

For the \qsq bins in which the leptons are considered massless, which are those where $\qsq > 1.1\gevgev$, there are 19 \CP-averaged (8 P-wave, 1 S-wave and 10 interference split into real and imaginary parts) angular observables to be fitted as listed in Eq.\eqref{eq:massless_obs_list}, plus the total P+S-wave rate. In these bins $\beta^{2} = 1$. For the \qsq bins in which the lepton is treated as massive, where $\qsq < 1.1\gevgev$, there are 24 \CP-averaged observables, as per Eq.\eqref{eq:massive_obs_list}; and one may again choose to fit the total rate as well. For these bins \qsq is evaluated at the centre of the bins.

\subsection{Results with massless leptons}
\label{sec:experiment:massless}

\begin{figure}[htp]
    \centering
    \includegraphics[width = 0.48\textwidth, clip, trim = 0mm 0mm 14mm 14mm]{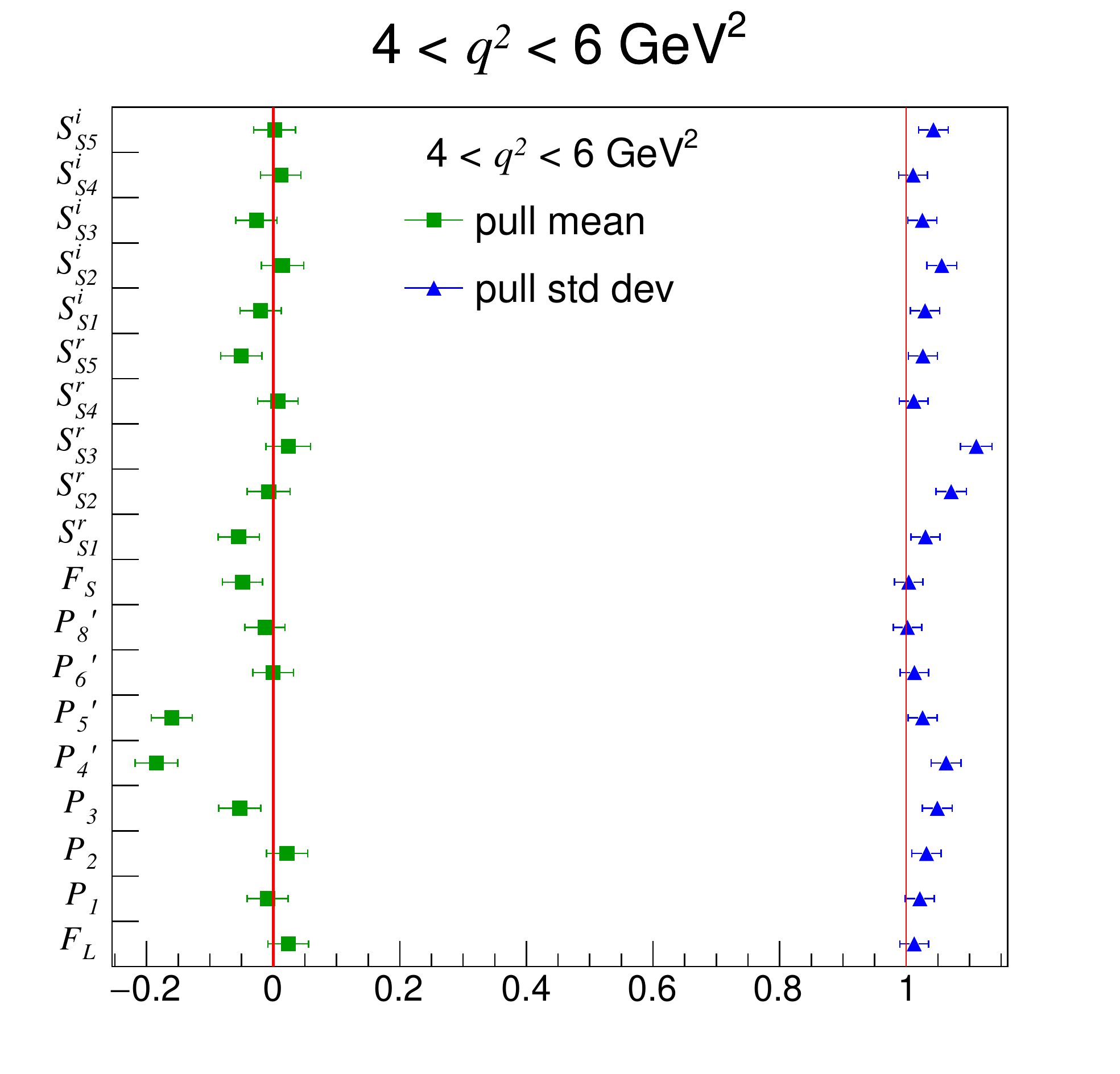}
    \includegraphics[width = 0.48\textwidth, clip, trim = 0mm 0mm 14mm 14mm]{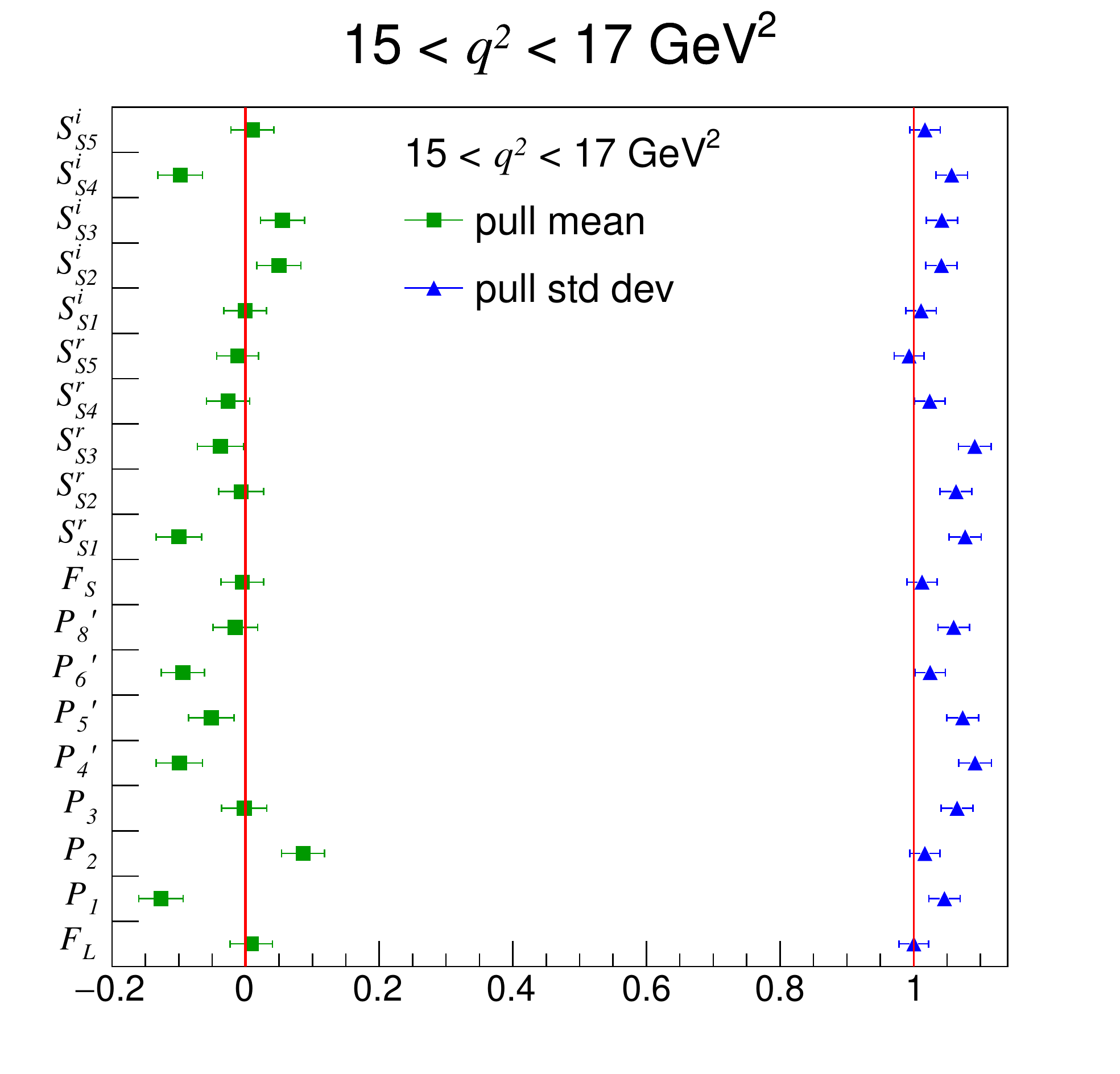}
    \caption{Summaries of the (green) means and (blue) standard deviations of the pull distribution for the optimised P-wave observables and normal interference observables in two bins of \qsq. The red lines are references at 0 and 1.}
    \label{sec:experiment:massless:fig:standard}
\end{figure}

Initially only the \CP-averaged observables are free to vary in the fit; those for the \CP asymmetry are fixed to 0. For both the $P^{(')}_i$ and $S_i$ basis, despite the inclusion of all the S-P-wave interference terms, $>99\%$ of the pseudoexperiment fits for massless leptons in \qsq bins with $\qsq>1.1\gevgev$ converge and the $P$-wave \CP-averaged observables are determined without any significant bias (approximately 20\% of the statistical uncertainty or less) and with good statistical coverage. A summary of the distribution of the pulls resulting from the pseudoexperiment fits of the angular observables in two \qsq bins is shown in Fig.~\ref{sec:experiment:massless:fig:standard}, including fits using the optimised ($P^{(')}_i$) P-wave observable basis. For an observable the pull is defined as the difference between the fitted value and true value, divided by the statistical uncertainty estimated in the fit. It should be noted that both the real and imaginary parts of the interference observables can be determined by the fit.

\begin{figure}[htp]
    \centering
    \includegraphics[width = 0.48\textwidth, clip, trim = 0mm 0mm 14mm 14mm]{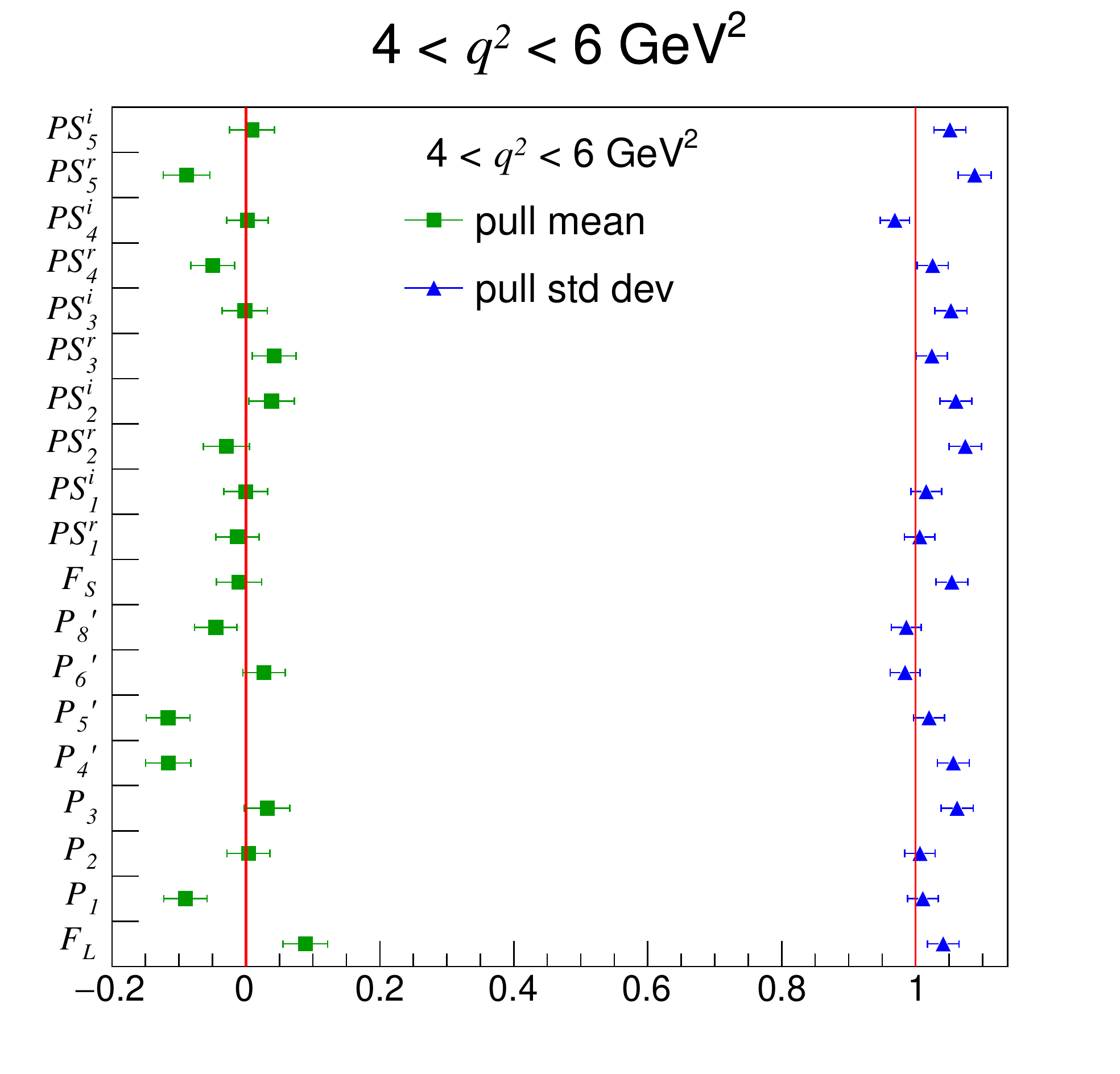}
    \includegraphics[width = 0.48\textwidth, clip, trim = 0mm 0mm 14mm 14mm]{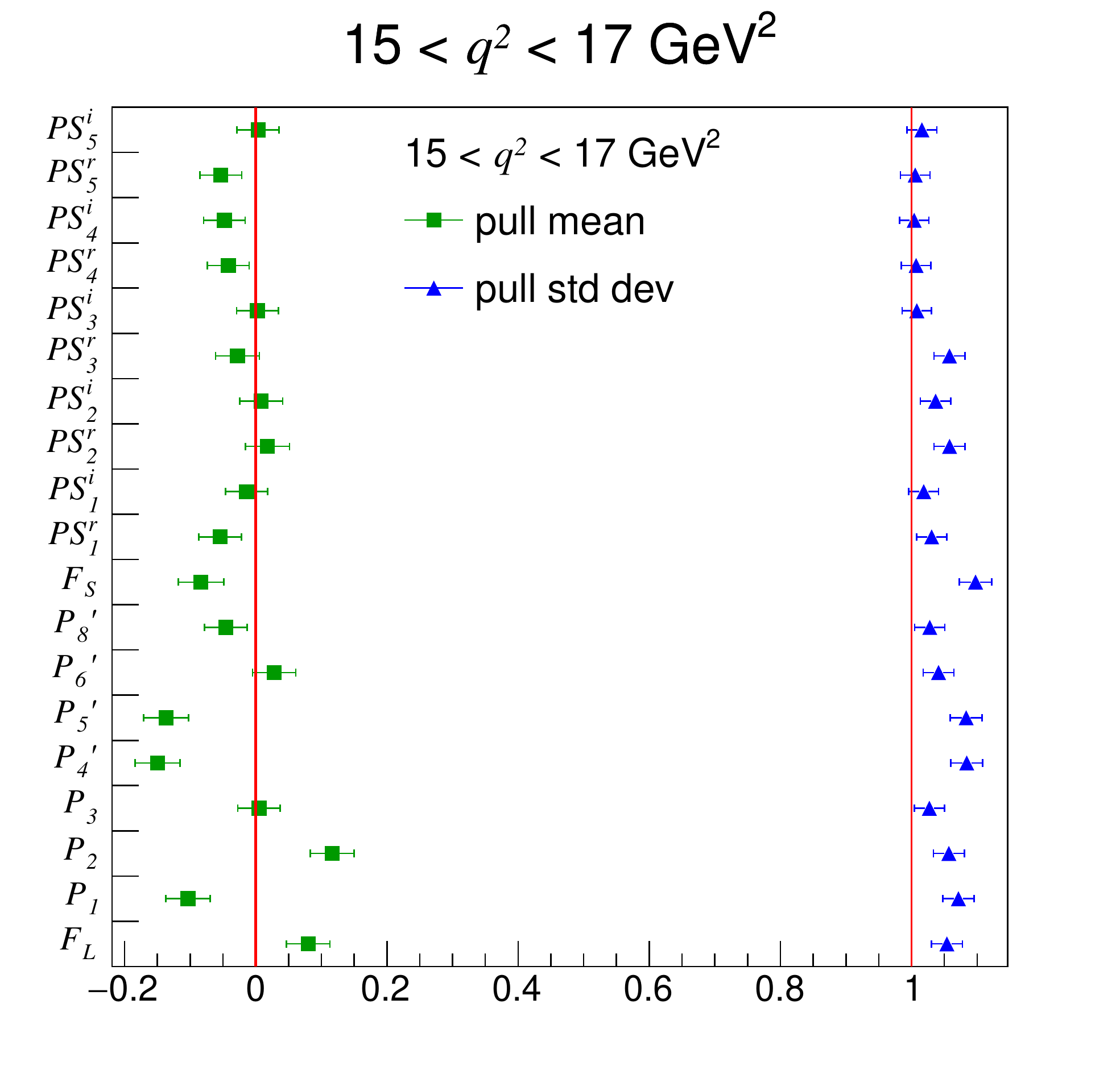}
    \caption{Summaries of the (green) means and (blue) standard deviations of the pull distribution for the optimised P-wave observables and optimised S-wave observables, $PS_{i}$, in two bins of \qsq. The red lines are references at 0 and 1.}
    \label{sec:experiment:massless:fig:optimis_swave}
\end{figure}

\begin{figure}[htp]
    \centering
    \includegraphics[width = 0.48\textwidth]{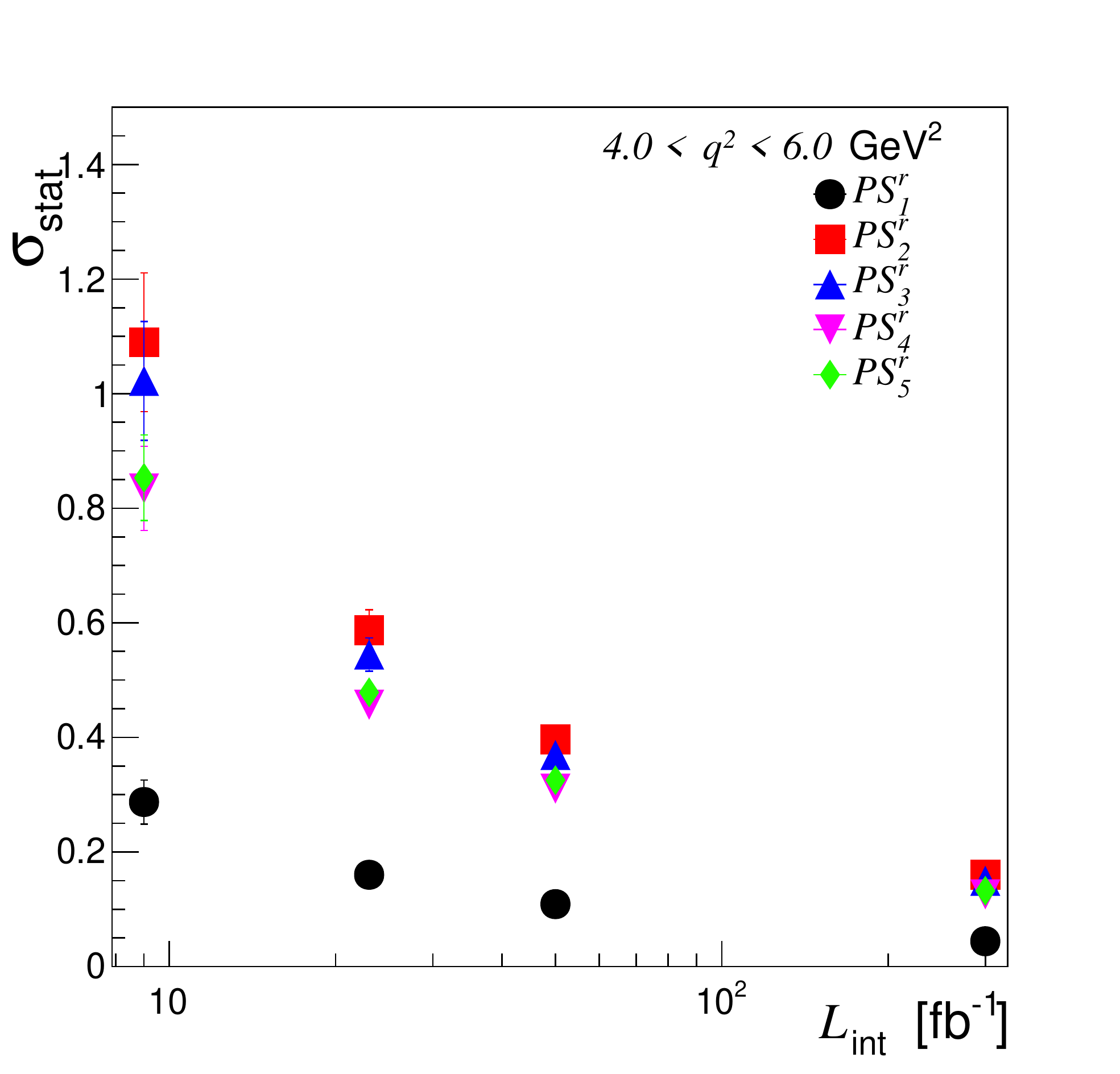}
    \includegraphics[width = 0.48\textwidth]{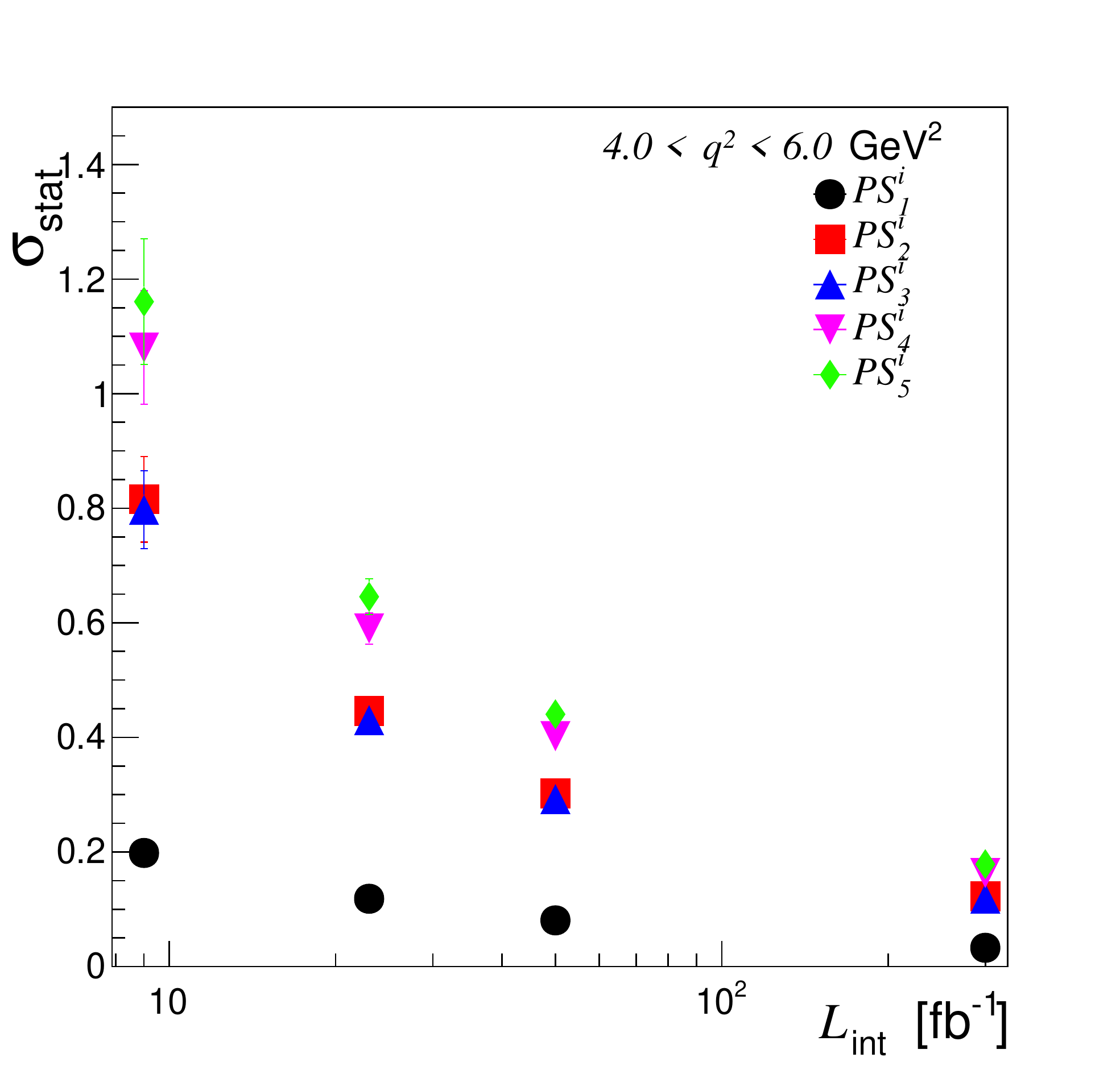}
    \caption{Estimated statistical uncertainty for the (left) real and (right) imaginary optimised interference observables $PS_{i}$ as a function of integrated luminosity for the \qsq bin $4.0<\qsq<6.0\gev^{2}$.}
    \label{sec:experiment:massless:fig:psi_precision}
\end{figure}

Furthermore the optimised interference observables, $PS_{i}$, may also be readily determined with the data that LHCb already has in hand. Summaries of the fit behaviour with this configuration are shown in Fig.~\ref{sec:experiment:massless:fig:optimis_swave}. The estimated statistical uncertainties for these new observables as a function of the integrated luminosity collected are shown in Fig.~\ref{sec:experiment:massless:fig:psi_precision}. The points show the expected luminosity for future LHCb runs.

\begin{figure}
    \centering
    \includegraphics[width = 0.48\textwidth, clip, trim = 0mm 0mm 14mm 14mm]{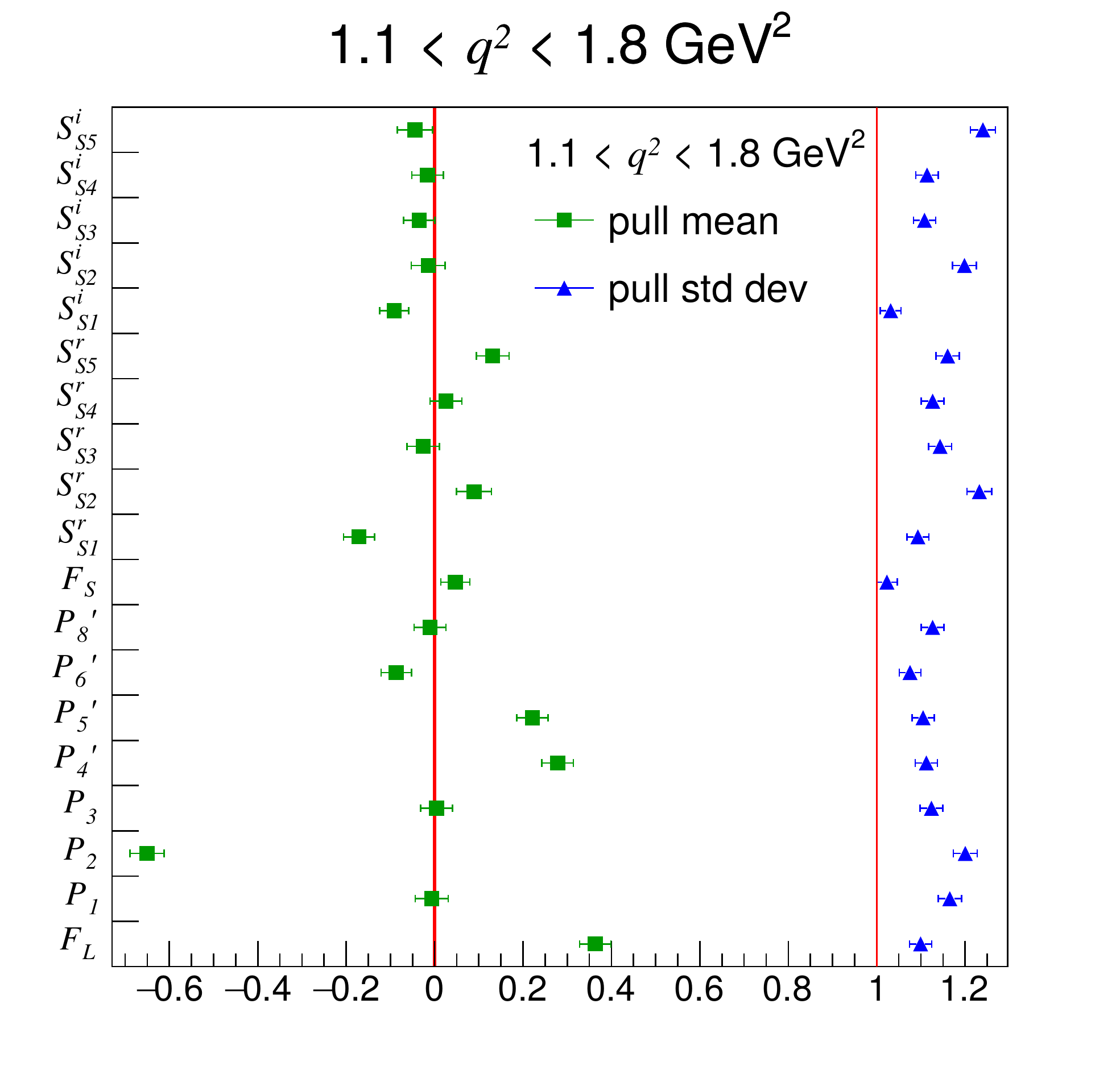}
    \includegraphics[width = 0.48\textwidth, clip, trim = 0mm 0mm 14mm 14mm]{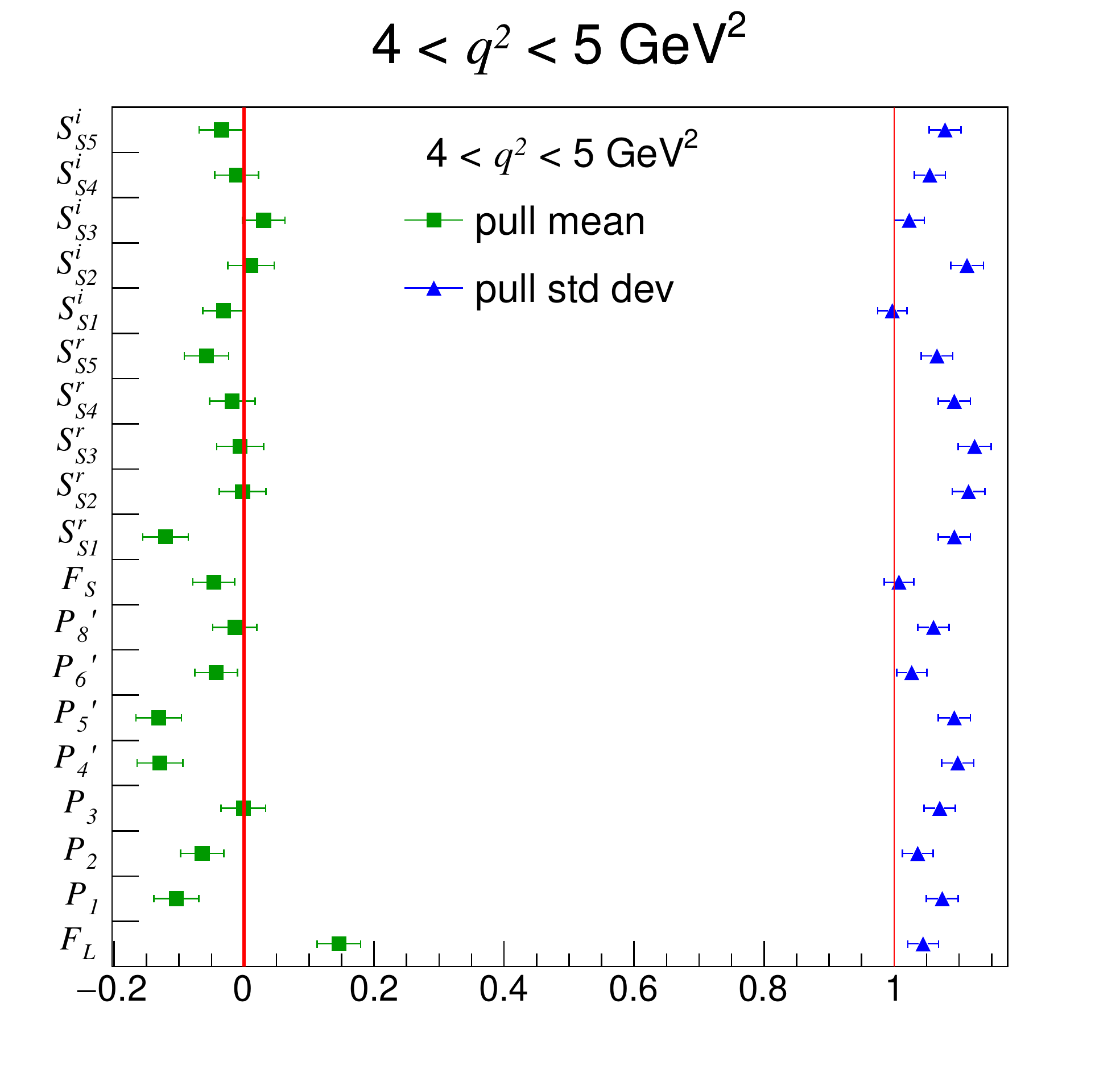}    
    \caption{Summaries of the (green) means and (blue) standard deviations of the pull distribution for the optimised P-wave observables, $PS_{i}$, in two narrow bins of \qsq. The red lines are references at 0 and 1.}
    \label{sec:experiment:massless:fig:narrow_pulls}
\end{figure}

For the alternative narrower \qsq binning, the situation is not so ideal; summaries for two \qsq bins are in Fig.~\ref{sec:experiment:massless:fig:narrow_pulls}. In general, the central fit values for all the variables do not show biases above 20\% of statistical uncertainties. The exception is the two bins in the region $1.1<\qsq<2.5\gevgev$ (the bins are $1.1-1.8\gevgev$, shown on the left of Fig.~\ref{sec:experiment:massless:fig:narrow_pulls} and $1.8-2.5\gevgev$), where the predicted values of $F_L$ and $A_{\rm FB}$ lie close to the edge of the physically allowed parameter space. For the narrower bins this boundary distorts the likelihood close to where the minimum should be. The result is an imperfect determination of these variables, or $P_{2}$ in the optimised basis, as the fit crosses into the unphysical region. However, as the other observables in these bins behave well and all the other bins behave well, there is motivation to use the finer \qsq bins even with the Run~2 data set. As the uncertainties are shown to be too small by the pull distributions the Feldman-Cousins method~\cite{Feldman_1998} will need to be employed to establish confidence intervals. The problems of bias and error determination are readily ameliorated with more data and even by the end of Run~3 the fit behaviour will be much improved.

\begin{figure}
    \centering
    \includegraphics[width = 0.48\textwidth, clip, trim = 0mm 0mm 14mm 1mm]{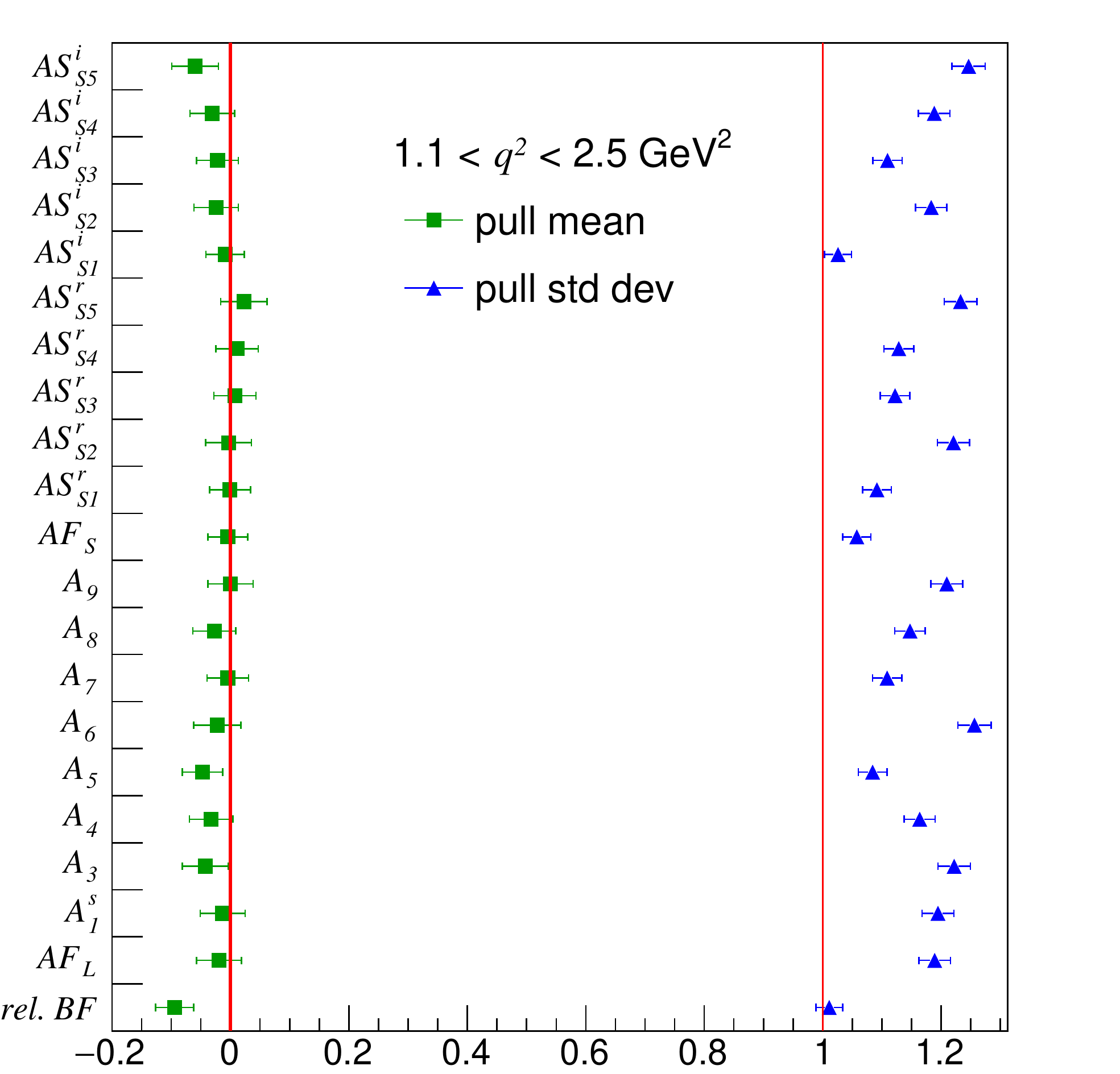}
    \includegraphics[width = 0.48\textwidth, clip, trim = 0mm 0mm 14mm 1mm]{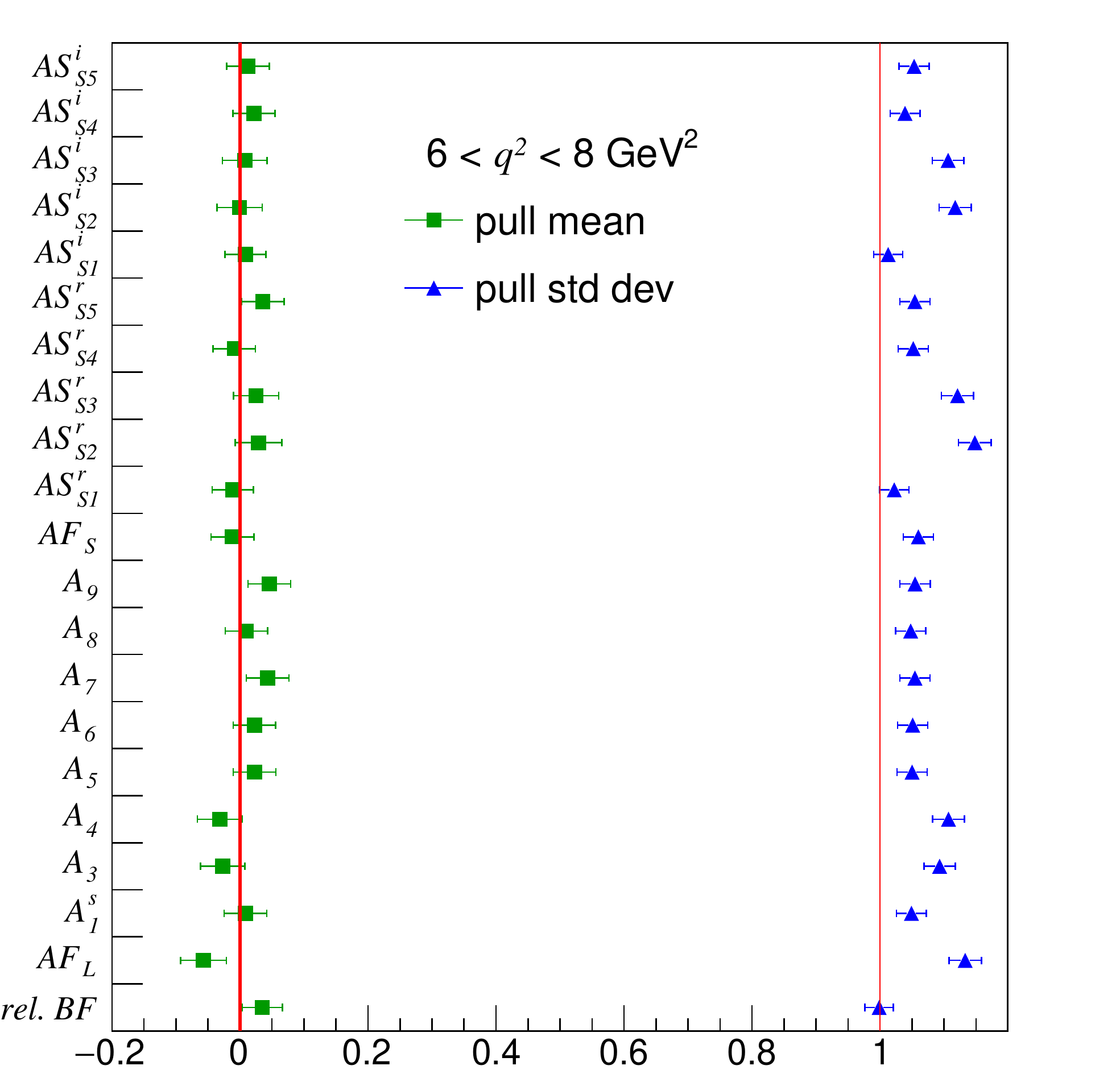}
    \caption{Summaries of the (green) means and (blue) standard deviations of the pull distribution for the \CP-asymmetry observables and the relative branching fraction, denoted $rel.\, BF$, in two bins of \qsq. The red lines are references at 0 and 1.}
    \label{sec:experiment:massless:fig:standard_ai}
\end{figure}

When including the \CP-asymmetry observables as free parameters in the fit it is again found that the fits converge successfully. These parameters themselves are found to be unbiased, although the estimated uncertainties are in general too small. Examples are shown in Fig.~\ref{sec:experiment:massless:fig:standard_ai}. The extracted relative branching fraction is also found to be unbiased. However, the extra free parameters lead to larger biases in the \CP-averaged observables of up to 40-50\% in some cases. With more data the situation will be improved and after 50\invfb it will be possible to extract all \CP-averaged and \CP-asymmetry observables in a single fit with minimal biases and good coverage.

\subsection[Results for massive leptons]{Results for massive leptons in $0.1<q^2<0.98\gevgev$}
\label{sec:eperiment:massive}
In the $0.1<\qsq < 0.98\gevgev$ bin for massive leptons, the situation is more complex. For the basis fitting only unoptimised observables (both P- and S-wave) the fit in general gives unbiased pull distributions. The exception is for the observables $F_L$ and $S_{1c}$ which have a large anticorrelation between them. For the regular optimised P-wave observables the fit also does not converge well, likely due to the small value of $F_L$ in this bin, which appears in the denominator of the optimised observables. The Feldman-Cousins method will therefore be required to obtain the correct confidence intervals for all observables in this \qsq bin.

\begin{figure}[htp]
    \centering
    \includegraphics[width = 0.48\textwidth]{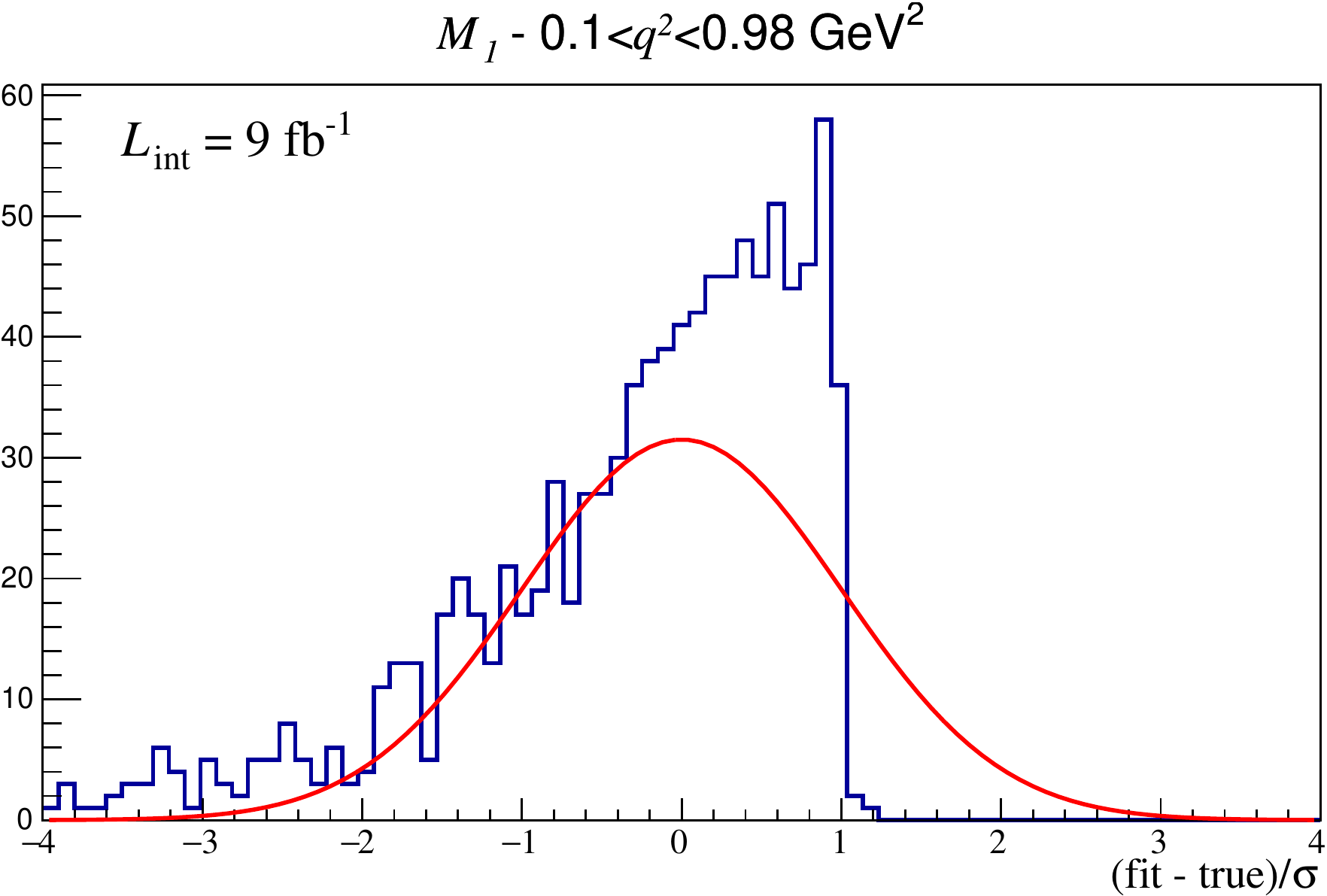}
    \includegraphics[width = 0.48\textwidth]{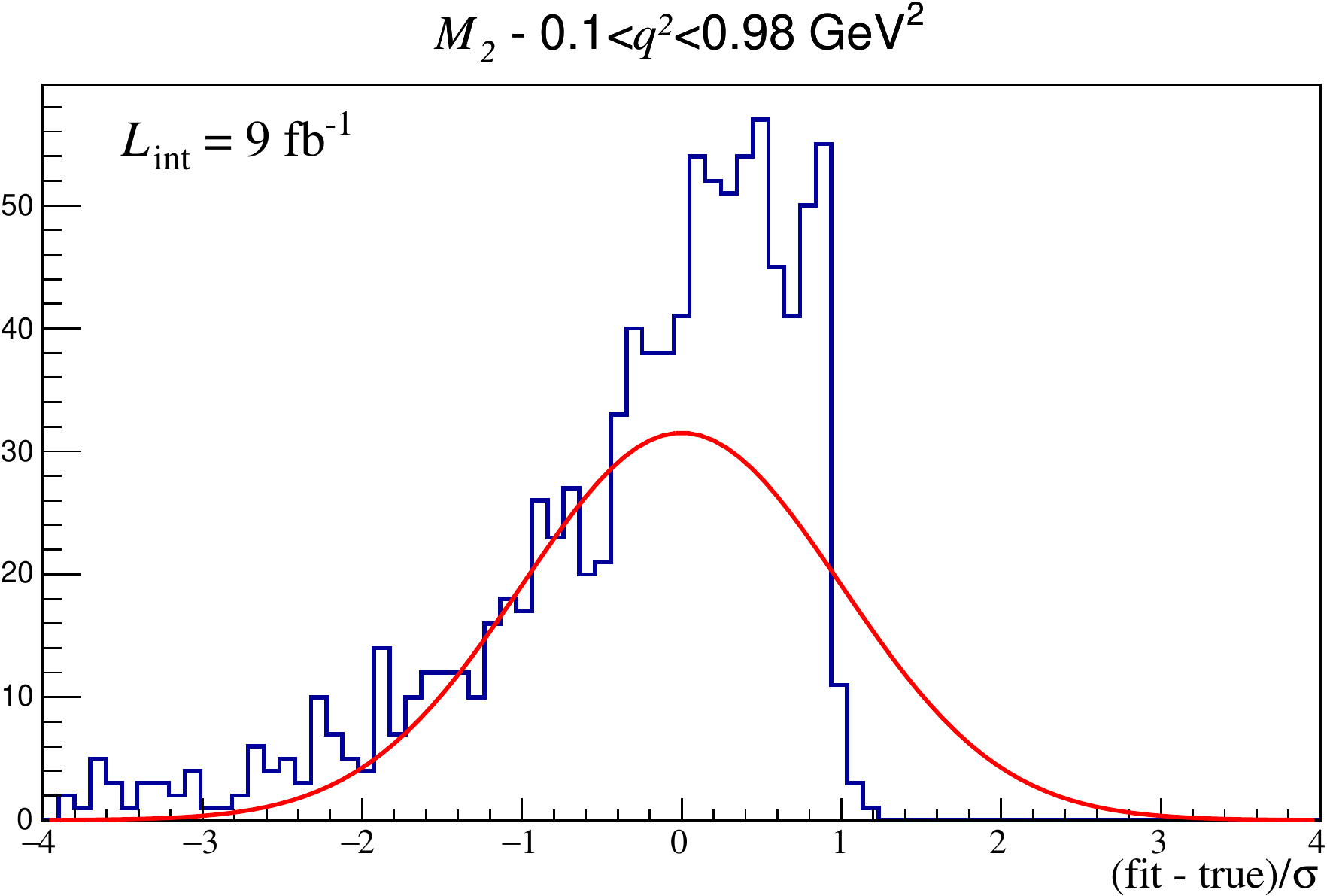}
    \caption{The pull distributions for the optimised P-wave observables (left) $M_{1}$ and (right) $M_2$ from pseudoexperiments with the estimated LHCb Run~2 yields. The overlaid red curve is a reference of the ideal distribution for the pulls, a Gaussian function with $\sigma=1$.}
    \label{fig:m1_m2_run2}
\end{figure}

\begin{figure}[htp]
    \centering
    \includegraphics[width = 0.48\textwidth]{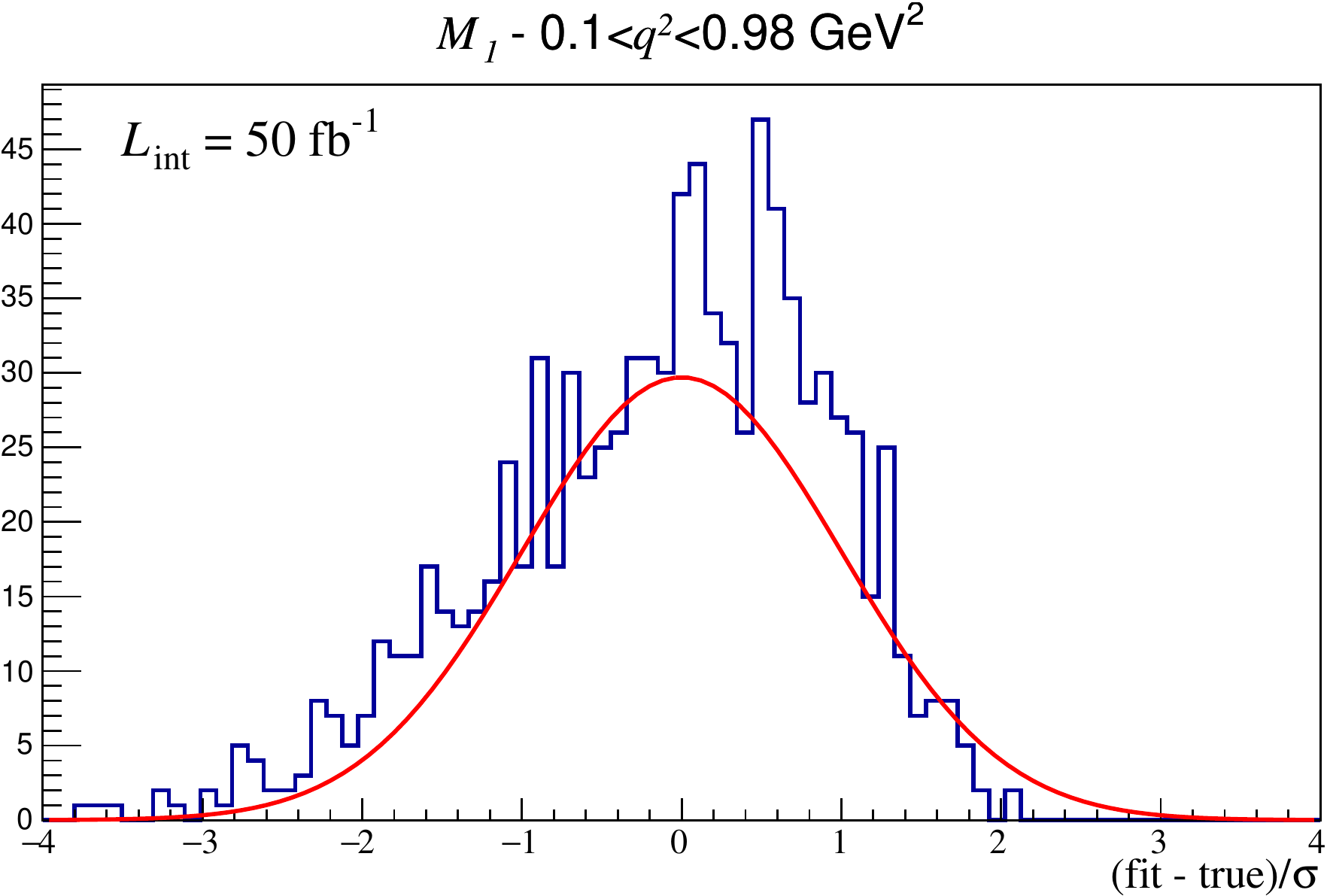}
    \includegraphics[width = 0.48\textwidth, clip]{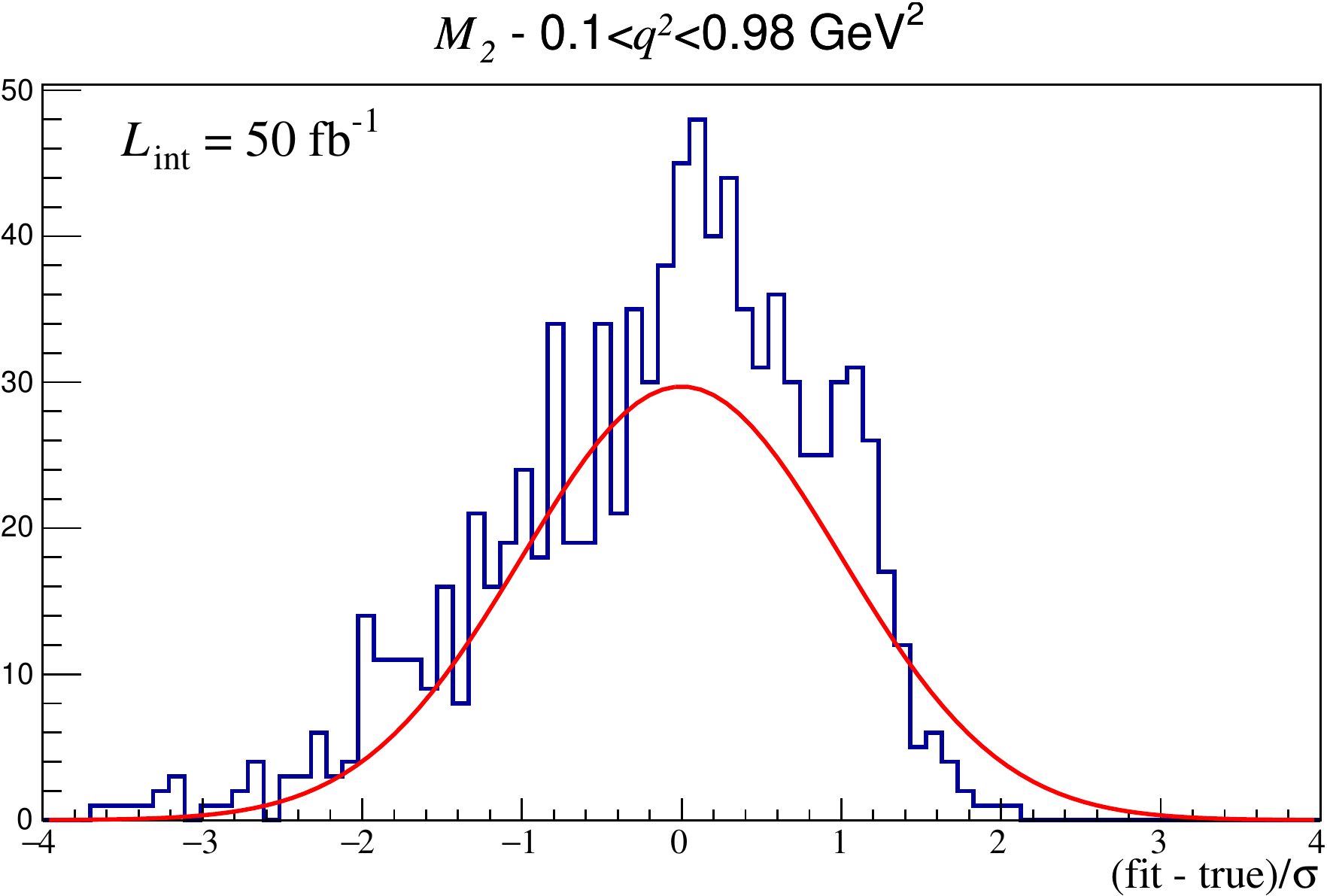}
    \caption{The pull distributions for the optimised P-wave observables (left) $M_{1}$ and (right) $M_2$ from pseudoexperiments with the expected LHCb Upgrade~I yield from $50\invfb$ integrated luminosity. The overlaid red curve is a reference of the ideal distribution for the pulls, a Gaussian function with $\sigma=1$.}
    \label{fig:m1_m2_run4}
\end{figure}

For the new optimised P-wave observables in the massive lepton \qsq bin, $M_{1}$ and $M_{2}$, good behaviour is only obtained with the integrated luminosities expected from the LHCb upgrade. These observables are problematic for the fits as they are essentially the ratio of two angular coefficients with the same \mkpi dependence. Therefore they are almost completely anti-correlated and the fit struggles to converge, as shown in Fig.~\ref{fig:m1_m2_run2}.  However, with enough data the fit will improve and even by the end of LHCb Upgrade~I reasonable behaviour for these observables can be expected, as shown in Fig.~\ref{fig:m1_m2_run4}.

\begin{figure}[htp]
    \centering
    \includegraphics[width = 0.48\textwidth]{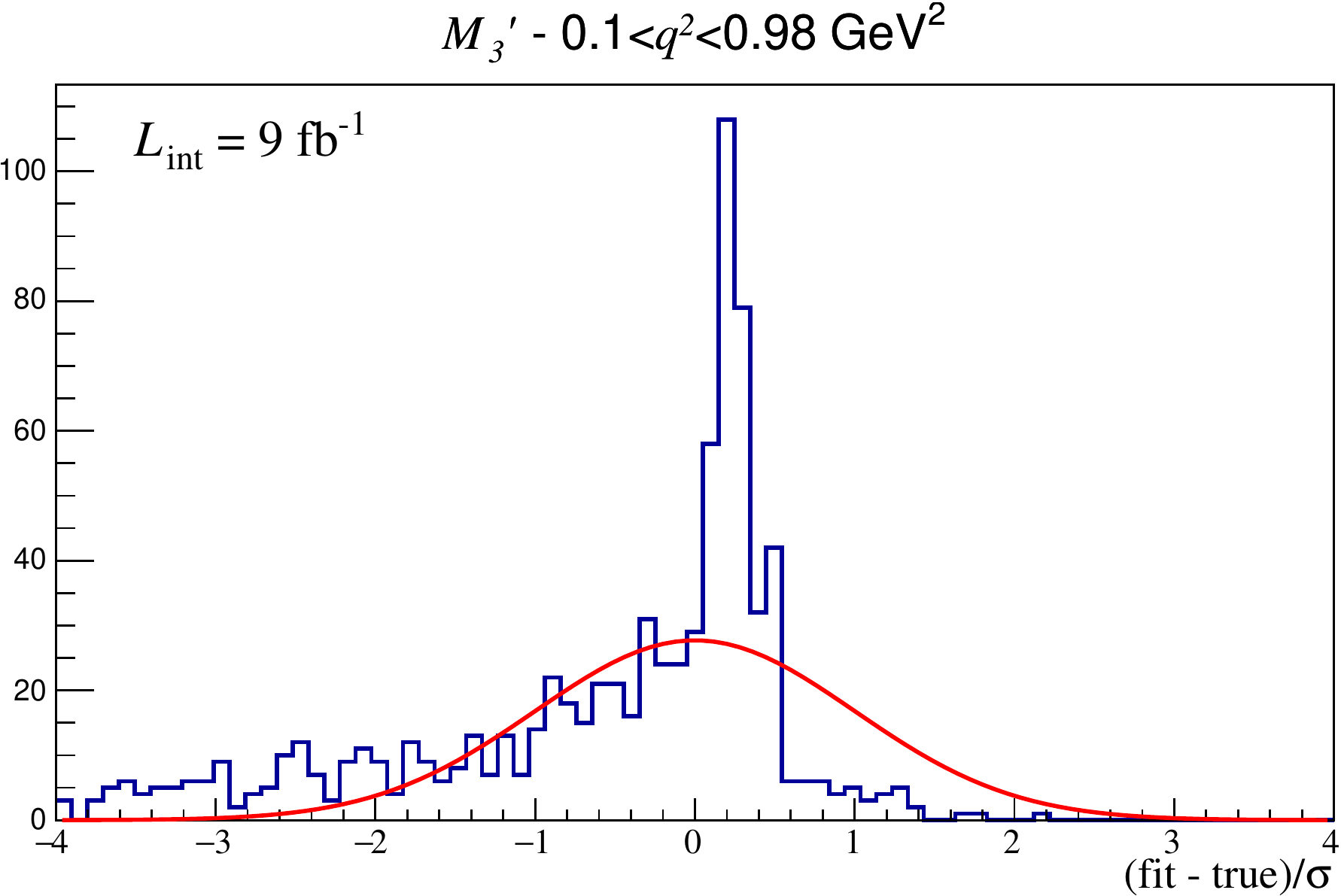}
    \includegraphics[width = 0.48\textwidth]{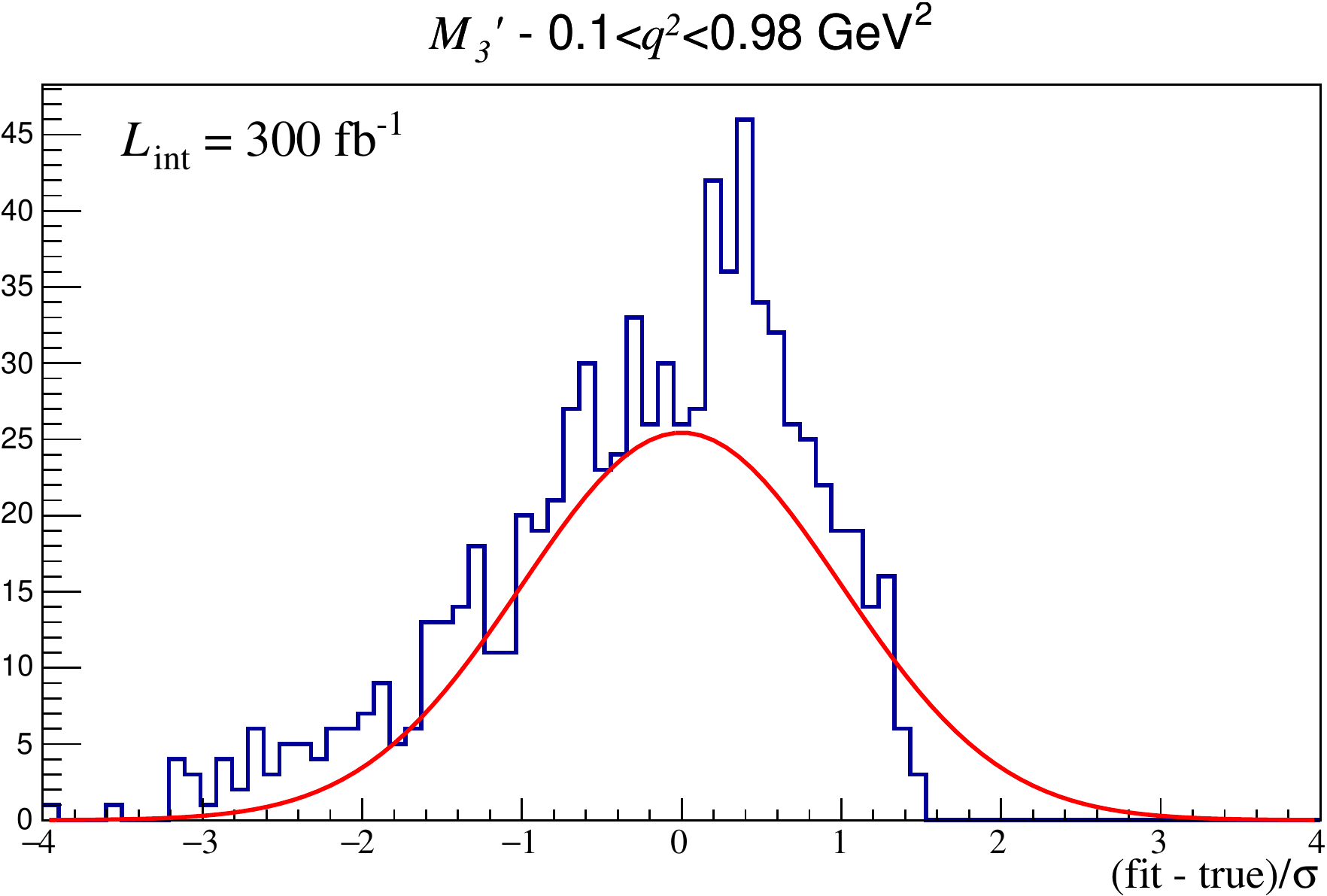}    
    \caption{The pull distributions for the optimised S-wave observable $M_{3}'$ from pseudoexperiments with (left) the expected LHCb Run~2 yields and (right) the yields estimated after 300\invfb integrated luminosity.}
    \label{fig:m3p}
\end{figure}

For similar reasons the S-wave only optimised observable $M_{3}'$ is poorly behaved. Due to the small S-wave contribution an even larger data set, such as the 300\invfb expected with LHCb Upgrade~II, would be required for its successful extraction. This is shown in Fig.~\ref{fig:m3p}.

\begin{figure}[htp]
    \centering
    \includegraphics[width = 0.48\textwidth, clip]{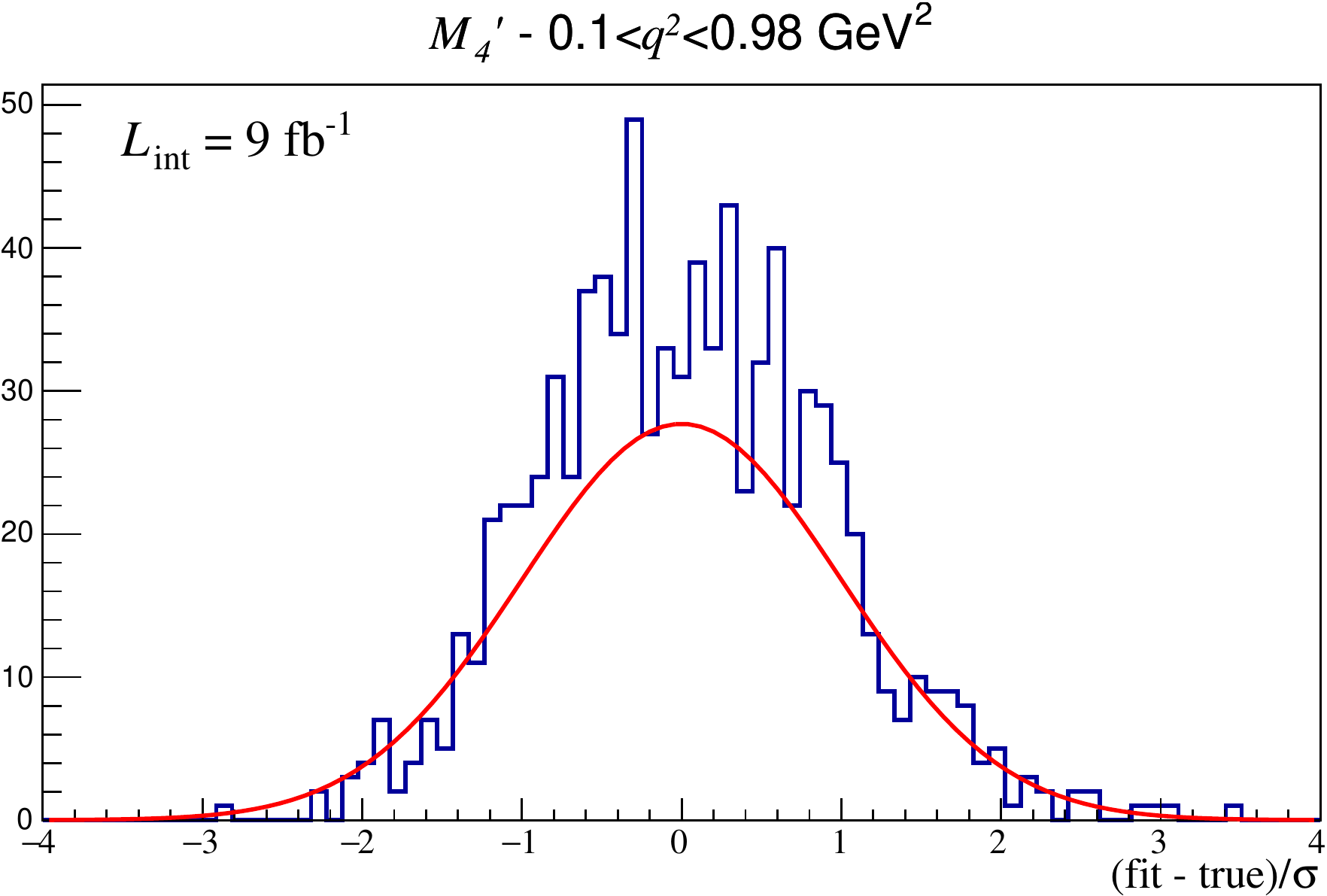}
    \includegraphics[width = 0.48\textwidth, clip]{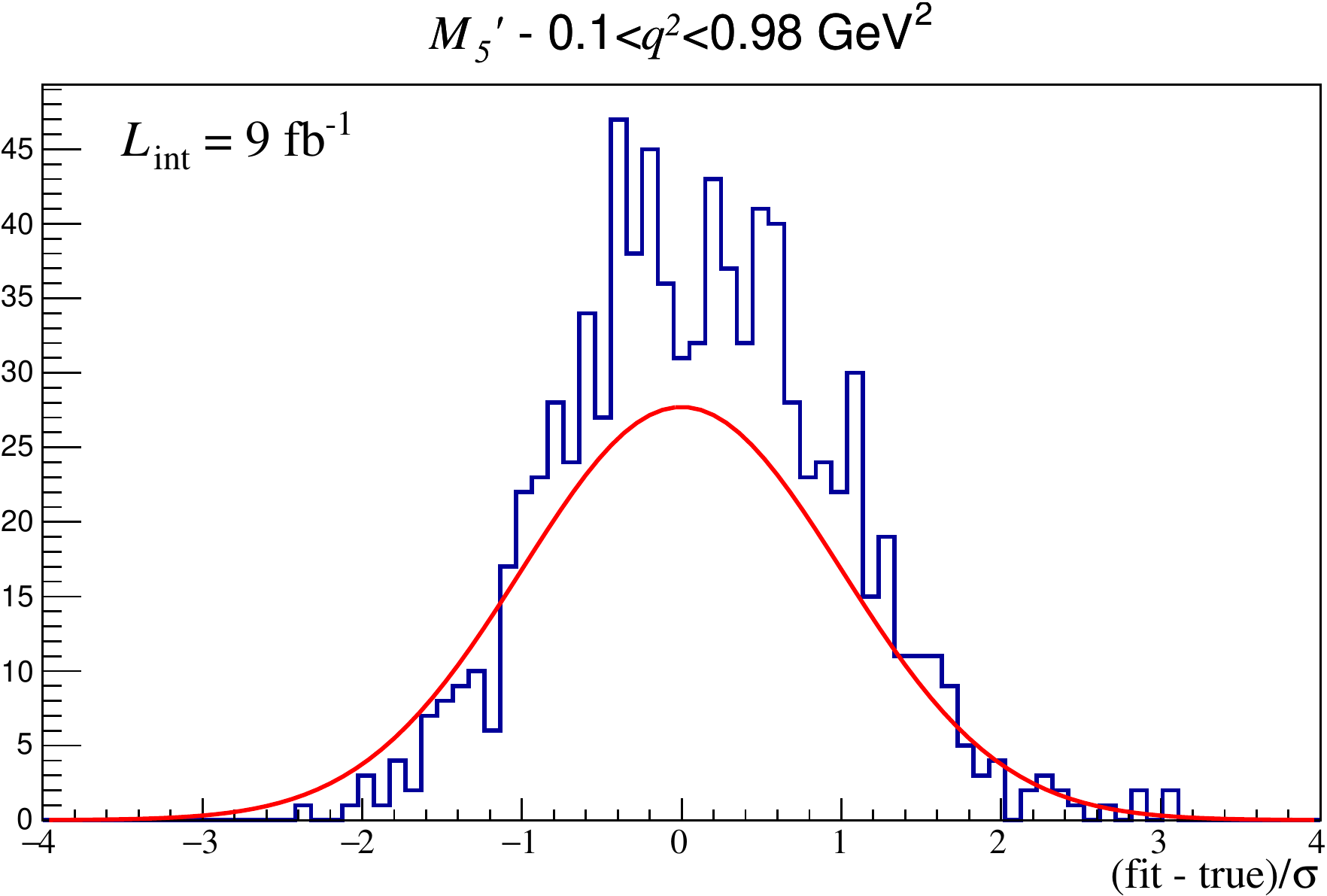}
    \caption{The pull distributions for the optimised P-wave observables (left) $M_{4}'$ and (right) $M_{5}'$ from pseudoexperiments with the expected LHCb Run~2 yields. The overlaid red curve is a reference of the ideal distribution for the pulls, a Gaussian function with $\sigma=1$.}
    \label{fig:m4p_m5p}
\end{figure}

Finally the optimised P- and S-wave interference observables that occur when accounting for the lepton mass, $M_{4}'$ and $M_{5}'$, have been considered. These can be extracted with the Run~2 data set as displayed in Fig.~\ref{fig:m4p_m5p}. The observables are not straightforward ratios of two other observables, which lessens the correlations in the fit. Furthermore they are functions of P-wave and S-wave observables, which are likely to be well constrained by the rest of the angular PDF; in particular they have different \mkpi shapes.

For a judicious choice of observable quantities, future experimental analyses should therefore be able to use the full angular distribution, including both the additional S-wave terms, and assuming massive leptons. 

\subsubsection{Results with a possible scalar amplitude}
\label{sec:experiment:massive:scalar}
\begin{figure}[htp]
    \centering
    \includegraphics[width = 0.99\textwidth]{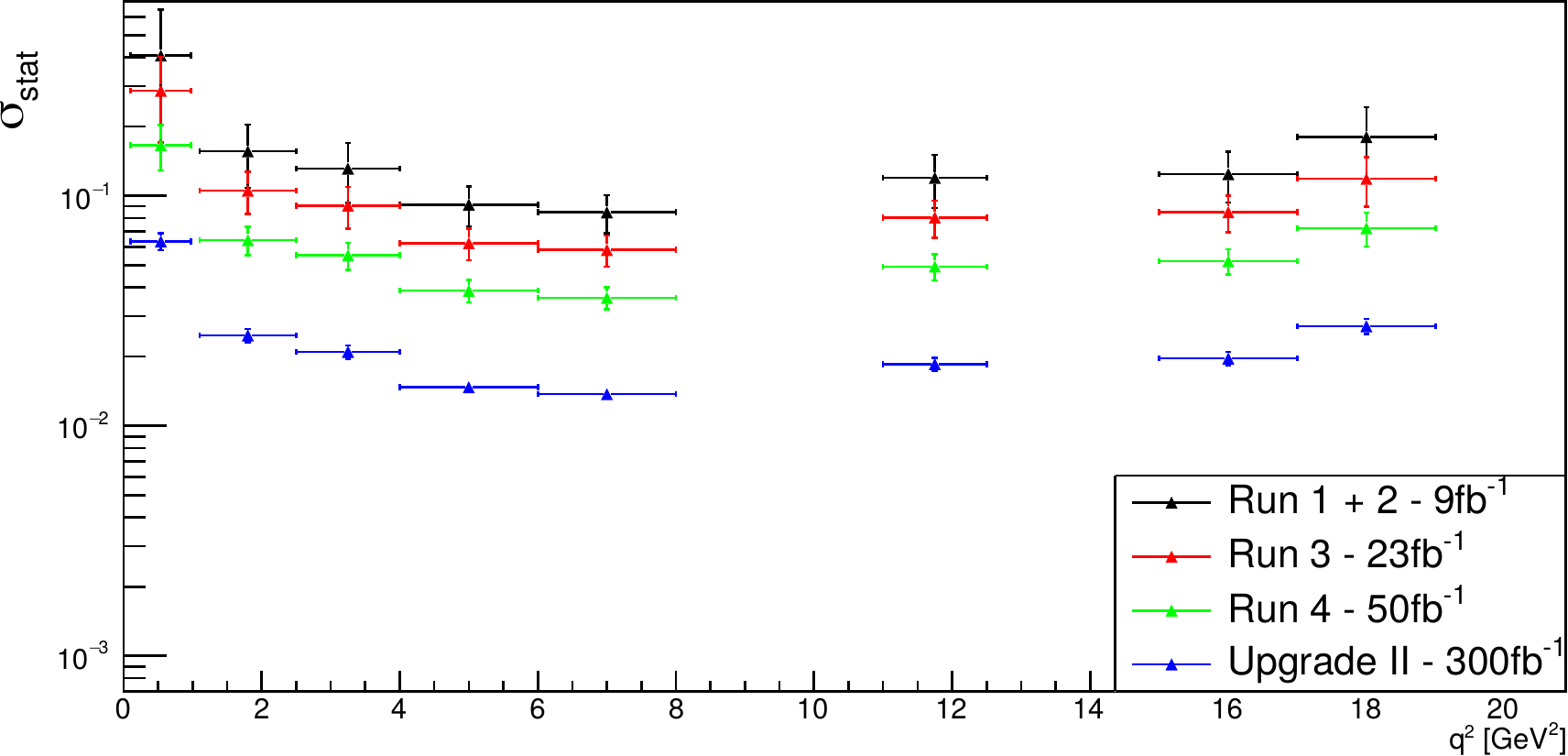}
    \caption{The expected statistical uncertainty of the observable $M_{2}$ as a function of \qsq for various future integrated luminosities.}
    \label{fig:m2_sensitivity}
\end{figure}

If one wanted to fit the data without assuming the absence of scalar amplitudes one must introduce the observables $S_{6}^{c}$ and $S_{1}^{c}$ in all bins. For maximal theoretical reach $S_{1}^{c}$ would ideally be replaced with its optimised equivalent $M_{2}$
(see the discussion in section~\ref{sec21}). The precision on this has been estimated for various future integrated luminosity scenarios, as shown in Fig.~\ref{fig:m2_sensitivity}. Even with 300\invfb of data, the expected statistical uncertainty is much larger than that required to measure significant scalar new physics. 

\subsection{Symmetry relations}
\label{sec:experiment:relations}

\begin{figure}
    \centering
    \includegraphics[width = 0.48\textwidth, page = 1]{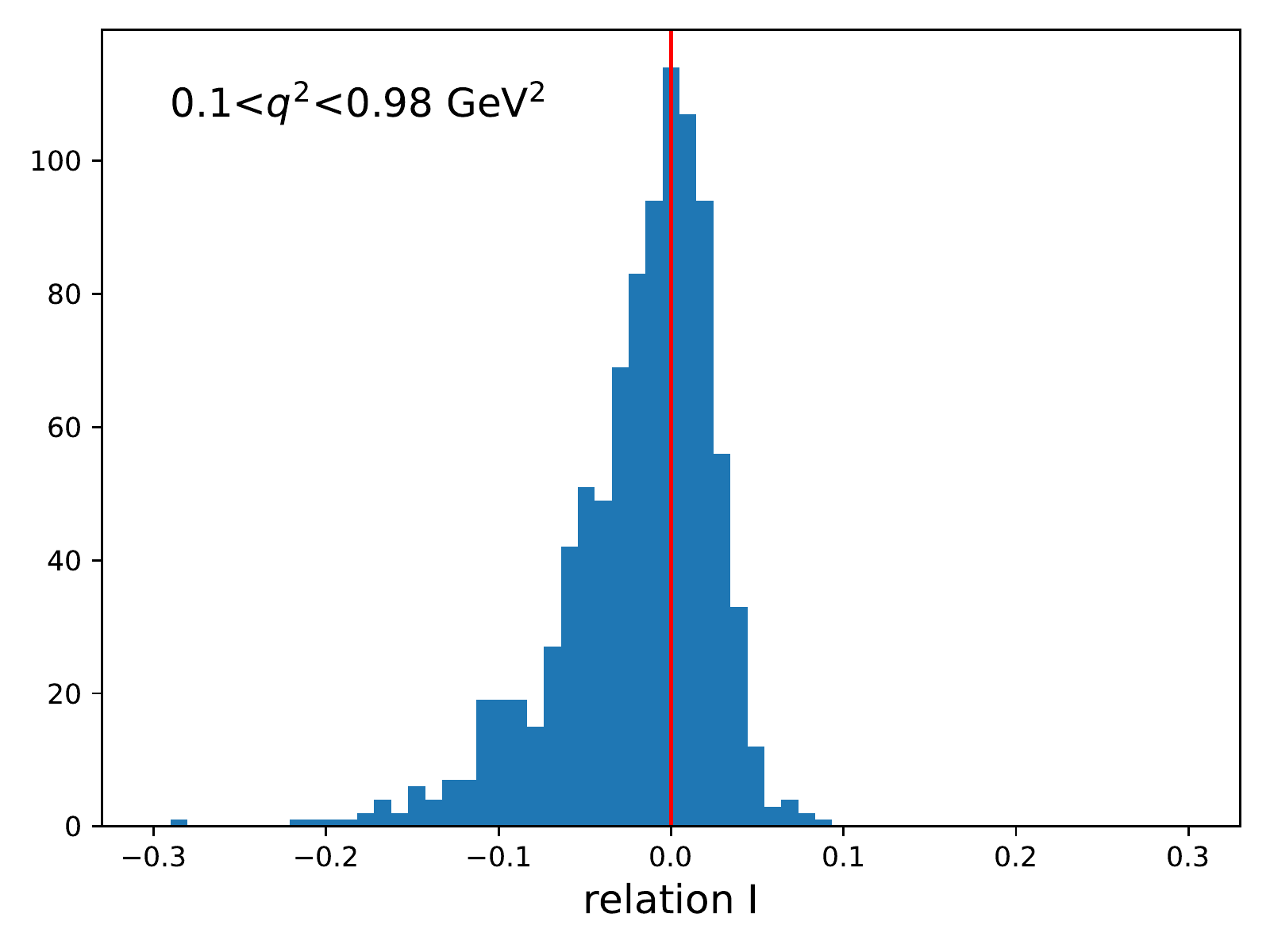}
    \includegraphics[width = 0.48\textwidth, page = 2]{experimental_figs/relations_paper.pdf}\\
    \includegraphics[width = 0.48\textwidth, page = 3]{experimental_figs/relations_paper.pdf}
    \includegraphics[width = 0.48\textwidth, page = 4]{experimental_figs/relations_paper.pdf}\\
    \includegraphics[width = 0.48\textwidth, page = 5]{experimental_figs/relations_paper.pdf}
    \includegraphics[width = 0.48\textwidth, page = 6]{experimental_figs/relations_paper.pdf}
    \caption{Example distributions of the six symmetry relations for the various \qsq bins. The red line is a reference at 0 for the case when the relations are exact. The spread of the distributions is a reflection of the statistical precision of the fit.}
    \label{sec:experiment:relations:fig:summary}
\end{figure}

The six symmetry relations may be applied to the results of the binned fits as an independent check of the robustness of the experimental methodology. As the fitted observables are averaged over a \qsq bin the relations are not exact in this experimental context. This is particularly apparent in the lowest \qsq bin, where the changes in the variables with \qsq are most notable. Furthermore, as only the bins for $\qsq<1\gevgev$ are treated as having massive leptons there is some small imprecision in the symmetry relations for the bins immediately above $1\gevgev$ due to residual effects of the massless lepton treatment. Example distributions of the relations are shown in Fig.~\ref{sec:experiment:relations:fig:summary}.

These distributions of the symmetry relations may be used for a ready check by an experimenter of their fit to real data. If the relation calculated from the data lies outside these distributions the fit can be discounted and the experimenter invited to check their method. Care must be taken however as the experimental relations are calculated with \qsq averaged observables. This introduces some model dependence in the distributions of the pseudo-experiments.

\begin{figure}
    \centering
    \includegraphics[width = 0.45\textwidth, page = 1]{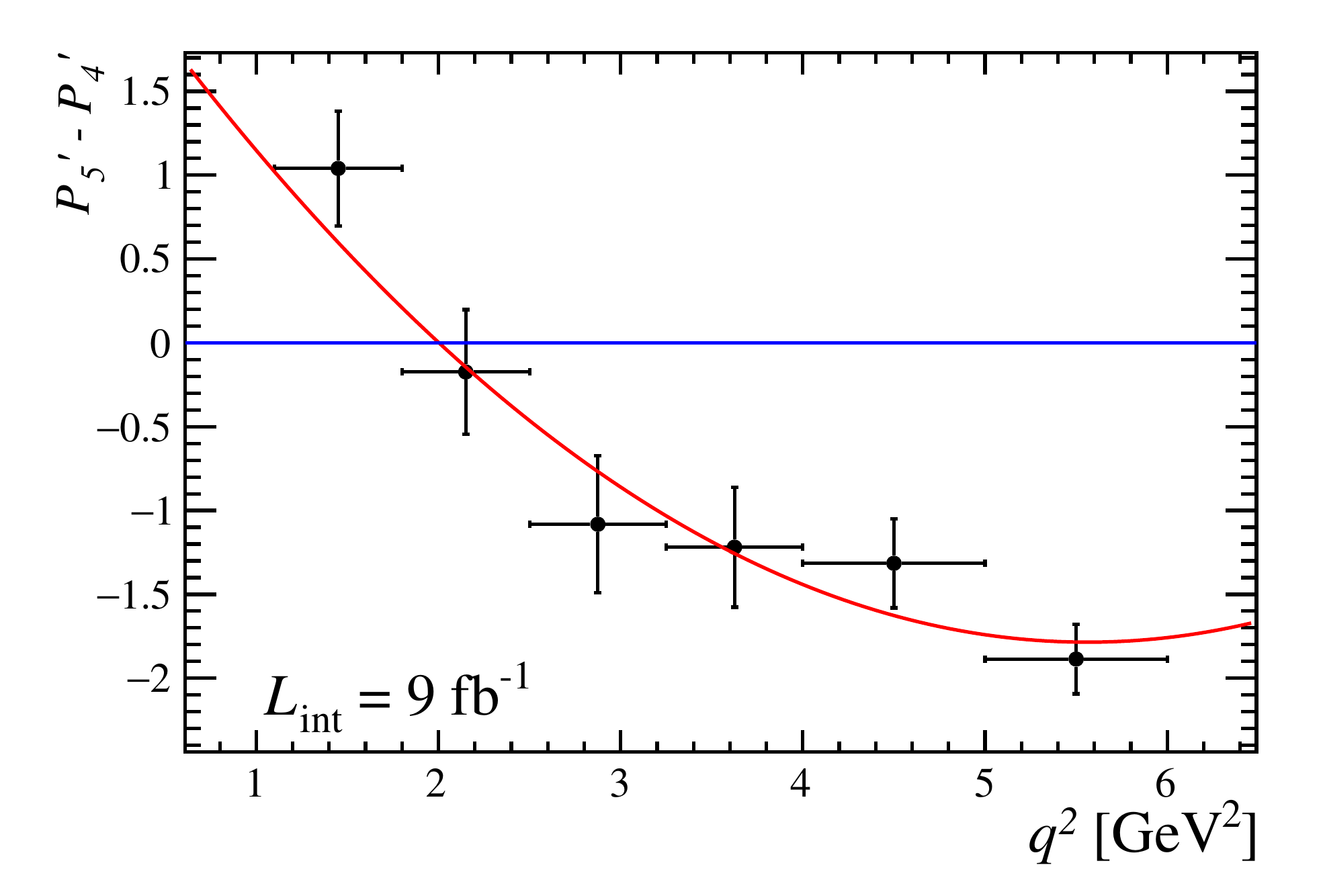}
    \includegraphics[width = 0.45\textwidth, page = 2]{experimental_figs/zcp_narrow_SM_run2_fits.pdf}
    \includegraphics[width = 0.45\textwidth, page = 3]{experimental_figs/zcp_narrow_SM_run2_fits.pdf}    
    \caption{Example distributions of three observables in \qsq. Overlaid in red is the result of fits of second polynomials with a common zero between the three observables. The blue line is a reference for 0. The observables are taken from pseudo-experiment fits for the estimated LHCb Run~2 yields. The value of $\beta$ is taken to be 1.}
    \label{sec:experiment:zcp:fig:q2_fits}
\end{figure}

\subsection{Zero points}
\label{sec:experiment:zcp}
The zero points of the observables $X_{2}$, $X_{3}$ and $X_{4}$ provide a good test of the SM. From the experimental results they are found by taking the independent fit results from each \qsq bin for the relevant observables. The three observables are plotted in \qsq and a $\chi^{2}$ fit of second-order polynomials is carried out simultaneously for each observable. The point at which the polynomials are zero is a common fit parameter. The correlations between the fitted observables within a \qsq bin are included in the $\chi^{2}$ fit.

As the \qsq dependence of the observables is of most interest, it makes sense to employ the half-sized binning, doubling the number of \qsq points. The fits are found to behave well in these finer \qsq bins with the expected yield for the LHCb Run~2 data set for those variables of interest.

\begin{figure}
    \centering
    \includegraphics[width = 0.48\textwidth, page = 1]{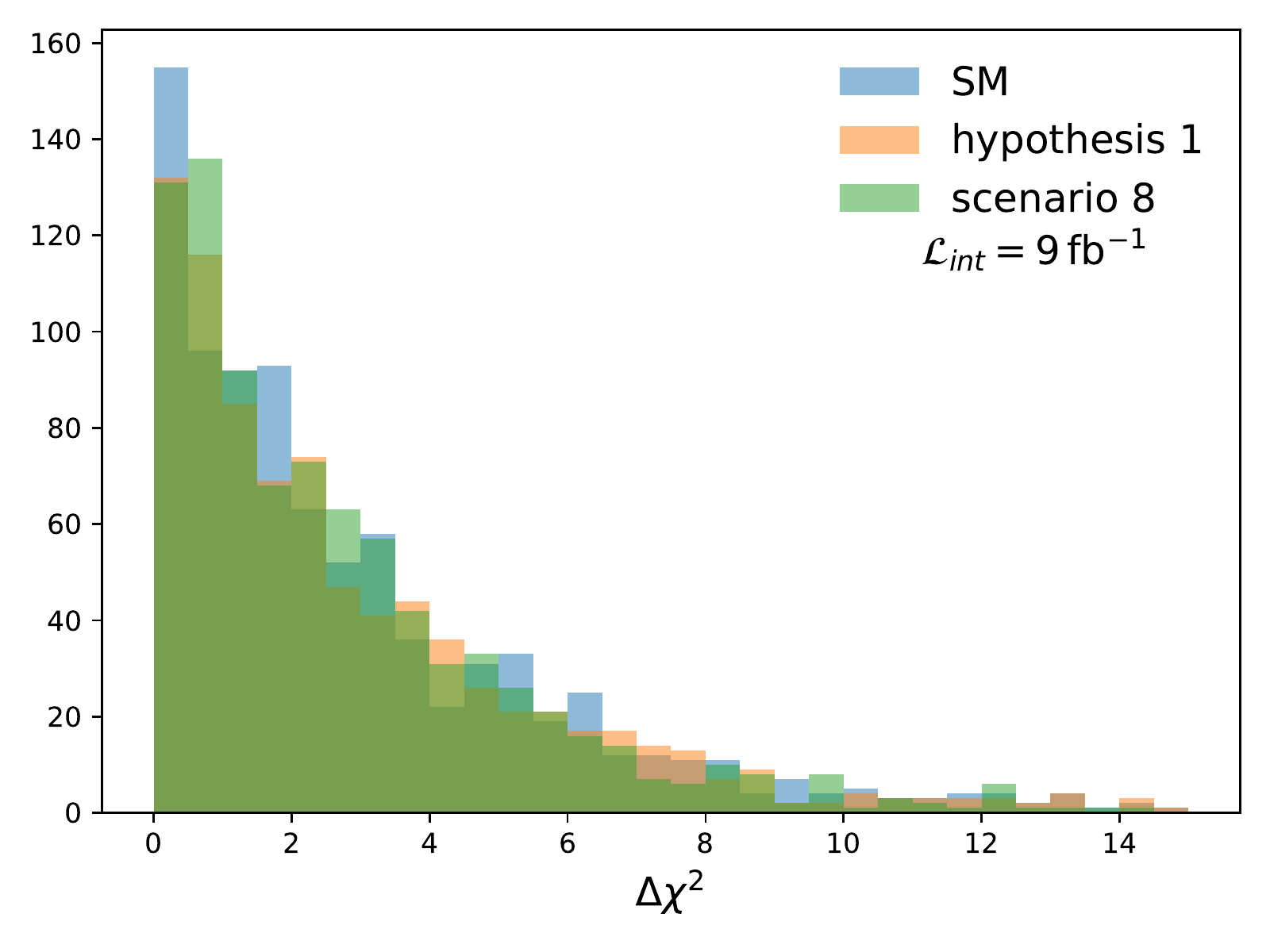}
    \includegraphics[width = 0.48\textwidth, page = 2]{experimental_figs/zcp_plots_paper.pdf}
    \caption{$\Delta\chi^{2}$ distributions for three physics scenarios for (left) 9\invfb and (right) 300\invfb of data.}
    \label{sec:experiment:zcp:fig:delta_chi2}
\end{figure}

Example fits of the \qsq distributions for the expected LHCb Run~2 data set are shown in Fig.~\ref{sec:experiment:zcp:fig:q2_fits}. Alternative fits are performed with no common zero between the observables and the change in $\chi^{2}$ determined in order to test the hypothesis of a common zero crossing point. Three hypotheses have been tested: the SM and two NP models from the fits to current experimental results in Ref.~\cite{quim_moriond}. The two NP scenarios are: i) `Scenario 8', which corresponds to only left-handed new physics and includes a LFU new physics contribution; and ii) `Hypothesis 1', which introduces right handed currents that do not satisfy condition$_R$ (defined by Eq.\eqref{rhcextra2}), and should lead to the three $X$ observables not having a common zero crossing point. See Tab.~\ref{tab1} and  Fig.~\ref{fig:w1w2} for the definitions of these scenarios in terms of the Wilson coefficients.
For each scenario, 900 pseudo-experiments are carried out and the expected $\Delta\chi^{2}$ distributions ascertained. It is found that the distribution is indistinguishable between these three physics simulations with 9\invfb of data, as shown in the left of Fig.~\ref{sec:experiment:zcp:fig:delta_chi2}. With 300\invfb, as displayed in the right of Fig.~\ref{sec:experiment:zcp:fig:delta_chi2}, it can clearly be seen that the $\chi^{2}$ of the fit with a common zero is worse than that for independent zeroes, giving discrimination between Hypothesis 1 and the SM.

\begin{figure}
    \centering
    \includegraphics[width = 0.48\textwidth, page = 3]{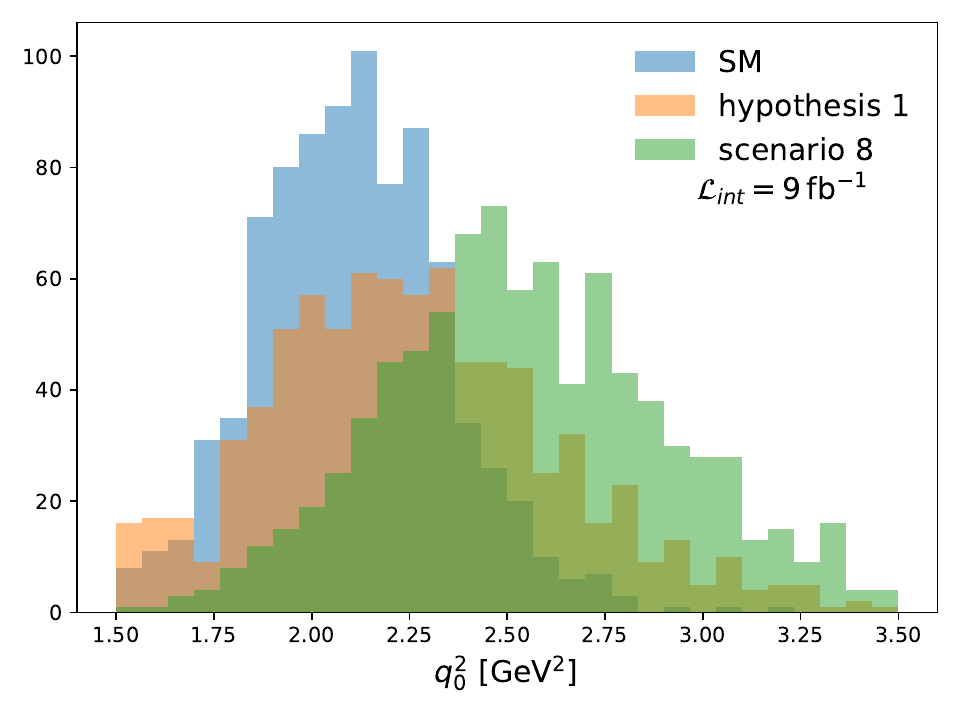}
    \includegraphics[width = 0.48\textwidth, page = 4]{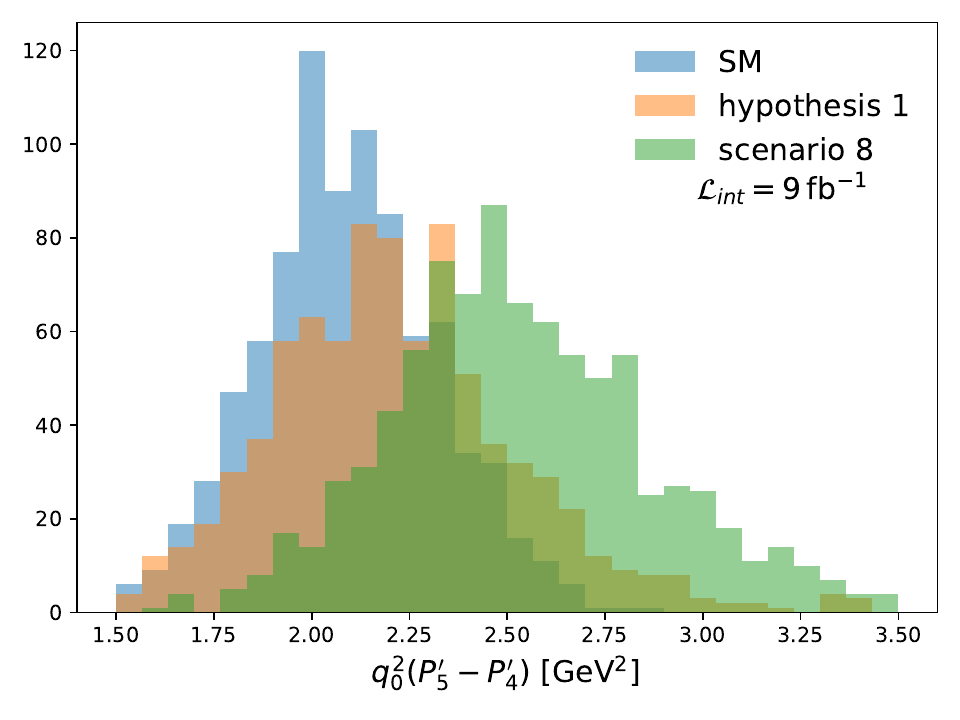}\\
    \includegraphics[width = 0.48\textwidth, page = 5]{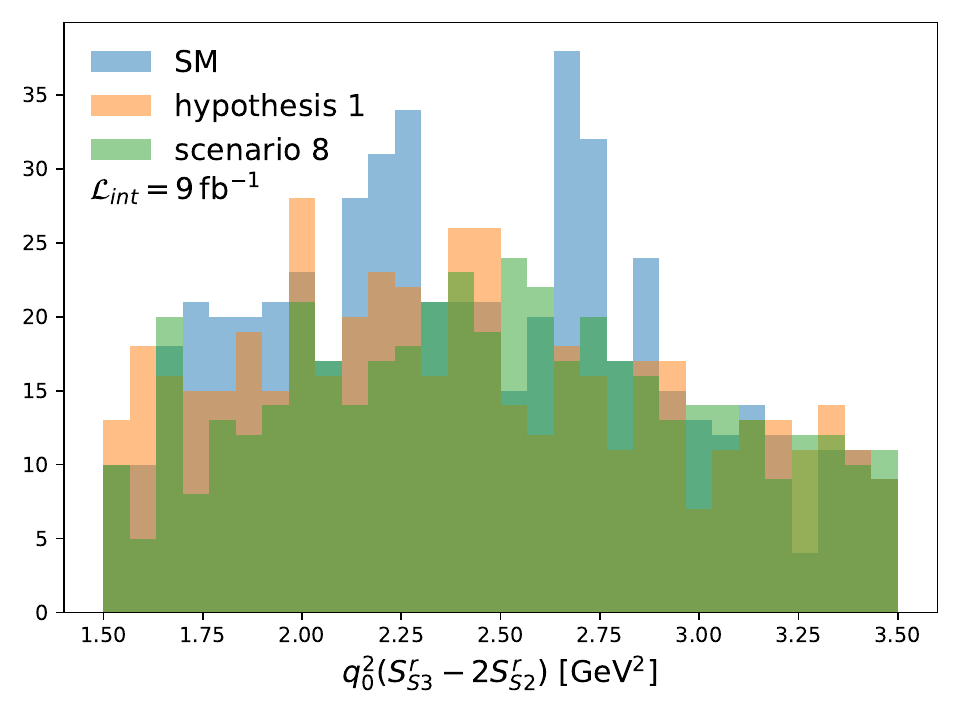}
    \includegraphics[width = 0.48\textwidth, page = 6]{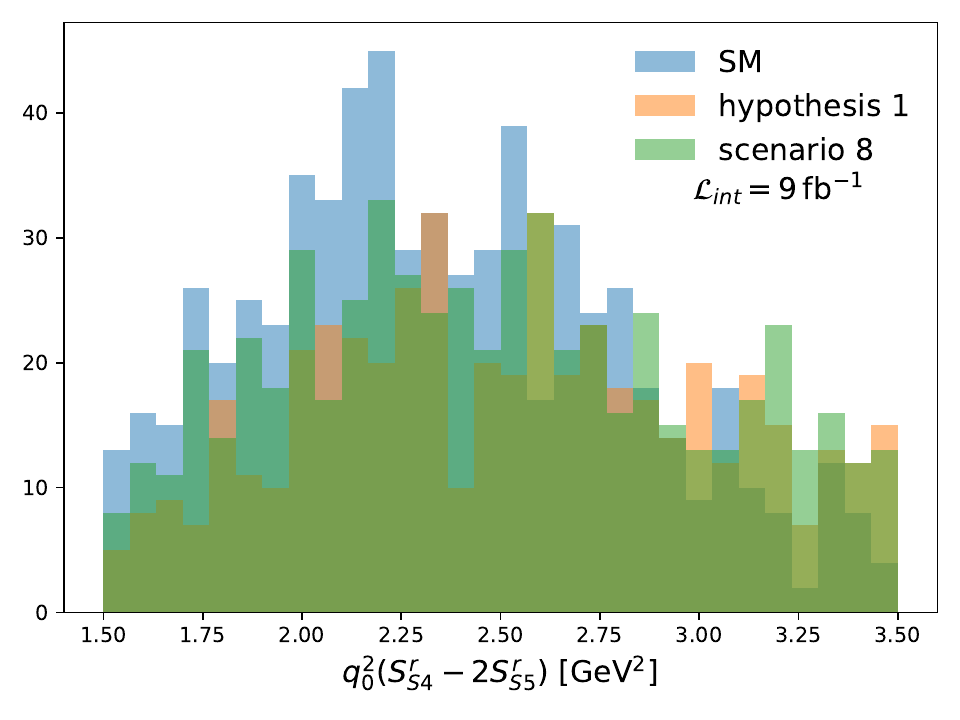}
    \caption{Distributions of measured zero crossing points with 9\invfb of data. Shown are the (top left) common fitted zeros and the independent zeros of (top right) $P_{5}' - P_{4}'$, (bottom left) $S_{S3}^{r} - 2S_{S2}^{r}$ and (bottom right) $S_{S4}^{r} - 2S_{S5}^{r}$.}
    \label{sec:experiment:zcp:fig:zeros}
\end{figure}

\begin{figure}
    \centering
    \includegraphics[width = 0.48\textwidth, page = 7]{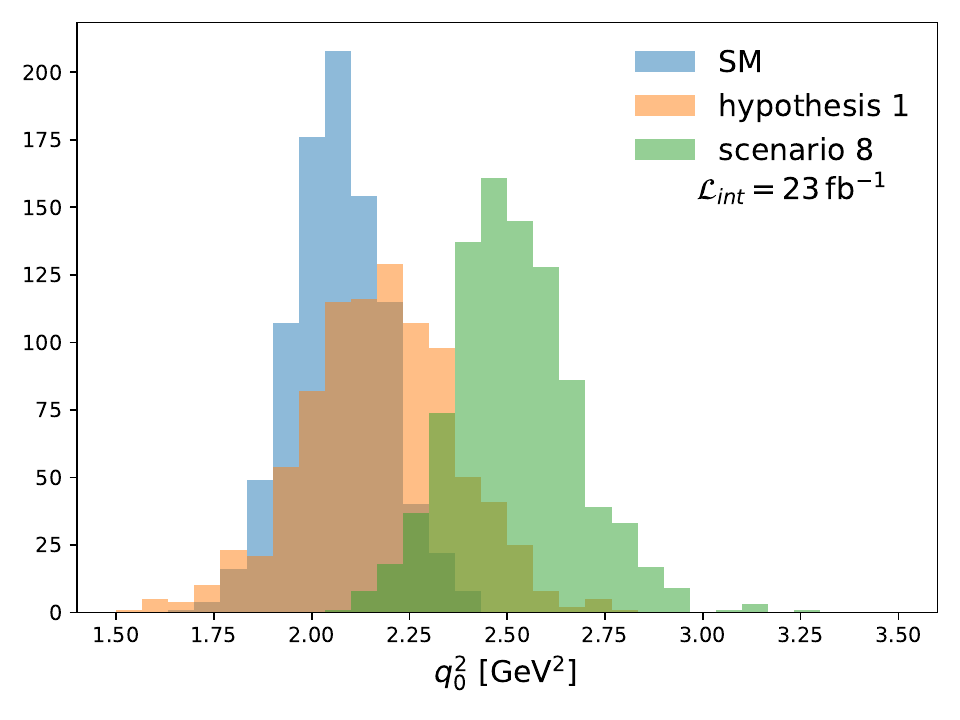}
    \includegraphics[width = 0.48\textwidth, page = 8]{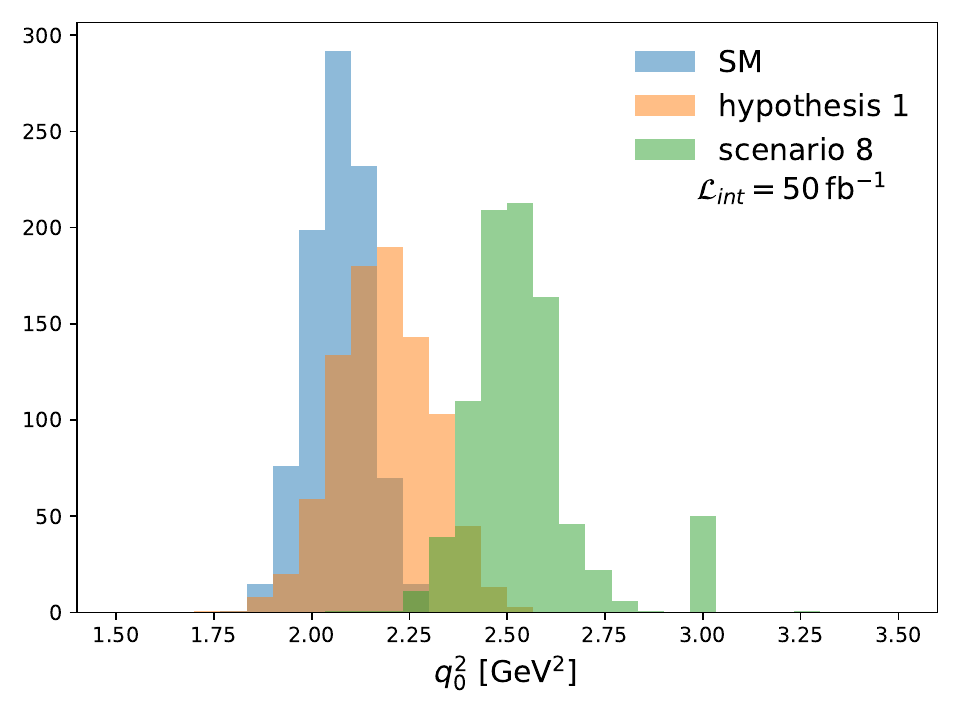}    
    \caption{Distribution of the common fitted zero crossing points with (left) 23\invfb and (right) 50\invfb of data.}
    \label{sec:experiment:zcp:fig:future_zeros}
\end{figure}

Even if the common-zero $\chi^{2}$ fit is unable to distinguish between the three physics hypotheses with the available data, the position of the zero may enable them to be separated. The expected precision on the common zero crossing point with 9\invfb of data is $\sim 0.18\gevgev$, becoming $0.07\gevgev$ with 50\invfb. For comparison, the estimated uncertainty on the zero using the regular \qsq binning is found to be marginally worse: $\approx 0.19\gevgev$ for the Run~2 data set. The uncertainty is completely dominated by the precision with which the $P_{5}' - P_{4}'$ observable is determined. The distribution of measured zeros for the three observables fitted independently is shown in Fig.~\ref{sec:experiment:zcp:fig:zeros}. It is clear that the S-wave interference observables are comparatively imprecise, which is to be expected given their small simulated contributions of $\approx 10\%$. Fig.~\ref{sec:experiment:zcp:fig:zeros} suggests that with the just the Run~2 data set there is little discrimination between the SM and the trialled NP hypotheses from the position of the zero point. However, Fig.~\ref{sec:experiment:zcp:fig:future_zeros} demonstrates that with 50\invfb there is clear distinction between the SM and the scenario 8 NP model.

\subsection{S wave in the global fits}
\label{sec:experiment:global}

Equations~\eqref{eq54} and~\eqref{eq55} allow us to include S-wave interference observables in Wilson coefficient fits for new physics without having to calculate the S-wave form-factors. The expected precision for $W_{1}$ and $W_{2}$ with only the P-wave observables, with the interference observables, and the combination of the two has been assessed. Pseudo-experiments are run with the SM hypothesis and using the new optimised interference observables, $PS_{i}^{r/i}$ introduced in section ~\ref{sobservables}. For each of the 1000 pseudo experiments used, $W_{1}$ and $W_{2}$ are calculated along with their uncertainties, accounting for the correlations between the fitted parameters. The correlation between the expressions involving only P-wave observables and that including the interference observables is assessed for each of $W_{1}$ and $W_{2}$. Subsequently the average and statistical uncertainty when combining the P wave only part with the interference part is found for each observable.

\begin{figure}
    \centering
    \includegraphics[page = 1, width = 0.49\textwidth]{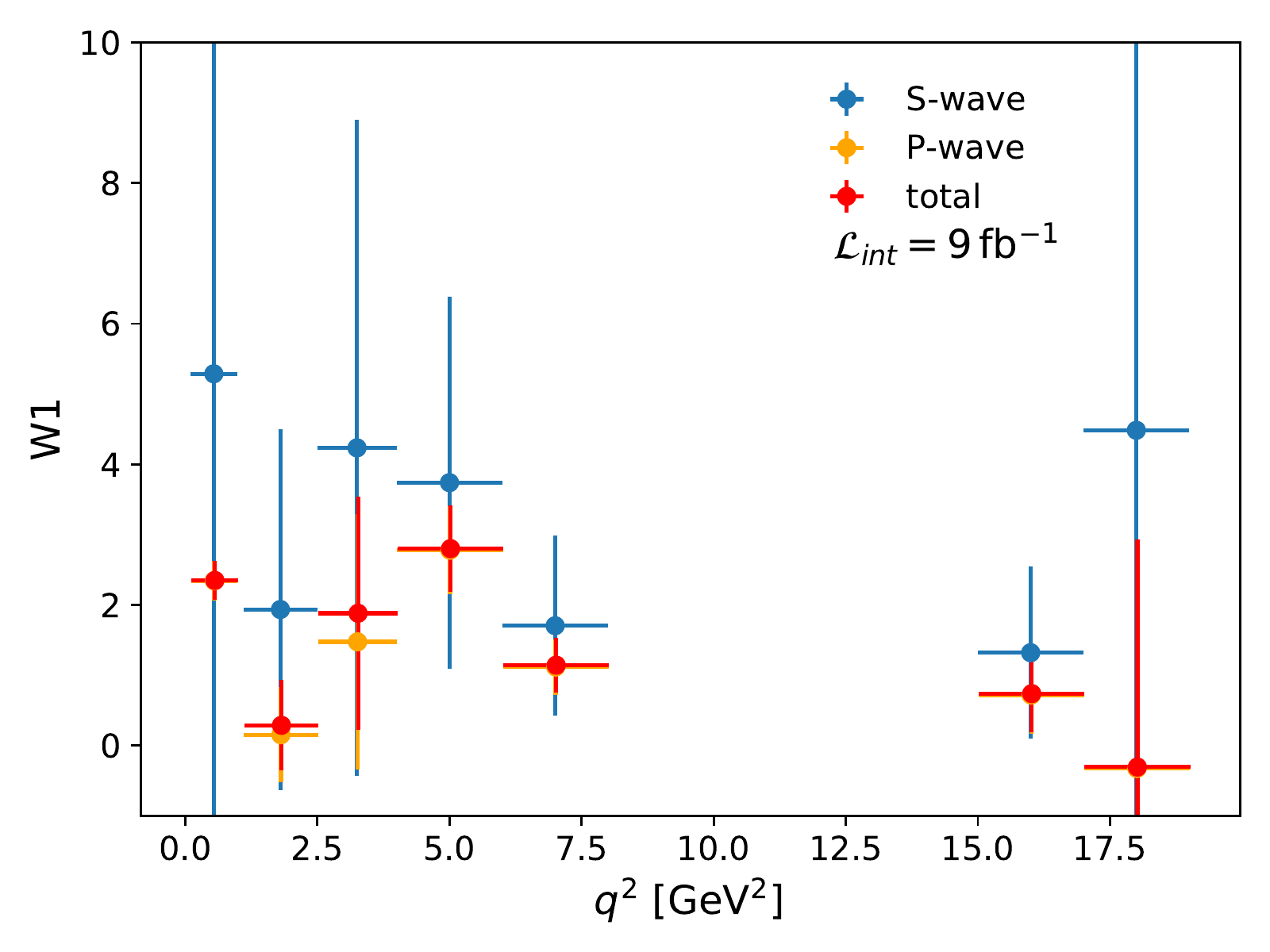}
    \includegraphics[page = 2, width = 0.49\textwidth]{experimental_figs/Wi_observables_wide_bins_run2_results.pdf}
    \caption{Pseudo-experiment results for (left) $W_{1}$ and (right) $W_{2}$ with the LHCb Run~2 data set.}
    \label{sec:experiment:global:fig:run2}
\end{figure}

For the Run~2 data set the narrow bins cannot reliably be used to extract the optimised observables. Therefore here the wider \qsq bins are used. The expected precision of $W_{1}$ and $W_{2}$ is shown in Fig.~\ref{sec:experiment:global:fig:run2}. It can be seen that the combination of P-wave only with the P- and S-wave observables is only marginally more precise than for the P-wave only alone. This is to be expected due to the small contribution of the S wave that is simulated and the presence of P-wave parameters in the combination with the interference observables such that the contribution of the S wave is not statistically independent.

\begin{figure}
    \centering
    \includegraphics[width = 0.49\textwidth, page = 1]{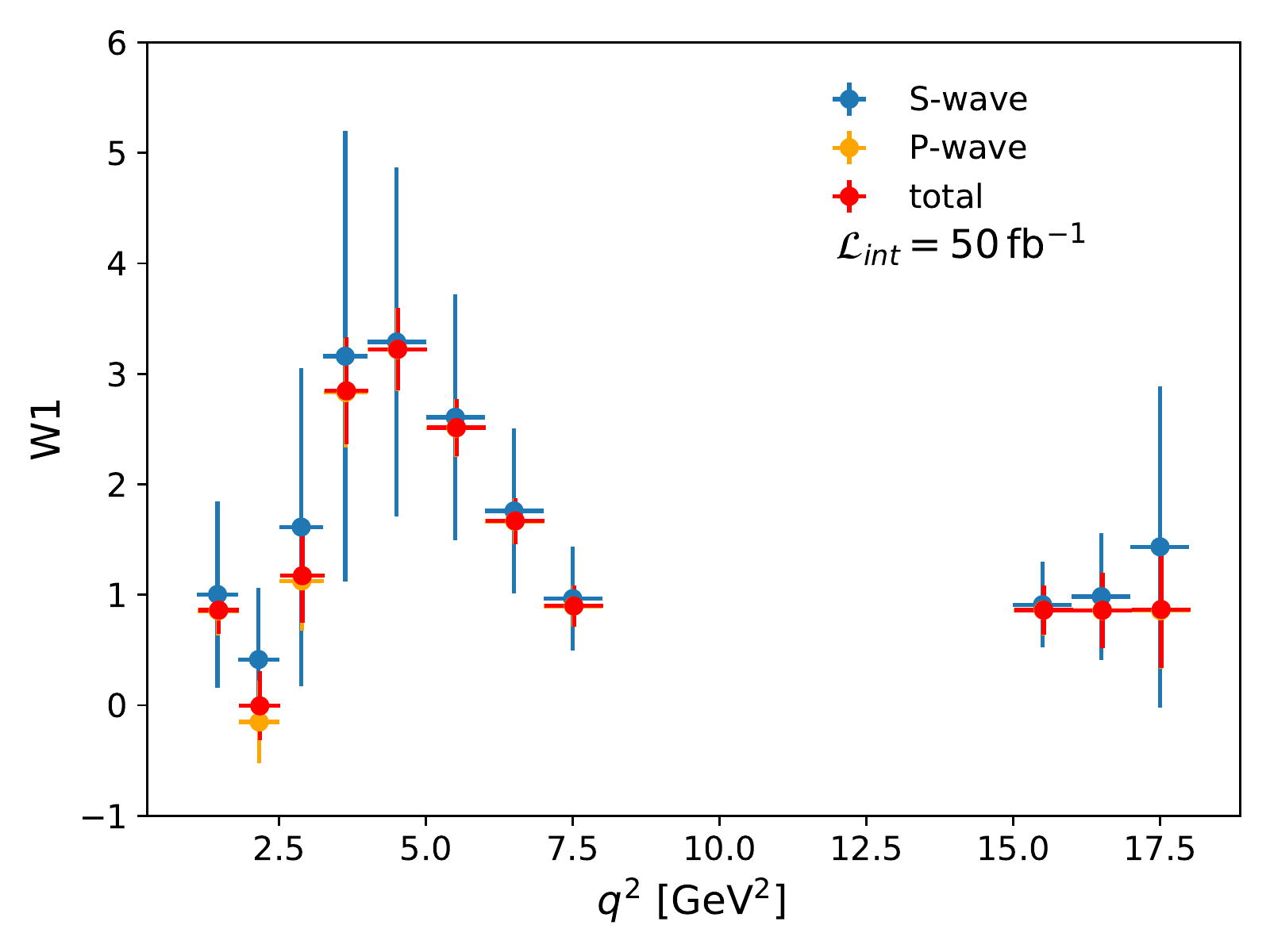}
    \includegraphics[width = 0.49\textwidth, page = 2]{experimental_figs/Wi_observables_narrow_bins_run4_results.pdf}
    \caption{Pseudo-expriment results for (left) $W_{1}$ and (right) $W_{2}$ with the expected LHCb Run~4 data set of 50\invfb.}
    \label{sec:experiment:global:fig:run4}
\end{figure}

In the future the size of the data sets will become sufficient for the narrower bins to be readily used. An example is shown in Fig.~\ref{sec:experiment:global:fig:run4} of the putative LHCb Run~4 data set with 50\invfb.

\section{Summary and Conclusions}
\label{sec:summary}

This paper presents the fully differential decay rate of \BdKpill transitions, with the $\Kp\pim$ system in a P- or S-wave configuration, which can be used to analyse such decays in current and future experiments. This work paves the way for the next step in the analysis of this decay, going beyond previous analyses by identifying and exploring the experimental prospects of massive and S-wave observables that were previously neglected or treated as nuisance parameters. Our analysis relies on a complete description 
 of the symmetries that apply to the full distribution. This enables us to define the complete set of observables that describe the decay and the relations between them, excluding only the presence of NP scalar or tensor contributions. 

Our study shows, in particular, that the symmetries of the \BdKpill decay rate give rise to relations 
that allow a combination of S-wave observables, $W_{1,2}$, to be expressed in terms of P-wave only observables. These combined observables then have no dependence on the poorly known S-wave form factors and therefore offer genuine probes of physics beyond the SM. This opens a new seam in the  phenomenology and, for the first time, will allow S-wave events in the data to contribute to global fits for the underlying physics coefficients.

We also present strong bounds on the set of new S-wave observables using two different methods, the relations themselves and Cauchy-Schwartz inequalities relying only on the structure  of the observables in terms of 2D complex vectors. They serve as important experimental cross-checks.

From the point of view of experimental analyses, it has been shown that all of the P- and S-wave angular observables for the $\Bz\to\Kp\pim\mumu$ decay may be extracted with a five-dimensional fit to the data sample that the LHCb collaboration already has in hand. Our analysis includes the complete description of the \mkpi dependence of the differential decay rate for the first time, as well as the treatment of the leptons as massive at low values of \qsq. The exploitation of the symmetry relations for the observables will allow an immediate test of the veracity of the fits to data without resorting to theoretical predictions.
Finally, the common zero crossing point of a set of P-wave and S-wave interference observables may contribute to discrimination between the SM and NP independently of global fits, and can offer insight into the hadronic contributions.
\bigskip

 \section*{Acknowledgements}
 This work received financial support from the Spanish Ministry of Science, Innovation and Universities (FPA2017-86989-P) and the Research Grant Agency of the Government of Catalonia (SGR 1069) [MA, JM]. IFAE is partially funded by the CERCA program of the Generalitat de Catalunya. JM acknowledges  the financial support by ICREA under the ICREA Academia programme. PAC, AMM, MAMC, MP, KAP and MS  acknowledge support from the UK Science and Technology Facilities Council (STFC).

\appendix

\section{Appendix: The 7th massive relation}\label{sec:appendix}

In this Appendix we will provide the necessary steps to determine the last 
relation. This relation vanishes in the massless limit and is particularly lengthy. For both reasons, specially the latter, is of limited practical use. Therefore we will present here the steps to derive this relation but will not write it out explicitly. The derivation is based on five steps:

Step 1: Our starting point will be a particular combination of the 2D vectors that will allow us to introduce the structure of the observable $M_1$ for the first time. 
\begin{eqnarray} \label{eq1system}
&\phantom{+}&(n_\|^\dagger n_S + n_\|^\dagger n_S^\prime) \times (n_\|^\dagger n_S - n_\|^\dagger n_S^\prime)\nonumber \\&+&
(n_\perp^\dagger n_S + n_\perp^\dagger n_S^\prime) \times (n_\perp^\dagger n_S^\prime
-n_\perp^\dagger n_S)=
+4 (A_\|^{L*} A_\|^R + A_\perp^{L*} A_\perp^R) A_0^{\prime L} A_0^{\prime R*}
\end{eqnarray}

In order to avoid repeating the coefficient $4 m_\ell^2/q^2$ of $M_1$, we introduce a reduced version, that we will call $m_1$ defined by
 \begin{equation}  {m_1}=\frac{q^2}{4 m_\ell^2 \beta_\ell^2} (\beta^2_\ell J_{1s}-(2+\beta_\ell^2) J_{2s})=
   {\rm Re}(A_\perp^L A_\perp^{R*}+A_{\|}^L A_{\|}^{R*})\,. \end{equation}

We will use the freedom given by the symmetry (see section~\ref{sec3}) to choose the phase such that $A_0^{\prime L}$ has only a real component. Then we solve Eq.(\ref{eq1system}) for $m_1$ and its imaginary counterpart:
\begin{eqnarray} \label{m1}
{m_1}&=&\frac{-b\, {\rm Im}[A_0^{\prime R}]+a\, {\rm Re}[A_0^{\prime R}]}{4 |A_0^{\prime R}|^2 {\rm Re}[A_0^{\prime L}]} 
 \\
{ {\rm Im}[A_\|^{L*} A_\|^{R}+ A_\perp^{L*} A_\perp^{R}]}&=& \frac{a\, {\rm Im}[A_0^{\prime R}]+b\, {\rm Re}[A_0^{\prime R}]}{ 4|A_0^{\prime R}|^2 {\rm Re} [A_0^{\prime L}]}\quad
\end{eqnarray}
where 
\begin{eqnarray} \label{coefabobs}
a=&+& \frac{1}{6 \beta^4} \left(\frac{4 \Gamma^\prime}{3}\right)^2 \left(-\beta^2[{(S_{S3}^i)}^2+{(S_{S3}^r)}^2 +{(S_{S4}^i)}^2+{(S_{S4}^r)}^2]
\right. \nonumber \\
 &+&4[ ({S_{S2}^i)}^2+{(S_{S2}^r)}^2+{(S_{S5}^i)}^2+{(S_{S5}^r)}^2] \big) 
\nn \\
b= &+&\frac{2}{3\beta^3} \left(\frac{4 \Gamma^\prime}{3}\right)^2 \big( S_{S2}^r S_{S4}^i-S_{S2}^i S_{S4}^r - S_{S3}^r S_{S5}^i  + S_{S3}^i S_{S5}^r \big)
\end{eqnarray}

Step 2: Using $n_0=e n_S + f n_S^\prime$ and multiplying this equation by $\sigma.n_S$, $\sigma.n_S^\prime$ and $\sigma.n_0$, where $\sigma=((0,1),(1,0))$ one can show that all  terms $A_0^{(')L} A_0^{(')R}$ can be written in terms of $A_0^{\prime L} A_0^{\prime R*}$. 
\begin{eqnarray}\label{eqA0sx}
A_0^L A_0^{\prime R*} + A_0^{\prime L} A_0^{R*}&=&2 e A_0^{\prime L} A_0^{\prime R*} \nn \\
-A_0^L A_0^{\prime R*} + A_0^{\prime L} A_0^{R*}&=&-2 f A_0^{\prime L} A_0^{\prime R*} \nn \\
A_0^L A_0^{R*}&=&(e^2-f^2) A_0^{\prime L} A_{0}^{\prime R*}
\end{eqnarray}
where
\begin{eqnarray}
e&=& \frac{ (n_\|^\dagger n_S^\prime) (n_\perp^\dagger n_0) - (n_\|^\dagger n_0) (n_\perp^\dagger n_S^\prime)}{(n_\|^\dagger n_S^\prime) (n_\perp^\dagger n_S)-(n_\|^\dagger n_S) (n_\perp^\dagger n_S^\prime)}
\nn \\
f&=&
\frac{ (n_\|^\dagger n_S) (n_\perp^\dagger n_0) - (n_\|^\dagger n_0) (n_\perp^\dagger n_S)}{(n_\|^\dagger n_S) (n_\perp^\dagger n_S^\prime)-(n_\|^\dagger n_S^\prime) (n_\perp^\dagger n_S)} 
\end{eqnarray}
Both coefficients $e$ and $f$ can be trivially rewritten in terms of P- and S-wave observables, as in Eq.(\ref{coefabobs}).

Step 3: We  define a set of reduced observables related to the corresponding remaining massive observables:
\begin{eqnarray}
{m_2}&=&|A_t|^2+2 {\rm Re}(A_0^L A_0^{R*})
\nonumber \\
{m_3^\prime}&=&|A_t^\prime|^2+2 {\rm Re}(A_0^{\prime L} A_0^{\prime R*})
\nonumber \\
{m_4^\prime}&=&{\rm Re}(A_t^\prime A_t^*)+ {\rm Re}(A_0^{\prime L} A_0^{R*}+A_0^L A_0^{\prime R*})
\nonumber \\
{m_5^\prime}&=&{\rm Im}(A_t^\prime A_t^*)+ {\rm Im}(A_0^{\prime L} A_0^{R*}+A_0^L A_0^{\prime R*})
\end{eqnarray}
We can combine them in one single equation cancelling the dependence on $A_t^{(\prime)}$:
\begin{eqnarray}
({ m_2}- 2 {\rm Re}[A_0^L A_0^{R*}]) ({ m_3^\prime}- 2 {\rm Re}[A_0^{\prime L} A_0^{\prime R*}])
=&\!\!\!+\!\!\!&({ m_4^\prime}-  {\rm Re}[A_0^L A_0^{\prime R*}+ A_0^{\prime L} A_0^{R*}])^2
\nonumber \\
&\!+\!&({ m_5^\prime}-  {\rm Im}[A_0^L A_0^{\prime R*}+ A_0^{\prime L} A_0^{R*}])^2 \quad \quad \quad
\end{eqnarray}
and using Eqs.(\ref{eqA0sx}) we can rewrite this equation in terms of only $A_0^{\prime L} A_0^{\prime R *}$:
\begin{eqnarray}\label{therelation}
({ m_2}- 2 {\rm Re}[(e^2-f^2) A_0^{\prime L} A_0^{\prime R*}]) ({ m_3^\prime}- 2 {\rm Re}[A_0^{\prime L} A_0^{\prime R*}])
=&\!+\!&({m_4^\prime}-  {\rm Re}[
2 e A_0^{\prime L} A_0^{\prime R*}
])^2
\nonumber \\
&\!+\!&({ m_5^\prime}-  {\rm Im}[
2 e A_0^{\prime L} A_0^{\prime R*}])^2\,, \quad \quad \quad
\end{eqnarray}
giving the desired relation but involving $A_0^{\prime L}$ and $A_0^{\prime R}$ amplitudes that still need to be expressed in terms of observables. 

Step 4: 
Using the decomposition $n_\perp=g n_S + h n_S^\prime$ and after determining $g$ and $h$ by multiplying by $n_\perp$ and $n_\|$, we find the following relation:
\begin{equation}
(h^{*2}-g^{*2}) {n_S^\prime}^\dagger n_S= h^* n_\perp^\dagger n_S- g^* n_\perp^\dagger n_S^\prime\,,
\end{equation}
where 
\begin{eqnarray}
g&=&\frac{|n_\perp|^2 (n_\|^\dagger n_S^\prime) - (n_\|^\dagger n_\perp) (n_\perp^\dagger n_S^\prime)}{(n_\|^\dagger n_S^\prime)(n_\perp^\dagger n_S)-(n_\|^\dagger n_S)(n_\perp^\dagger n_S^\prime)}\,, 
\nn \\
h&=&\frac{|n_\perp|^2 (n_\|^\dagger n_S) - (n_\|^\dagger n_\perp) (n_\perp^\dagger n_S)}{(n_\|^\dagger n_S)(n_\perp^\dagger n_S^\prime)-(n_\|^\dagger n_S^\prime)(n_\perp^\dagger n_S)}\,. 
\end{eqnarray}
Then combining the previous equation with the observable $F_S$, one can determine $|A_0^{\prime L}|^2$ and $|A_0^{\prime R}|^2$ (remember that $A_0^{L\prime}$ is taken to be real using the symmetry properties) by solving the system: 
\begin{eqnarray} \label{dif}
|A_0^{\prime L}|^2- |A_0^{\prime R}|^2=\frac{h^* n_\perp^\dagger n_S - g^* n_\perp^\dagger n_S^\prime}{h^{*2}-g^{*2}}=\Delta\,,%
\end{eqnarray}
\begin{eqnarray} \label{sum}
|A_0^{\prime L}|^2+ |A_0^{\prime R}|^2\equiv { F_S \Gamma^\prime}.
\end{eqnarray}
Now we have all the necessary ingredients to arrive at the relation. If we define  \begin{eqnarray}
x&=&{\rm Re}[A_0^{\prime L}] {\rm Re}[A_0^{\prime R}]\,, \nn \\ y&=&{\rm Re}[A_0^{\prime L}] {\rm Im}[A_0^{\prime R}]\,,
\end{eqnarray}
we have two equations in terms of $x$ and $y$ (using Eq.~\ref{m1} and Eqs.~\ref{dif} and~\ref{sum}):
\begin{eqnarray} \label{above}
{ m_1}&=&\frac{-b y + a x}{4(x^2+y^2)} \nn \\
x^2+y^2&=&\frac{1}{4}\left(({ F_S \Gamma^\prime})^2-\Delta^2\right)
\end{eqnarray}
These two equations can be solved to determine $x$ and $y$ in terms of observables.

Step 5: Finally, the last step consists of trivially expressing $A_0^{\prime L}$, $A_0^{\prime R}$ in Eq.(\ref{therelation}) in terms of $x$ and $y$ (all other quantities like the $m_i$ and the coefficients $e$ and $f$ are already direct functions of observables). Then after solving the system   for $x$ and $y$ using Eq.(\ref{above}) insert the result in Eq.(\ref{therelation}) to get a final lengthly expression written entirely in terms of observables.

Notice that in order to relate the reduced observables to the measured massive observables $M_{1,2,3^\prime,4^\prime,5^\prime}$ one needs to multiply the previous relations involving the $m_i$'s  on both sides by factors of $4 m_\ell^2/q^2$. For this reason in particular Eq.(\ref{therelation})   vanishes exactly in the massless limit.

\bibliographystyle{JHEP}
\bibliography{main}

\end{document}